\pdfoutput=1
\documentclass[11pt,twoside,a4paper,cmspaper,final,collab]{cms-tdr}
\def\svnVersion{7d57522}\def\svnDate{2022/05/06}\def\cmsCernNoTag{CERN-EP-2021-192}\def\cmsCernDate{\today}\def\cmsMessage{Published in the Journal of High Energy Physics as \href{http://dx.doi.org/10.1007/JHEP05(2022)091}{\doi{10.1007/JHEP05(2022)091}.}}
\begin{document}\cmsNoteHeader{TOP-21-004}

\providecommand{\cmsTable}[1]{\resizebox{\textwidth}{!}{#1}}
\newlength\cmsTabSkip\setlength{\cmsTabSkip}{1ex}

\newcommand{\ttG}{\ensuremath{\ttbar\PGg}\xspace}
\newcommand{\sqrts}[1][13]{\ensuremath{\sqrt{s}=#1\TeV}\xspace}
\newcommand{\emu}{\ensuremath{\Pepm\PGmmp}\xspace}
\newcommand{\ee}{\ensuremath{\Pep\Pem}\xspace}
\newcommand{\mumu}{\ensuremath{\PGmp\PGmm}\xspace}

\newcommand{\pp}{\ensuremath{\Pp\Pp}\xspace}

\newcommand{\ttH}{\ensuremath{\ttbar\PH}\xspace}
\newcommand{\tW}{\ensuremath{\PQt\PW}\xspace}
\newcommand{\GGtoZZ}{\ensuremath{\Pg\Pg\to\PZ\PZ}\xspace}
\newcommand{\PYTHIAeight}{\ensuremath{\PYTHIA8}\xspace}
\newcommand{\Zjets}{\ensuremath{\PZ\text{+jets}}\xspace}
\newcommand{\ZG}{\ensuremath{\PZ\PGg}\xspace}
\newcommand{\tZq}{\ensuremath{\PQt\PZ\PQq}\xspace}
\newcommand{\tchannel}{\ensuremath{t\text{ channel}}\xspace}
\newcommand{\schannel}{\ensuremath{s\text{ channel}}\xspace}
\newcommand{\abseta}{\ensuremath{\abs{\eta}}\xspace}
\newcommand{\dR}{\ensuremath{\DR}\xspace}
\newcommand{\dEta}{\ensuremath{\Delta\eta}\xspace}
\newcommand{\dPhi}{\ensuremath{\Delta\varphi}\xspace}
\newcommand{\mll}{\ensuremath{m(\Pell\Pell)}\xspace}
\newcommand{\pptottG}{\ensuremath{\pp\to\ttG}\xspace}
\newcommand{\sigSM}{\ensuremath{\sigma_\text{SM}(\pptottG)}\xspace}
\newcommand{\Nleps}{\ensuremath{N_{\Pell}}\xspace}
\newcommand{\Nphotons}{\ensuremath{N_{\PGg}}\xspace}
\newcommand{\Nbjets}{\ensuremath{N_{\PQb}}\xspace}
\newcommand{\dRgl}{\ensuremath{\dR(\PGg,\Pell)}\xspace}
\newcommand{\dRjetl}{\ensuremath{\dR(\text{jet},\Pell)}\xspace}
\newcommand{\dRjetg}{\ensuremath{\dR(\text{jet},\PGg)}\xspace}

\newcommand{\sieie}{\ensuremath{\sigma_{\eta\eta}}\xspace}
\newcommand{\pizerodecay}{\smash{\ensuremath{\PGpz\to\PGg\PGg}}\xspace}
\newcommand{\chIso}{\ensuremath{I_{\text{chg}}}\xspace}
\newcommand{\mZ}{\ensuremath{m_{\PZ}}\xspace}
\newcommand{\dMllZ}{\ensuremath{\abs{\mll-\mZ}}\xspace}
\newcommand{\mllg}{\ensuremath{m(\Pell\Pell\PGg)}\xspace}
\newcommand{\dMllgZ}{\ensuremath{\abs{\mllg-\mZ}}\xspace}
\newcommand{\jetone}{\ensuremath{\mathrm{j}_1}\xspace}

\newcommand{\tG}{\ensuremath{\PQt\PGg}\xspace}
\newcommand{\Njets}{\ensuremath{N_{\mathrm{j}}}\xspace}
\newcommand{\NumSR}{\ensuremath{N_{\text{SR}}}\xspace}
\newcommand{\NumSB}{\ensuremath{N_{\text{SB}}}\xspace}
\newcommand{\TF}{\ensuremath{f}\xspace}

\newcommand{\fullcorr}{\ensuremath{\checkmark}\xspace}
\newcommand{\partcorr}{\ensuremath{\sim}\xspace}
\newcommand{\nocorr}{\ensuremath{\times}\xspace}
\newcommand{\ptgamma}{\ensuremath{\pt(\PGg)}\xspace}
\newcommand{\muF}{\ensuremath{\mu_{\mathrm{F}}}\xspace}
\newcommand{\muR}{\ensuremath{\mu_{\mathrm{R}}}\xspace}

\newcommand{\likelihood}{\ensuremath{L}\xspace}
\newcommand{\sigstr}{\ensuremath{r}\xspace}
\newcommand{\nuisan}{\ensuremath{\theta}\xspace}
\newcommand{\sigstrmax}{\ensuremath{\hat{\sigstr}}\xspace}
\newcommand{\nuisanmax}{\ensuremath{\hat{\nuisan}}\xspace}
\newcommand{\nuisanmaxr}{\ensuremath{\nuisanmax_r}\xspace}
\newcommand{\sigfid}{\ensuremath{\sigma_\text{fid}(\pptottG)}\xspace}

\newcommand{\etagamma}{\ensuremath{\abseta(\PGg)}\xspace}
\newcommand{\drgammaclose}{\ensuremath{\min\dRgl}\xspace}
\newcommand{\drgammafirst}{\ensuremath{\dR(\PGg,\Pell_1)}\xspace}
\newcommand{\drgammasecond}{\ensuremath{\dR(\PGg,\Pell_2)}\xspace}
\newcommand{\drgammabjet}{\ensuremath{\min\dR(\PGg,\PQb)}\xspace}
\newcommand{\drlepjet}{\ensuremath{\min\dR(\Pell,\mathrm{j})}\xspace}
\newcommand{\detall}{\ensuremath{\abs{\dEta(\Pell\Pell)}}\xspace}
\newcommand{\dphill}{\ensuremath{\dPhi(\Pell\Pell)}\xspace}
\newcommand{\ptll}{\ensuremath{\pt(\Pell\Pell)}\xspace}
\newcommand{\sumptll}{\ensuremath{\pt(\Pell_1)+\pt(\Pell_2)}\xspace}
\newcommand{\ptjet}{\ensuremath{\pt(\jetone)}\xspace}
\newcommand{\TUnfold}{\ensuremath{\textsc{TUnfold}}\xspace}
\newcommand{\chisq}{\ensuremath{\chi^2}\xspace}
\newcommand{\dof}{\ensuremath{\text{dof}}\xspace}
\newcommand{\chisqdof}{\ensuremath{\chisq/\dof}\xspace}

\newcommand{\Lagrangian}[1]{\ensuremath{\mathcal{L}_{\text{#1}}}\xspace}
\newcommand{\ci}{\ensuremath{c_i}\xspace}
\newcommand{\Oi}{\ensuremath{\mathcal{O}_i}\xspace}
\newcommand{\CuB}{\ensuremath{c_{\PQu\PB}^{(33)}}\xspace}
\newcommand{\CuW}{\ensuremath{c_{\PQu\PW}^{(33)}}\xspace}
\newcommand{\ttZ}{\ensuremath{\ttbar\PZ}\xspace}
\newcommand{\RE}{\ensuremath{\mathrm{Re}}\xspace}
\newcommand{\IM}{\ensuremath{\mathrm{Im}}\xspace}
\newcommand{\thetaw}{\ensuremath{\theta_{\PW}}}
\newcommand{\sinw}{\ensuremath{\sin\thetaw}}
\newcommand{\cosw}{\ensuremath{\cos\thetaw}}
\newcommand{\ctZ}{\ensuremath{c_{\PQt\PZ}}\xspace}
\newcommand{\ctZI}{\ensuremath{c_{\PQt\PZ}^{\mathrm{I}}}\xspace}
\newcommand{\ctG}{\ensuremath{c_{\PQt\PGg}}\xspace}
\newcommand{\ctGI}{\ensuremath{c_{\PQt\PGg}^{\mathrm{I}}}\xspace}
\newcommand{\Wtb}{\ensuremath{\PW\PQt\PQb}\xspace}
\newcommand{\elljets}{\ensuremath{\Pell\text{+jets}}\xspace}
\newcommand{\LambTevSq}{\ensuremath{(\Lambda/\TeVns)^2}}

\newcommand{\XsecMeasuredFull}{\ensuremath{175.2\pm2.5\stat\pm6.3\syst\unit{fb}}\xspace}
\newcommand{\XsecTheoryFull}{\ensuremath{155\pm27\unit{fb}}\xspace}

\cmsNoteHeader{TOP-21-004}
\title{
    Measurement of the inclusive and differential \texorpdfstring{\ttG}{ttgamma} cross sections in the dilepton channel and effective field theory interpretation in proton-proton collisions at \texorpdfstring{\sqrts}{sqrt(s)=13 TeV}
}

\date{\today}

\abstract{The production cross section of a top quark pair in association with a photon is measured in proton-proton collisions in the decay channel with two oppositely charged leptons (\emu, \ee, or \mumu). The measurement is performed using 138\fbinv of proton-proton collision data recorded by the CMS experiment at \sqrts during the 2016--2018 data-taking period of the CERN LHC. A fiducial phase space is defined such that photons radiated by initial-state particles, top quarks, or any of their decay products are included. An inclusive cross section of \XsecMeasuredFull is measured in a signal region with at least one jet coming from the hadronization of a bottom quark and exactly one photon with transverse momentum above 20\GeV. Differential cross sections are measured as functions of several kinematic observables of the photon, leptons, and jets, and compared to standard model predictions. The measurements are also interpreted in the standard model effective field theory framework, and limits are found on the relevant Wilson coefficients from these results alone and in combination with a previous CMS measurement of the \ttG production process using the lepton+jets final state.}

\hypersetup{pdfencoding=pdfdoc,
pdfauthor={CMS Collaboration},
pdftitle={Measurement of the inclusive and differential ttgamma cross sections in the dilepton channel and effective field theory interpretation in proton-proton collisions at sqrt(s)=13 TeV},
pdfsubject={CMS},
pdfkeywords={CMS, top quark}}

\maketitle

\section{Introduction}

The cross section measurement of top quark pair production in association with a photon (\ttG) probes the coupling between the top quark and the photon, making it both a test of standard model (SM) predictions and providing sensitivity to new-physics phenomena beyond the SM.
With the large amount of data collected with the CMS detector in proton-proton (\pp) collisions at \sqrts during the data-taking period from 2016 to 2018 of the CERN LHC, precise measurements of relatively small cross sections such as for \ttG production are possible.
We present inclusive and differential measurements of the \ttG production cross section in final states with two oppositely charged (OC) leptons (\emu, \ee, or \mumu).
The inclusion of differential information improves the sensitivity of the measurement to new-physics modifications.
Additionally, we perform a model-independent interpretation of the results in terms of the SM effective field theory (SMEFT).

First evidence for \ttG production was found by the CDF Collaboration in $\Pp\PAp$ collisions~\cite{CDF:2011ccg}.
At the LHC, \ttG production in \pp collisions was first observed by the ATLAS Collaboration at \sqrts[7]~\cite{ATLAS:2015jos}.
Further measurements were performed by the ATLAS and CMS Collaborations at 8~\cite{ATLAS:2017yax, CMS:2017tzb} and 13\TeV~\cite{ATLAS:2018sos, ATLAS:2020yrp, CMS:2021klw}.
In the latest CMS measurement presented in Ref.~\cite{CMS:2021klw}, \ttG production is measured in events with exactly one lepton (electron or muon) and at least three jets (``\elljets''), using the same data set as the dilepton measurement presented here.

\begin{figure}[!ht]
\centering
\includegraphics[width=0.3\textwidth]{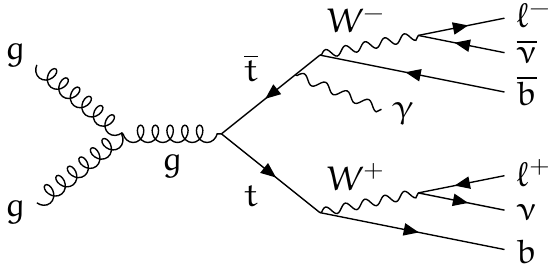}
\hfill
\includegraphics[width=0.3\textwidth]{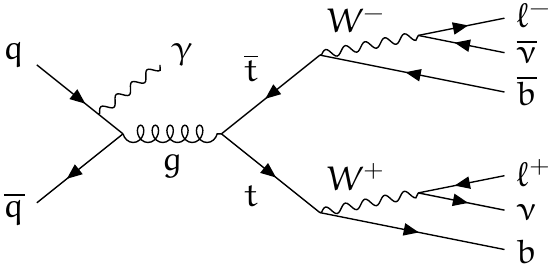}
\hfill
\includegraphics[width=0.3\textwidth]{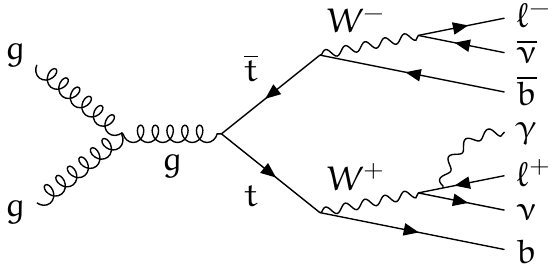}
\caption{Examples of leading-order Feynman diagrams for \ttG production with two leptons in the final state, where the photon is radiated by a top quark (left), by an incoming quark (middle), or by one of the charged decay products of a top quark (right).}
\label{fig:feynman}
\end{figure}

For this measurement, a fiducial phase space is defined for the \ttG signal process with criteria on the kinematic properties of the photon, leptons, and jets at the particle level.
Events are included where the photon is radiated from a top quark, an incoming quark, or any of the charged decay products of the top quarks.
Examples of leading-order (LO) Feynman diagrams for these processes are shown in Fig.~\ref{fig:feynman}.

Events are selected with two OC leptons, an isolated photon, and at least one jet.
Background contributions without top quarks are suppressed by applying additional \PQb tagging criteria on the selected jets.
After the event selection, the dominant source of background stems from events with nonprompt photons, \ie photons originating from particles inside hadronic jets or from additional \pp collisions, or hadronic jets misidentified as photons.
The nonprompt background contribution is estimated using control samples in data, while other background sources (mainly \PZ boson and single top quark production in association with a photon) are estimated from Monte Carlo (MC) event simulations.

The inclusive cross section is measured with a profile likelihood fit from the measured distribution of the transverse momentum \pt of the reconstructed photon, in which the sources of systematic uncertainty are treated as nuisance parameters.
The differential cross sections are measured by subtracting the estimated background contributions from the measured distributions and applying an unfolding method to correct for detector resolution effects.
To evaluate the sensitivity to possible modifications of the coupling between the top quark and the photon, the measured photon \pt distribution is used to constrain relevant Wilson coefficients in the SMEFT framework~\cite{Buchmuller:1985jz, Grzadkowski:2010es}.
Limits are derived on the relevant Wilson coefficients from these results alone and in combination with the \elljets measurement from Ref.~\cite{CMS:2021klw}.

The SM production of \ttG has been studied at next-to-LO (NLO) in quantum chromodynamics (QCD) and electroweak theory~\cite{Duan:2009kcr, Melnikov:2011ta, Duan:2011zxb, Kardos:2014zba, Maltoni:2015ena, Duan:2016qlc, Pagani:2021iwa}, including also the top quark decays and photons radiated from the final-state particles at full NLO accuracy~\cite{Bevilacqua:2018woc, Bevilacqua:2018dny, Bevilacqua:2019quz}.
The sensitivity to new-physics modifications has been studied for anomalous dipole moments of the top quark~\cite{Baur:2004uw, Bouzas:2012av, Schulze:2016qas} and in the SMEFT framework~\cite{BessidskaiaBylund:2016jvp}.
The possibility to study charge asymmetries in \ttG production has been investigated in Refs.~\cite{Aguilar-Saavedra:2014vta, Bergner:2018lgm}.

This paper is organized as follows.
The CMS detector is described briefly in Section~\ref{sec:detector}.
In Section~\ref{sec:simulation}, the data and simulated event samples are discussed.
The reconstruction and selection of events is detailed in Section~\ref{sec:selection}, followed by an estimation of the background contributions in Section~\ref{sec:backgrounds}.
Systematic uncertainties that affect the measurements are discussed in Section~\ref{sec:systematics}.
The results of the inclusive and the differential cross section measurements are presented in Sections~\ref{sec:inclusive} and~\ref{sec:differential}, respectively.
In Section~\ref{sec:eft}, the interpretation of the measurements in the SMEFT framework is provided.
The results are summarized in Section~\ref{sec:summary}.

Tabulated results are provided in the HEPData record for this analysis~\cite{hepdata}.

\section{The CMS detector}\label{sec:detector}

The central feature of the CMS apparatus is a superconducting solenoid of 6\unit{m} internal diameter, providing a magnetic field of 3.8\unit{T}.
Within the solenoid volume are a silicon pixel and strip tracker, a lead tungstate crystal electromagnetic calorimeter (ECAL), and a brass and scintillator hadron calorimeter, each composed of a barrel and two endcap sections.
Forward calorimeters extend the pseudorapidity ($\eta$) coverage provided by the barrel and endcap detectors.
Muons are measured in gas-ionization detectors embedded in the steel flux-return yoke outside the solenoid.
A more detailed description of the CMS detector, together with a definition of the coordinate system used and the relevant kinematic variables, can be found in Ref.~\cite{CMS:2008xjf}.

Events of interest are selected using a two-tiered trigger system.
The first level (L1), composed of custom hardware processors, uses information from the calorimeters and muon detectors to select events at a rate of around 100\unit{kHz} within a fixed latency of about 4\mus~\cite{CMS:2020cmk}.
The second level, known as the high-level trigger, consists of a farm of processors running a version of the full event reconstruction software optimized for fast processing, and reduces the event rate to around 1\unit{kHz} before data storage~\cite{CMS:2016ngn}.

\section{Data and simulated samples}\label{sec:simulation}

The data sample used in this measurement corresponds to an integrated luminosity of 138\fbinv of \pp collision events at \sqrts collected with the CMS detector between 2016 and 2018.
To account for the differences in the LHC running conditions and the CMS detector performance, the data collected in the three years of data taking are analyzed separately and appropriate per-year calibrations are applied, before the data are combined for the final cross section measurements.

Simulated MC events are used for the evaluation of the signal selection efficiency, the validation of the background estimates from control samples in data, and the prediction of other background contributions.
Three separate sets of simulated event samples are used, corresponding to the conditions of the three data-taking years.

For the \ttG signal process, simulated events are generated at LO accuracy in QCD using the \MGvATNLO2.6.0 event generator~\cite{Alwall:2014hca} for the $2\to7$ signal process, \ie as $\pp\to\PQb\Pell^+\PGn\PAQb\Pell^-\PAGn\PGg$ and similarly for the other \ttbar final states.
In this way, all contributions with photons radiated from incoming quarks, top quarks, or any of the charged top quark decay products are included.
For the normalization of the \ttG sample, a $K$-factor evaluated with \MGvATNLO at NLO accuracy for the $2\to3$ process \pptottG is applied.
The parton distribution functions (PDFs) used in the simulation of the hard process are the NNPDF3.1~\cite{NNPDF:2017mvq} sets.

For the simulation of background events from \ttbar, \ttH, and single top quark $t$-channel and \tW production, the $\POWHEG2$ event generator~\cite{Nason:2004rx, Frixione:2007nw, Frixione:2007vw, Alioli:2009je, Alioli:2010xd, Re:2010bp, Campbell:2014kua, Hartanto:2015uka} at NLO accuracy in QCD is used.
Samples for \GGtoZZ production are generated at LO accuracy in QCD with the \MCFM7.0.1 event generator~\cite{Campbell:2010ff,Campbell:2013una}.
For all other background processes, events are simulated with the \MGvATNLO event generator at LO or NLO accuracy in QCD.
In the simulation of the hard process, the NNPDF3.0~\cite{NNPDF:2014otw} (NNPDF3.1) sets are used for the 2016 (2017--2018) samples.

{\tolerance=500
All event generators are interfaced with the \PYTHIA8.226 (8.230) simulation~\cite{Sjostrand:2014zea} for parton showering and hadronization of the 2016 (2017--2018) samples.
The underlying event is modelled using the CP5 tune~\cite{CMS:2019csb}, except for some 2016 background samples for which the CUETP8M1, CUETP8M2, and CUETP8M2T4 tunes~\cite{Skands:2014pea, CMS:2015wcf, CMS:2016kle} are used.
For the LO (NLO) samples generated with \MGvATNLO and interfaced with \PYTHIAeight, jets from matrix element calculations are merged with those from the parton shower using the \textsc{MLM} (\textsc{FxFx})~\cite{Alwall:2007fs, Frederix:2012ps} matching scheme.
A summary of all simulated samples is given in Table~\ref{tab:samples}.
\par}

\begin{table}[!ht]
\centering
\topcaption{MC event generators used to simulate events for the signal and background processes.
For each simulated process, the order of the cross section normalization calculation, the MC event generator used, and the perturbative order in QCD of the generator calculation are shown.
The normalization and perturbative QCD orders are given as LO, NLO, next-to-NLO (NNLO), and including next-to-next-to-leading-logarithmic (NNLL) corrections.
The symbol \PV refers to \PW and \PZ bosons.}
\renewcommand{\arraystretch}{1.2}
\begin{tabular}{cccc}
    \multirow{2}{*}{Process} & Cross section & \multirow{2}{*}{Event generator} & Perturbative \\[-3pt]
    & normalization & & order in QCD \\ \hline
    \ttG & NLO & \MGvATNLO & LO \\
    \PZ & NNLO~\cite{Li:2012wna} & \MGvATNLO & LO \\
    \ZG, $\PW\PGg$, $\PV\PV$, $\PV\PV\PV$, & & & \\[-3pt]
    $\ttbar\PV$, \tZq, $\PQt\PW\PZ$, $\PQt\PH\Pq$, & NLO & \MGvATNLO & NLO \\[-3pt]
    $\PQt\PH\PW$, $\ttbar\PV\PV$, $\ttbar\ttbar$ & & & \\
    \ttbar & NNLO+NNLL~\cite{Czakon:2011xx} & \POWHEG & NLO \\
    single \PQt (\tchannel) & NLO~\cite{Aliev:2010zk, Kant:2014oha} & \POWHEG & NLO \\
    single \PQt (\schannel) & NLO~\cite{Aliev:2010zk, Kant:2014oha} & \MGvATNLO & NLO \\
    \tW & NNLO~\cite{Kidonakis:2015nna} & \POWHEG & NLO \\
    \ttH & NLO & \POWHEG & NLO \\
    \GGtoZZ & LO & \MCFM & LO \\
\end{tabular}
\label{tab:samples}
\end{table}

An overlap exists between the phase space modelled in the \ttG and \ttbar samples because of the addition of soft photons to simulated events by the \PYTHIAeight simulation.
To ensure orthogonal phase spaces, simulated events from the \ttG sample are only used if they contain a generated photon with $\pt>10\GeV$ and $\abseta<5$ that does not originate from the decay of a hadron and is separated from any other stable particle (\ie a generated particle with lifetime larger than 3\unit{ns}~\cite{CMS-NOTE-2017-004}) except neutrinos and photons by $\dR=\sqrt{(\dEta)^2+(\dPhi)^2}>0.1$, where \dEta and \dPhi are the differences in pseudorapidity and azimuthal angle, respectively, between the directions of the photon and the stable particle.
Conversely, simulated events from the \ttbar sample are only used if they do not contain such a photon.
Similarly, the overlap between the \ZG and \Zjets sample is removed, with modified requirements of $\pt>15\GeV$ and $\abseta<2.6$ for the generated photon and a minimum distance of $\dR>0.05$, corresponding to the different settings in the event simulation.

The \ttG cross section is measured in a fiducial phase space for the dilepton final state, defined at particle level after the event generation, parton showering, and hadronization of the \ttG event samples.
Electrons and muons, after adding all photons inside a cone of $\dR<0.1$ around the lepton direction, are required to have $\pt>15\GeV$ and $\abseta<2.4$.
Signal photons must have $\pt>20\GeV$ and $\abseta<1.44$, and be isolated from any other stable particle with $\pt>5\GeV$ except for neutrinos by $\dR>0.1$.
Additionally, photons must be separated from electrons and muons by $\dR>0.4$.
Jets are clustered with the anti-\kt algorithm~\cite{Cacciari:2008gp} with a distance parameter $R=0.4$, using all particles excluding neutrinos, and are required to have $\pt>30\GeV$ and $\abseta<2.4$.
Furthermore, jets must be isolated from all leptons and photons by $\dR>0.4$ and 0.1, respectively.
For each \PQb hadron, an additional collinear four-vector of infinitesimal magnitude is included in the jet clustering, and each jet that includes such a ``ghost'' is identified as \PQb jet~\cite{Cacciari:2008gn}.
The fiducial phase space is defined by requiring \ttG events to have exactly one photon, at least one \PQb jet, and exactly two OC leptons with an invariant mass $\mll>20\GeV$, of which at least one has $\pt>25\GeV$.
The fiducial phase space definition is summarized in Table~\ref{tab:fiducial}.
A small fraction of events from the \ttG event sample generated in the \elljets final state also passes the fiducial phase space requirements with an additional lepton created in the parton showering and hadronization, and is included as a signal contribution.
The SM prediction for the \ttG cross section in this fiducial phase space, evaluated from the LO event sample with the NLO $K$-factor as described above, is $\sigSM=\XsecTheoryFull$, where the uncertainty includes scale variations and the PDF choice.

\begin{table}[!ht]
\centering
\topcaption{Summary of the requirements at the particle level on the various physics objects in the fiducial phase space definition.
The two lepton \pt thresholds are applied to the highest and second-highest \pt lepton, respectively.
The ``isolated'' definition for the photon requires no stable particle with $\pt>5\GeV$ except neutrinos within a cone of $\dR=0.1$.
The parameters \Nleps, \Nphotons, and \Nbjets represent the numbers of leptons, photons, and \PQb jets, respectively, in the event.}
\renewcommand{\arraystretch}{1.2}
\cmsTable{\begin{tabular}{ccccc}
    Leptons           & Photons        & Jets          & \PQb jets              & Events          \\ \hline
    $\pt>25$ (15)\GeV & $\pt>20\GeV$   & $\pt>30\GeV$  & $\pt>30\GeV$           & $\Nleps=2$ (OC) \\
    $\abseta<2.4$     & $\abseta<1.44$ & $\abseta<2.4$ & $\abseta<2.4$          & $\Nphotons=1$   \\
                      & $\dRgl>0.4$    & $\dRjetl>0.4$ & $\dRjetl>0.4$          & $\Nbjets\geq1$  \\
                      & isolated       & $\dRjetg>0.1$ & $\dRjetg>0.1$          & $\mll>20\GeV$   \\
                      &                &               & matched to \PQb hadron &                 \\
\end{tabular}}
\label{tab:fiducial}
\end{table}

For all simulated events, the CMS detector response is subsequently modelled with the \GEANTfour toolkit~\cite{GEANT4:2002zbu}.
Additional minimum-bias \pp interactions in the same or nearby bunch crossing, referred to as pileup, are added from simulation as well.
All events in the data and simulated samples are reconstructed with the same algorithms described in Section~\ref{sec:selection}.
Corrections for differences in the selection performance between the data and simulated samples are applied to simulated events.

\section{Event reconstruction and selection}\label{sec:selection}

Information from the various subdetectors is used by the particle-flow algorithm~\cite{CMS:2017yfk} to reconstruct the particles (photons, electrons, muons, charged and neutral hadrons) produced in an event.
The candidate vertex with the largest value of summed physics-object $\pt^2$ is taken to be the primary \pp interaction vertex (PV).
The physics objects used for this determination are the jets, clustered using the jet finding algorithm~\cite{Cacciari:2008gp, Cacciari:2011ma} with the tracks assigned to candidate vertices as inputs, and the associated missing transverse momentum, taken as the negative vector sum of the \pt of those jets.

Electrons are identified as a primary charged particle track that can have bremsstrahlung photons emitted along its path through the silicon tracker material, and when extrapolated to the ECAL, can be associated with many ECAL energy deposits~\cite{CMS:2020uim}.
Muons are identified as tracks in the silicon tracker consistent with either a track or several hits in the muon system, and associated with compatible calorimeter deposits~\cite{CMS:2018rym}.

Electron and muon candidates with $\pt>15\GeV$ and $\abseta<2.4$ that pass the identification requirements described in Ref.~\cite{CMS:2021rvz} are selected.
These requirements are optimized to discriminate between ``prompt'' leptons originating from decays of heavy vector bosons and top quarks, and ``nonprompt'' leptons that originate from hadron decays, or are jets or hadrons misidentified as leptons.
To that end, discriminating variables are combined into a boosted decision tree discriminant trained with the TMVA toolkit~\cite{Hocker:2007ht}.
The discriminating variables are the kinematic properties of the lepton and the jet closest to the lepton; relative isolation variables defined as scalar \pt sums of all particles within cones around the lepton direction, divided by the lepton \pt; the impact parameter of the lepton tracks with respect to the PV; the muon segment compatibility~\cite{CMS:2018rym} in the case of muons; and the discriminant of a standard electron identification algorithm~\cite{CMS:2020uim} in the case of electrons.
Additionally, electrons with $1.44<\abseta<1.57$ in the transition region between the ECAL barrel and endcaps are removed due to the worse performance of electron reconstruction in this region.

Photons are identified based on the presence of an energy deposit in the ECAL with no charged particle tracks pointing towards this deposit.
The photon energy is obtained using this ECAL measurement, to which corrections for the energy scale and zero suppression are applied in both simulation and data.
In simulated events, smearing corrections are applied to photons to match the resolution observed in data~\cite{CMS:2020uim}.
Photon candidates are considered with $\pt>20\GeV$ and $\abseta<1.44$, \ie inside the barrel region of the detector, and are required to be isolated from selected leptons by requiring that $\dRgl>0.4$.
Identification criteria are imposed on photon candidates based on the shape of its electromagnetic shower, the isolation of the photon, and the absence of hits in the pixel tracker compatible with the photon direction, which improves the rejection of electrons, misidentified hadrons, and photons produced within jets~\cite{CMS:2020uim}.
One shower shape criterion is a requirement on \sieie, a measure of the electromagnetic shower width in units of the ECAL crystal spacing, of $\sieie<0.01$, thus rejecting \pizerodecay decays, which are typically reconstructed as a single wide shower.
One isolation criterion requires $\chIso<1.14\GeV$, where \chIso is the charged particle isolation parameter, computed as the scalar \pt sum of all charged hadrons compatible with the PV and inside a cone of $\dR<0.3$ around the photon direction.
The \sieie and \chIso criteria are not used to veto any additional reconstructed photons, and the inverted requirements are applied for the selection of control samples in data for the background estimation described in Section~\ref{sec:backgrounds}.

Jets are clustered using all particle-flow candidates with the anti-\kt algorithm~\cite{Cacciari:2008gp} and a distance parameter of $R=0.4$.
To mitigate the impact of pileup on the jet momentum, tracks identified as originating from pileup vertices are discarded, and a correction is applied for contributions from neutral pileup particles and those charged pileup particles that are not associated with a pileup vertex~\cite{CMS:2020ebo}.
Jet energy corrections are derived from simulation studies so that the average measured energy of jets are made equal to that of particle-level jets, and are applied in both data and simulation.
In situ measurements of the momentum balance in dijet, photon+jet, {\PZ}+jet, and multijet events are used to determine any residual differences between the jet energy scale in data and simulation, which are applied as corrections to the simulation~\cite{CMS:2016lmd}.
Additional selection criteria are applied to each jet to remove jets potentially dominated by instrumental effects or reconstruction failures~\cite{CMS:2020ebo}.
Reconstructed jets with $\pt>30\GeV$ and $\abseta<2.4$ are selected if they are separated from any selected lepton by $\dR>0.4$, and any selected photon by $\dR>0.1$.
Jets originating from the hadronization of \PQb quarks are identified (\PQb tagged) with the \textsc{DeepCSV} heavy-flavour jet tagging algorithm~\cite{CMS:2017wtu}, with an efficiency of about 70\% and a misidentification rate of 12 (1)\% for jets originating from \PQc quarks (light quarks or gluons).

A combination of trigger paths requiring the presence of one or two leptons is used to select events.
The \pt threshold of the single-lepton trigger paths for electrons and muons was 27 and 24\GeV in 2016, and 32 and 27\GeV in 2017--2018, respectively.
For the double-lepton trigger paths, the \pt thresholds were 23 and 17\GeV for the highest \pt (leading) and 12 and 8\GeV for the second-highest \pt (subleading) electron and muon, respectively.
The trigger efficiency for events passing the lepton selection described below is higher than 95\%, as measured from events selected with independent trigger paths.

Events are selected with exactly two OC leptons, exactly one photon, and at least one \PQb-tagged jet.
At least one lepton is required to have $\pt>25\GeV$.
The two leptons are required to have an invariant mass $\mll>20\GeV$ to reduce background contributions with nonprompt leptons and from low-mass quark resonances.
The \PQb jet requirement reduces background contributions from processes without top quarks.
To further reduce the background from \ZG production, events with same-flavour leptons where either the two leptons or the two leptons and the photon have an invariant mass close to the \PZ boson mass are removed.
Specifically, $\dMllZ>15\GeV$ and $\dMllgZ>15\GeV$ are required for same-flavour lepton pairs, where \mZ is the world-average \PZ boson mass~\cite{ParticleDataGroup:2020ssz}.

\begin{figure}[!htp]
\centering
\includegraphics[width=0.42\textwidth]{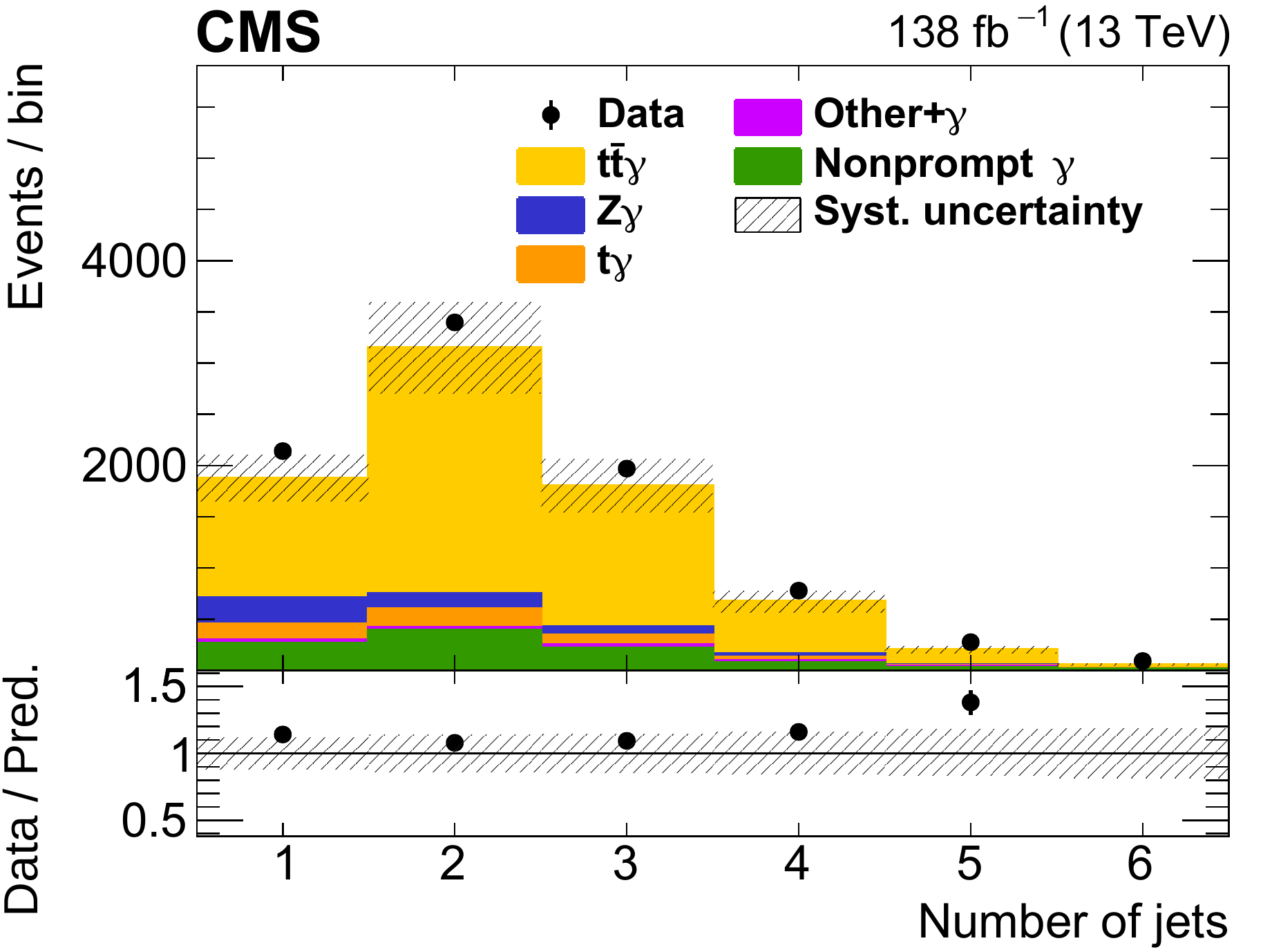}
\hspace{0.02\textwidth}
\includegraphics[width=0.42\textwidth]{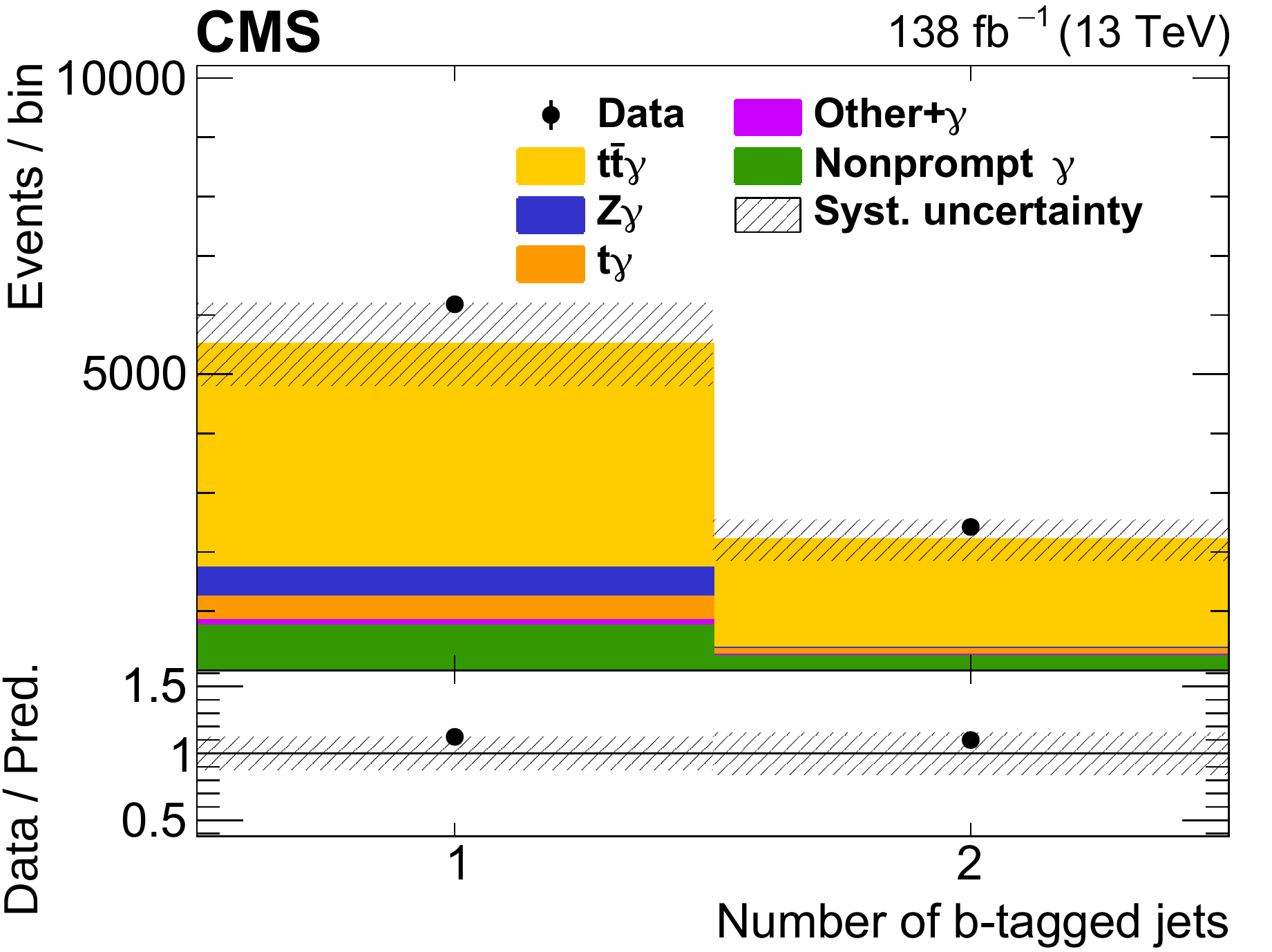}
\\
\includegraphics[width=0.42\textwidth]{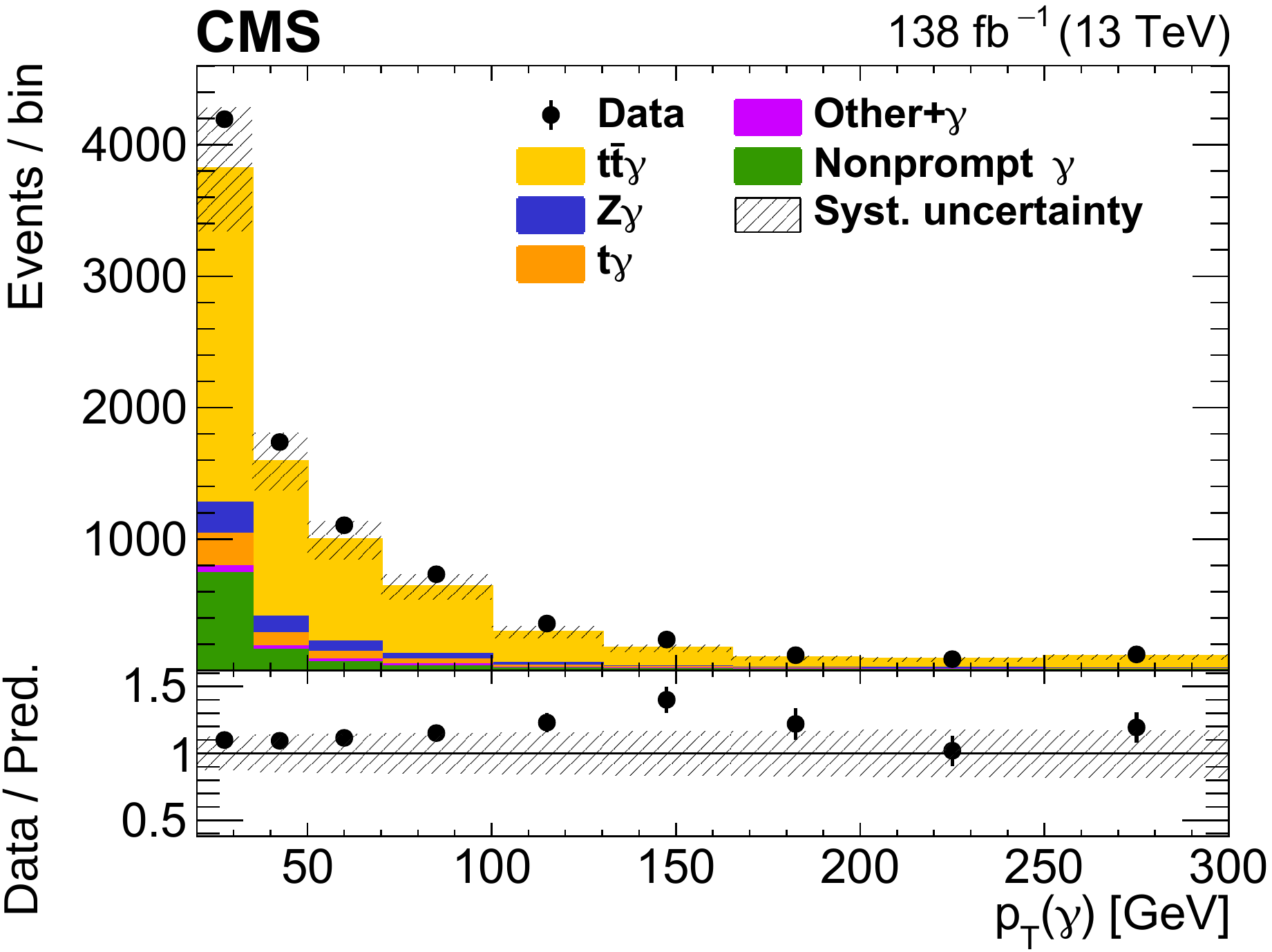}
\hspace{0.02\textwidth}
\includegraphics[width=0.42\textwidth]{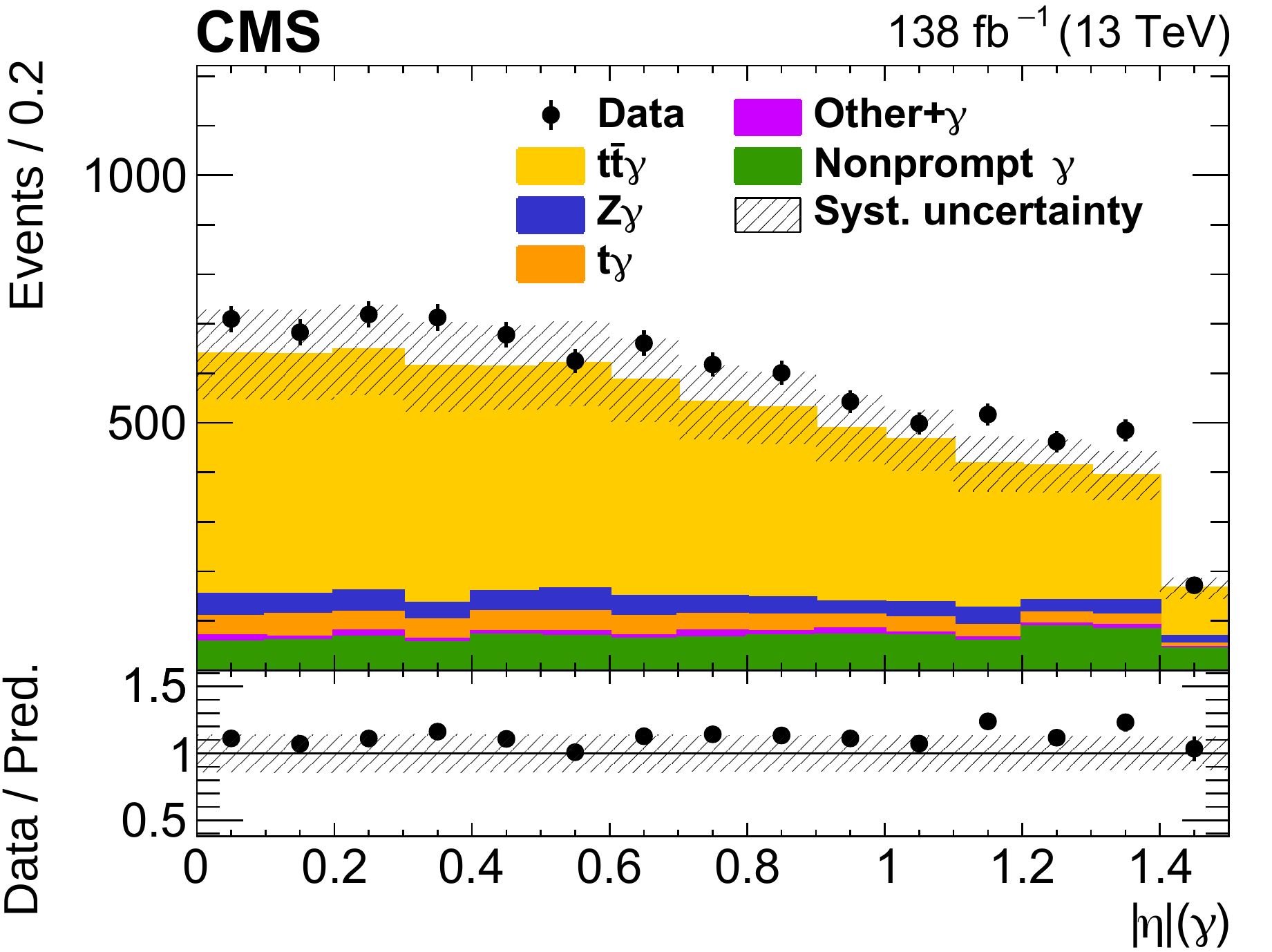}
\\
\includegraphics[width=0.42\textwidth]{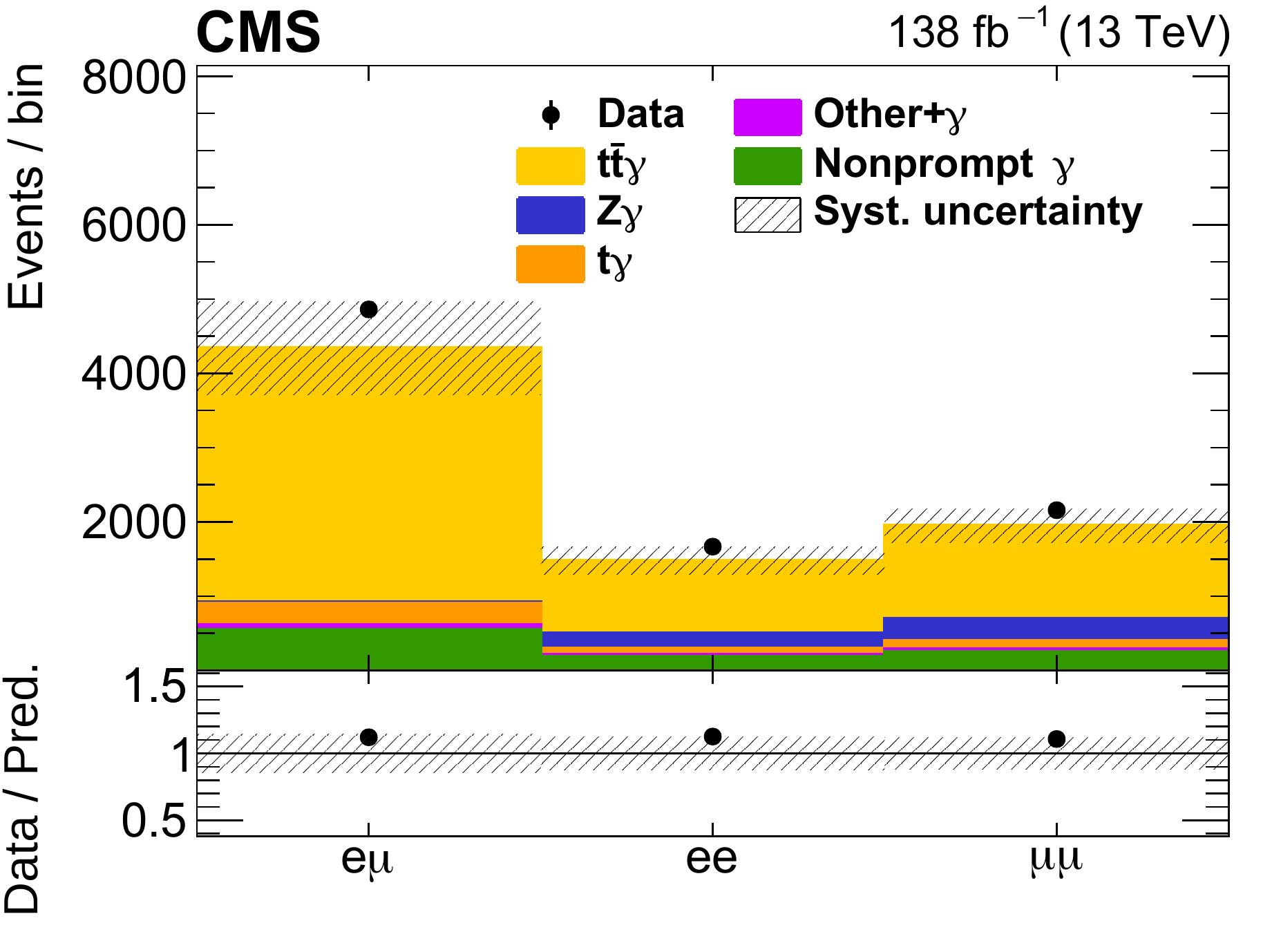}
\hspace{0.02\textwidth}
\includegraphics[width=0.42\textwidth]{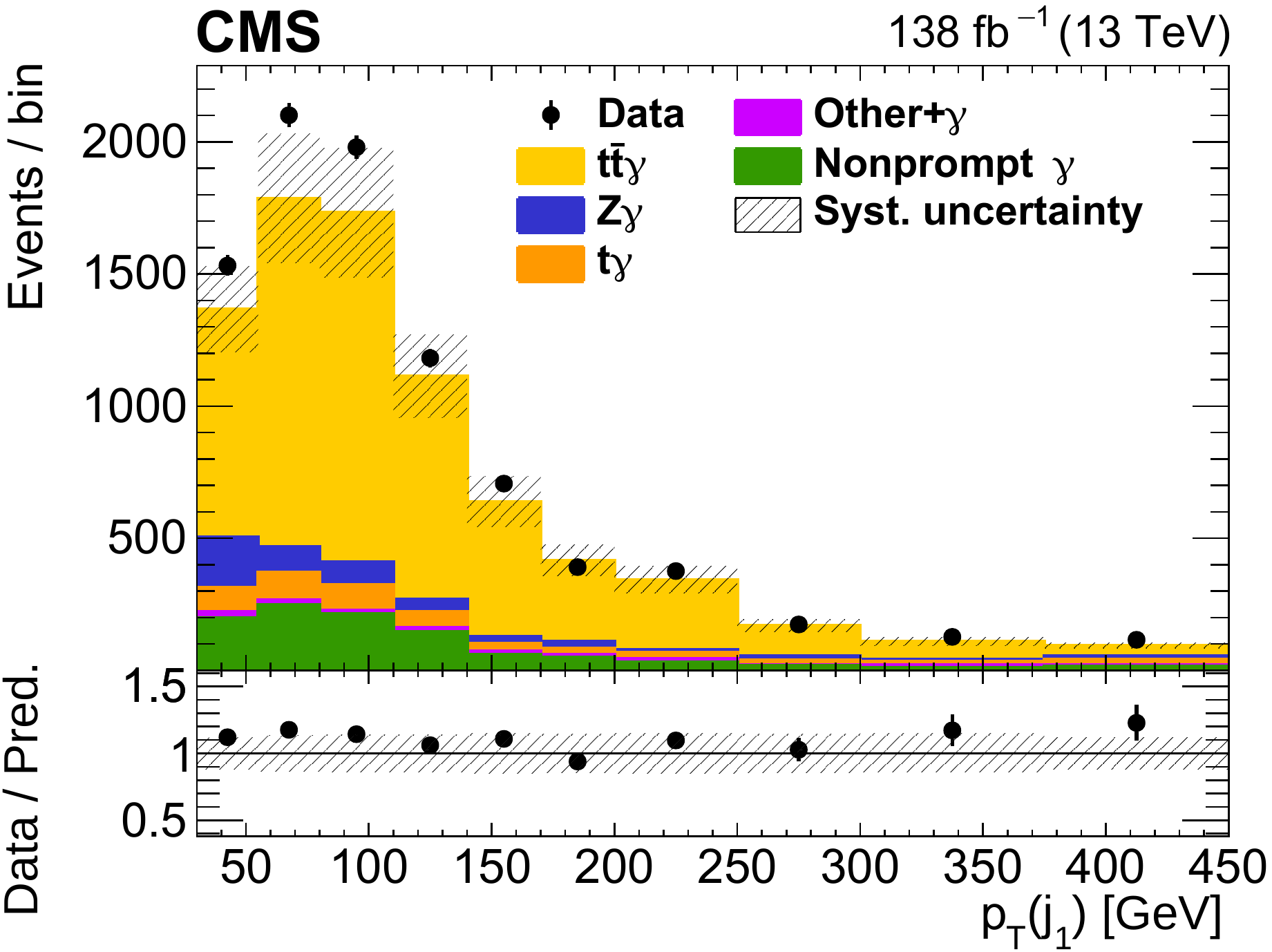}
\caption{The observed (points) and predicted (shaded histograms) signal and background yields as functions of the number of jets (upper left) and \PQb-tagged jets (upper right), the \pt (middle left) and \abseta (middle right) of the photon, the dilepton flavours (lower left), and the \pt of the leading jet \jetone (lower right), after applying the signal selection.
Distributions are shown for the three lepton flavour channels combined, with all relevant corrections applied.
The predictions are normalized to the expected yields, without taking the results of the fit to the data into account.
The vertical bars on the points show the statistical uncertainties in the data, and the hatched bands the systematic uncertainty in the predictions.
The lower panels show the ratio of the event yields in data to the overall sum of the predictions.}
\label{fig:yieldSR1}
\end{figure}

\begin{figure}[!ht]
\centering
\includegraphics[width=0.42\textwidth]{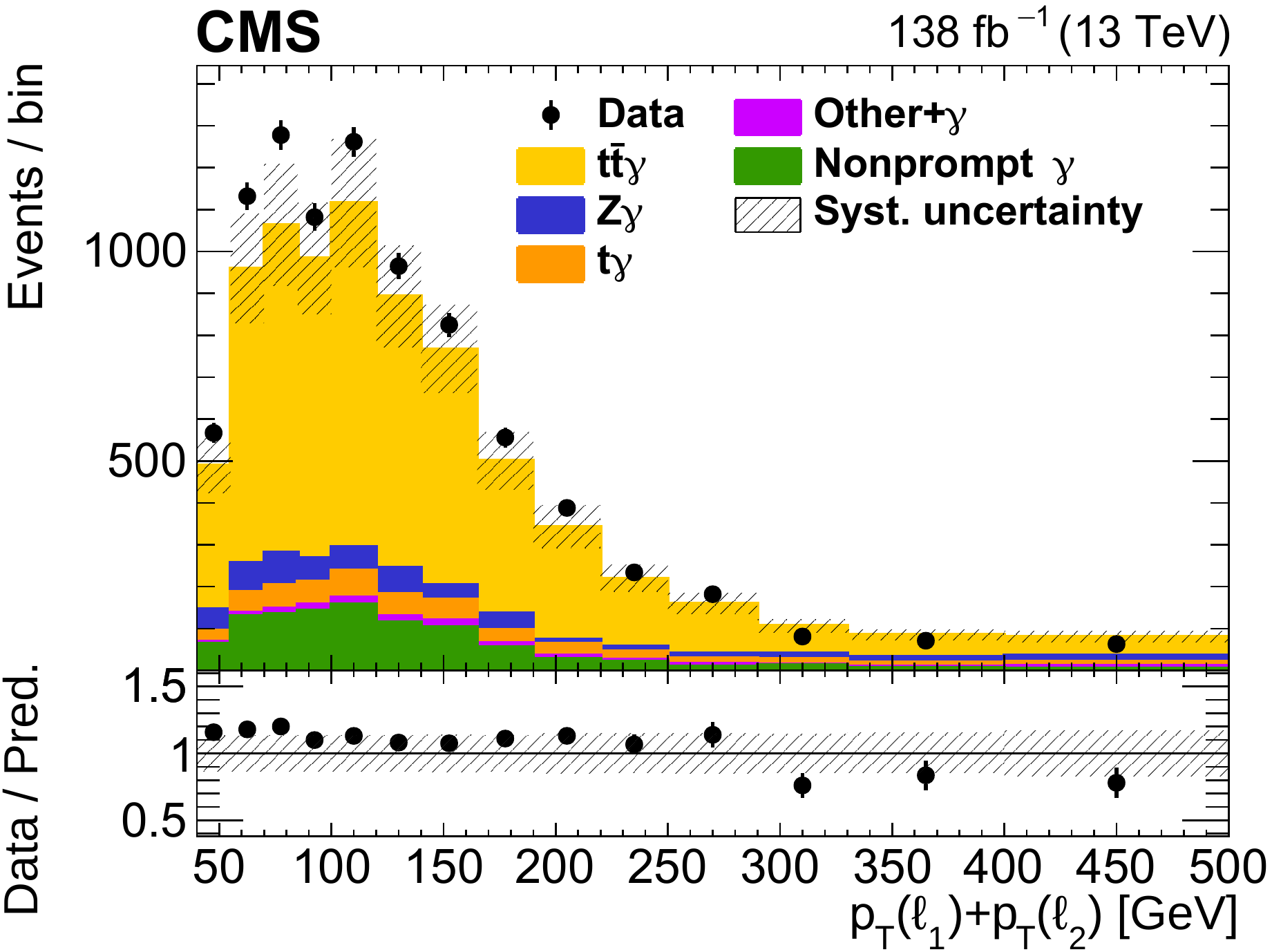}
\hspace{0.02\textwidth}
\includegraphics[width=0.42\textwidth]{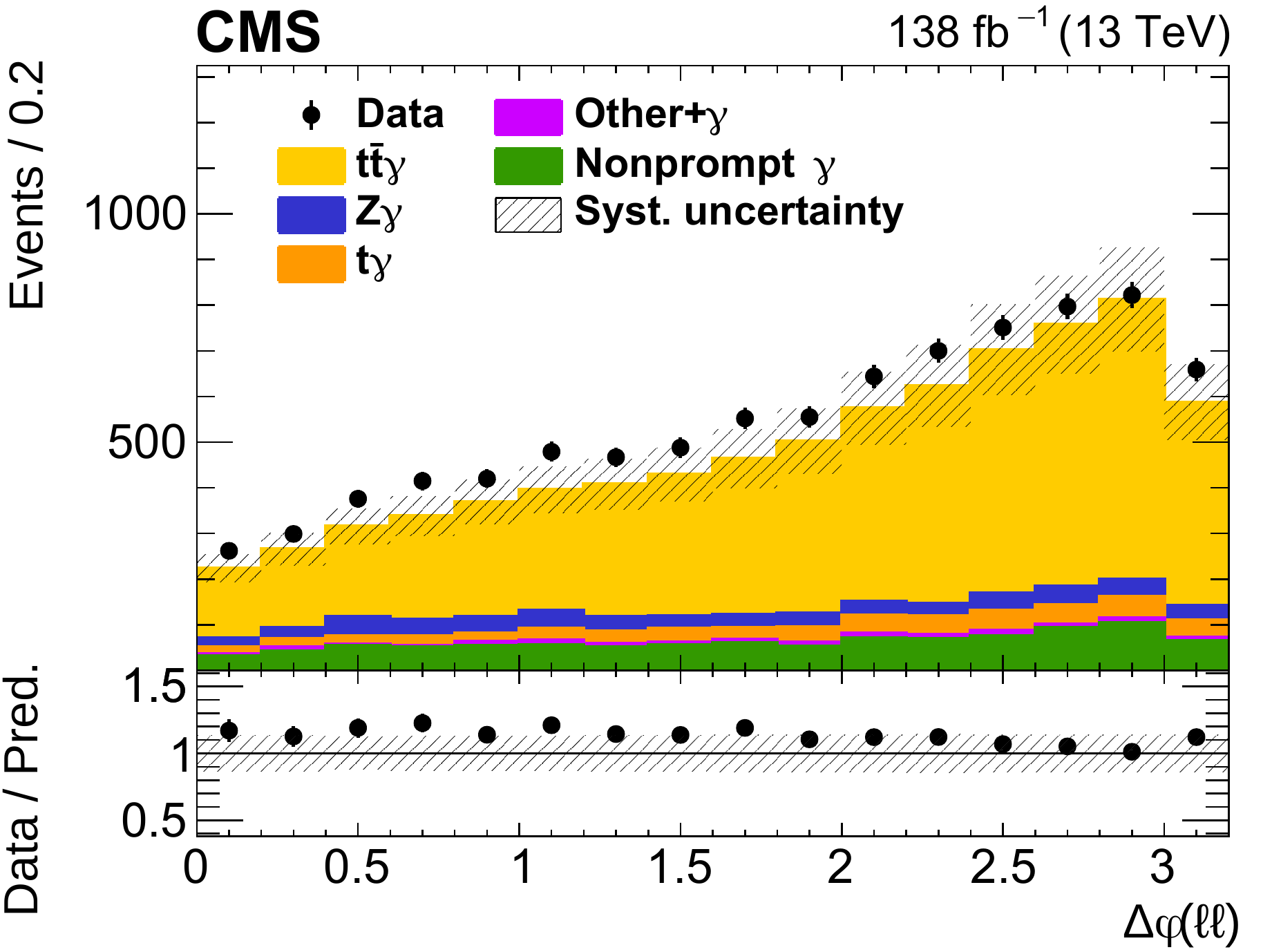}
\\
\includegraphics[width=0.42\textwidth]{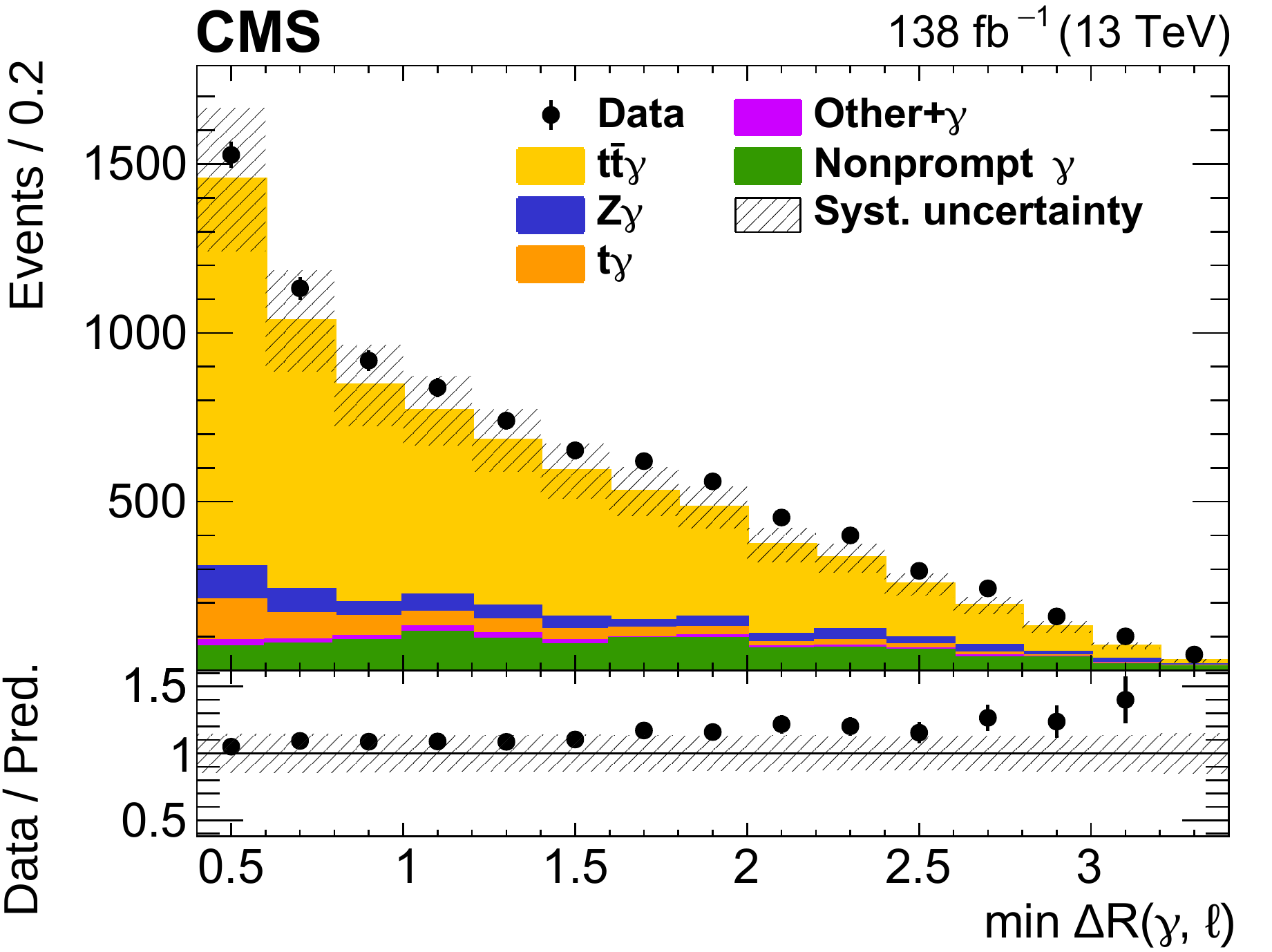}
\hspace{0.02\textwidth}
\includegraphics[width=0.42\textwidth]{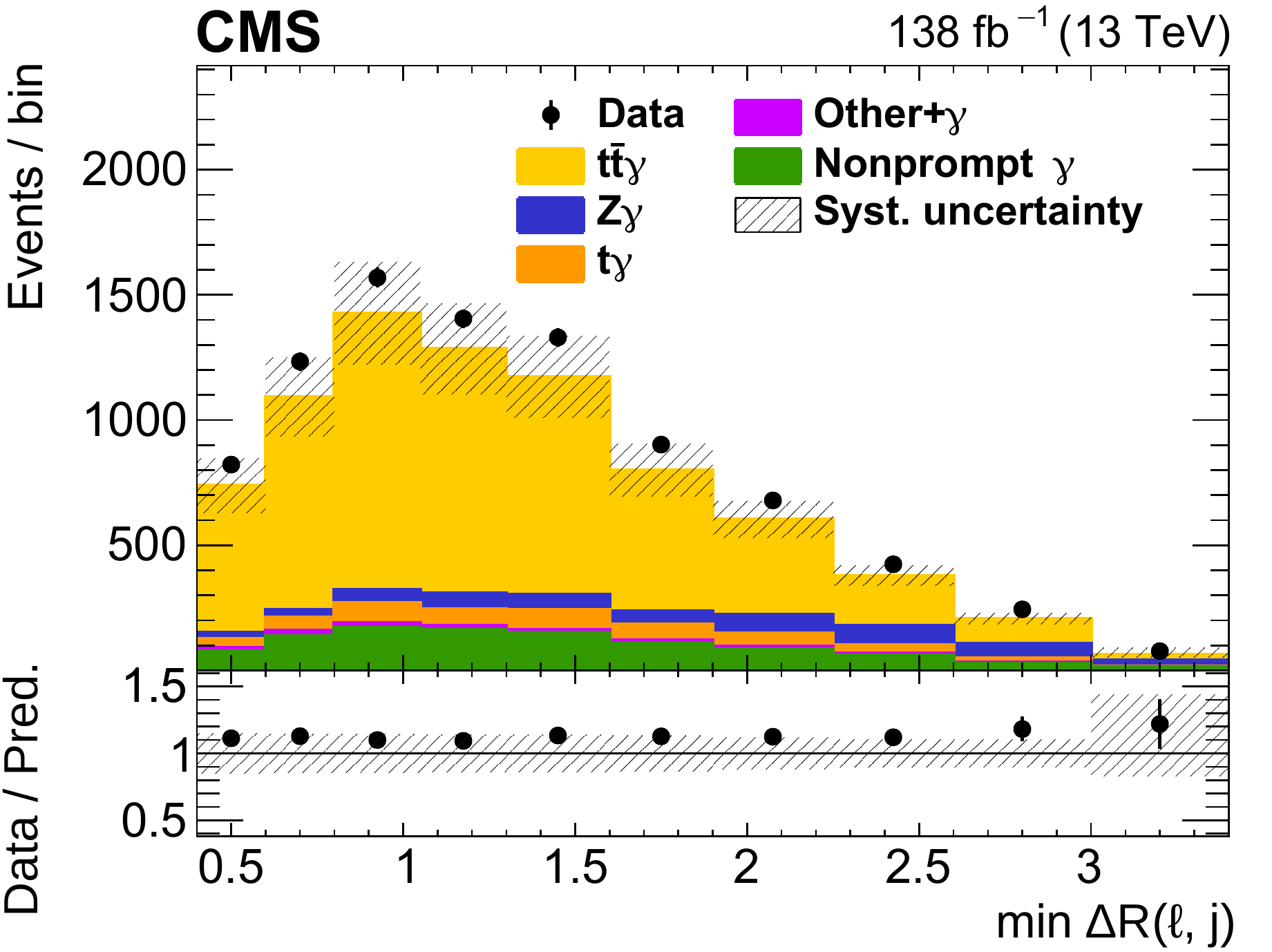}
\caption{The observed (points) and predicted (shaded histograms) signal and background yields as functions of the scalar \pt sum (upper left) and $\phi$ difference (upper right) of the two leptons, the smallest \dR between the photon and any lepton (lower left), and between any lepton and any jet (lower right), after applying the signal selection.
Details can be found in the caption of Fig.~\ref{fig:yieldSR1}.}
\label{fig:yieldSR2}
\end{figure}

The observed data yields for a selection of observables are shown in Figs.~\ref{fig:yieldSR1} and~\ref{fig:yieldSR2}, and compared to the signal yields expected from the simulated event samples and the predicted background yields, as described in Section~\ref{sec:backgrounds}.
The signal and background yields are normalized to the expected values, without taking the results of the fit to the data described in Section~\ref{sec:inclusive} into account.
The signal yields are normalized to the SM prediction and a normalization uncertainty of 18\% is included in the systematic uncertainty bands to account for the uncertainty in the prediction.

While an overall normalization difference, consistent with the result of the fit to the data performed for the inclusive cross section measurement discussed in Section~\ref{sec:inclusive}, is found between the observed and predicted yields, the shapes of the distributions are generally well described with two exceptions.
The observed jet multiplicity is larger than expected from the LO event samples for the \ttG signal process, likely explained by the missing simulation of additional radiation in the matrix element calculation.
Similarly, the observed minimal \dR separation between the photon and any lepton being larger than expected from the simulated \ttG event samples hints at a mismodelling of different photon origins in the LO simulation, as discussed in Ref.~\cite{Bevilacqua:2019quz}.

\section{Background estimation}\label{sec:backgrounds}

Background contributions to the final state with two leptons, one photon, and at least one \PQb-tagged jet arise from several SM processes.
A distinction is made between events with a prompt or a nonprompt photon.
For simulated events of all considered processes, the reconstructed photon is matched to the closest generated stable particle with $\dR<0.3$ for which the reconstructed and generated \pt agree within 50\%.
A reconstructed photon is classified as ``prompt'' if it is matched to a generated photon that is radiated from a lepton, quark, or heavy boson.
If the match is to a photon originating from other types of particles (\eg photons from $\pizerodecay$ decays or other hadronic sources), or it is matched to other types of particles (\eg electrons misidentified as a photon), or no match is found (\eg photons from pileup interactions), it is classified as ``nonprompt''.

Background contributions with prompt photons are estimated from simulated event samples.
The most important contributions arise from \Zjets and single top quark production (\tchannel, \schannel, and \tW) in association with a photon (referred to as \ZG and \tG, respectively).
These two categories are treated separately, while all other processes with prompt photons are grouped as ``other+\PGg''.
Normalization uncertainties of 5, 10, and 30\% are assigned to the \ZG, \tG, and other+\PGg background predictions, respectively, based on the precision of corresponding cross section measurements~\cite{CMS:2014cdf, ATLAS:2016qjc, CMS:2018amb}.
To improve the modelling of the \ZG prediction at high jet multiplicity, correction factors are derived from data samples orthogonal to the signal selection, as detailed in Section~\ref{sec:zgbackground}.
For all background contributions with nonprompt photons, an estimation based on data control samples is used, as described in Section~\ref{sec:nonprompt}.

After applying the event selection described in Section~\ref{sec:selection}, about 13\% of all selected events contain a nonprompt photon.
Considering only the estimated background contributions as determined from simulation and the data control samples, prompt and nonprompt photon events contribute about equally.

\subsection{The \texorpdfstring{\ZG}{Zg} control region}\label{sec:zgbackground}

Prompt photons in \Zjets production events originate either by initial-state radiation (ISR) from an incoming quark or final-state radiation (FSR) from one of the leptons in the \PZ boson decay.
To validate the prediction of this \ZG contribution from simulated event samples, a control region in data is defined using the inverted requirement $\dMllgZ<15\GeV$ for events with a same-flavour lepton pair (\ee or \mumu), which enriches the selection of events where the reconstructed photon originates from FSR in \Zjets production.
By keeping the requirement $\dMllZ>15\GeV$ unchanged, the contribution from \Zjets production events with nonprompt photons remains small.

\begin{figure}[!b]
\centering
\includegraphics[width=0.42\textwidth]{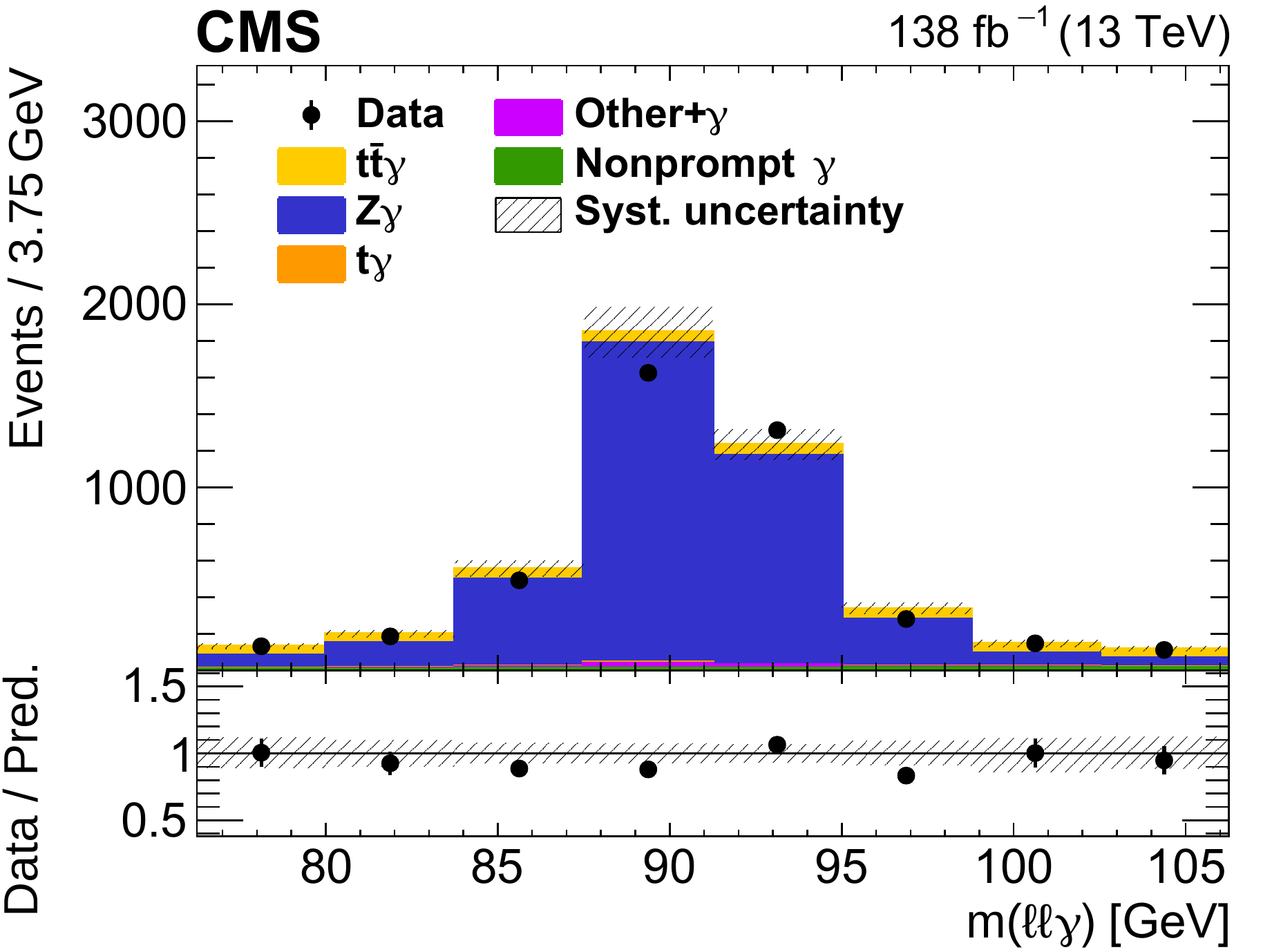}
\hspace{0.02\textwidth}
\includegraphics[width=0.42\textwidth]{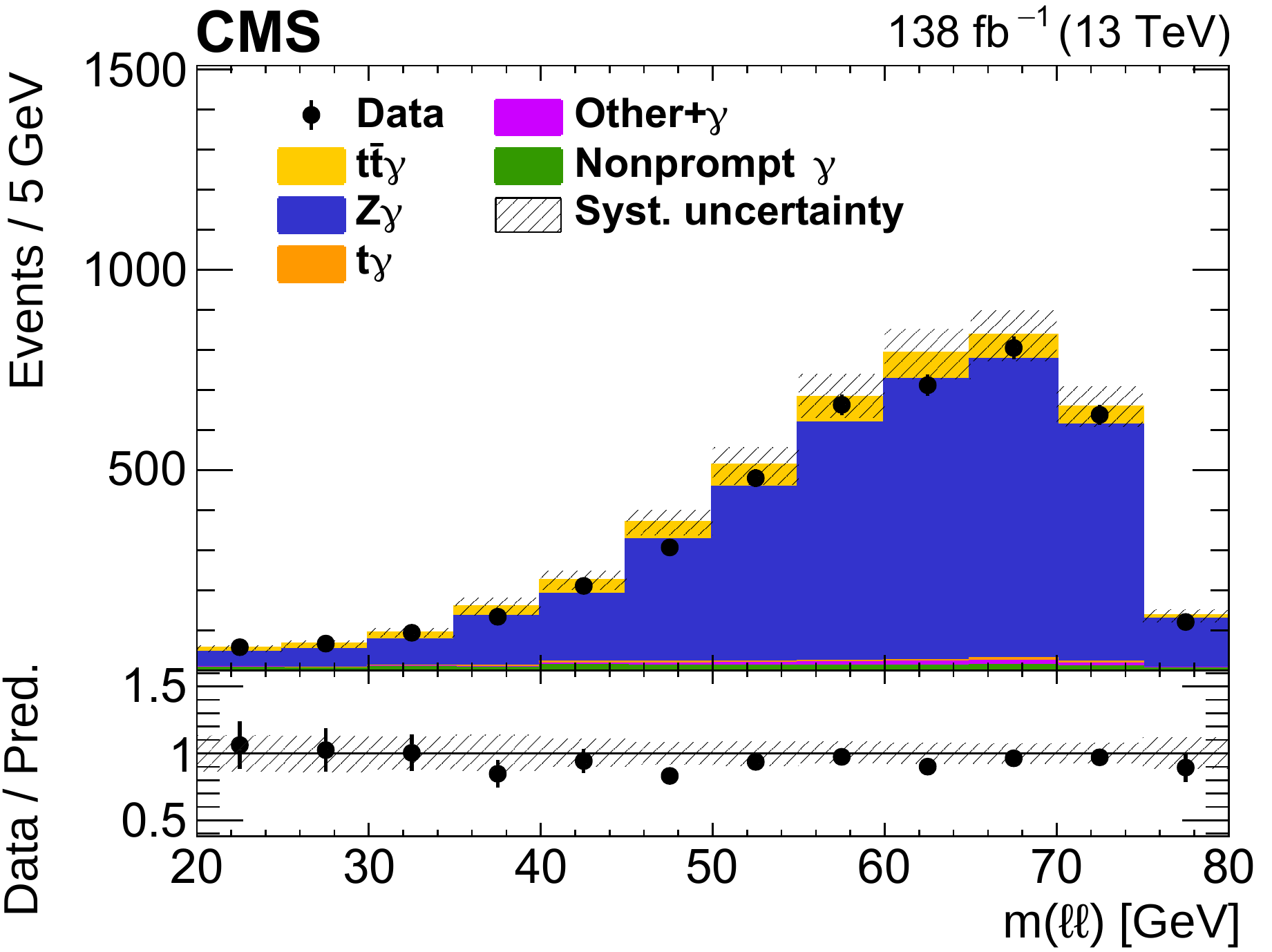}
\\
\includegraphics[width=0.42\textwidth]{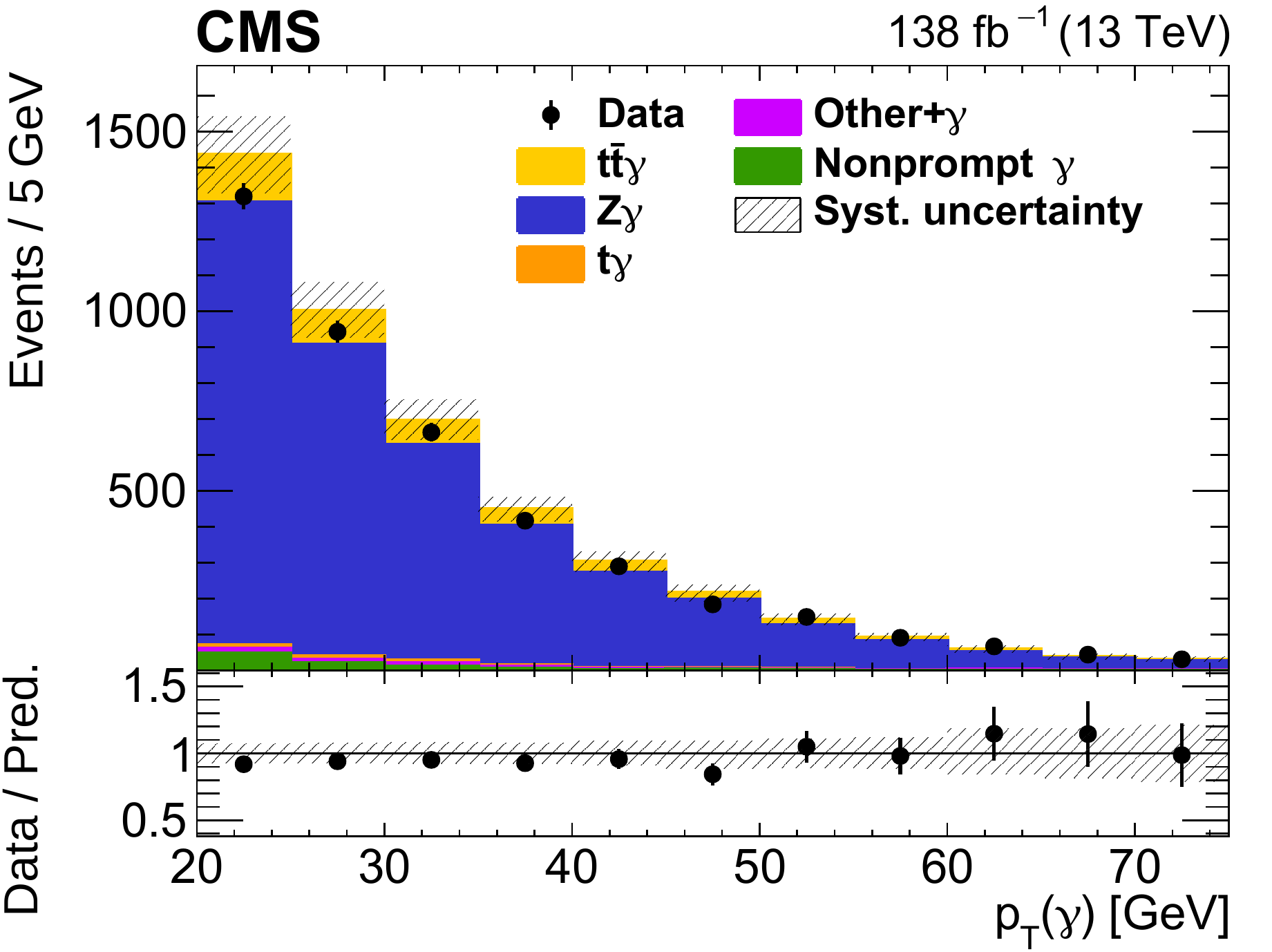}
\hspace{0.02\textwidth}
\includegraphics[width=0.42\textwidth]{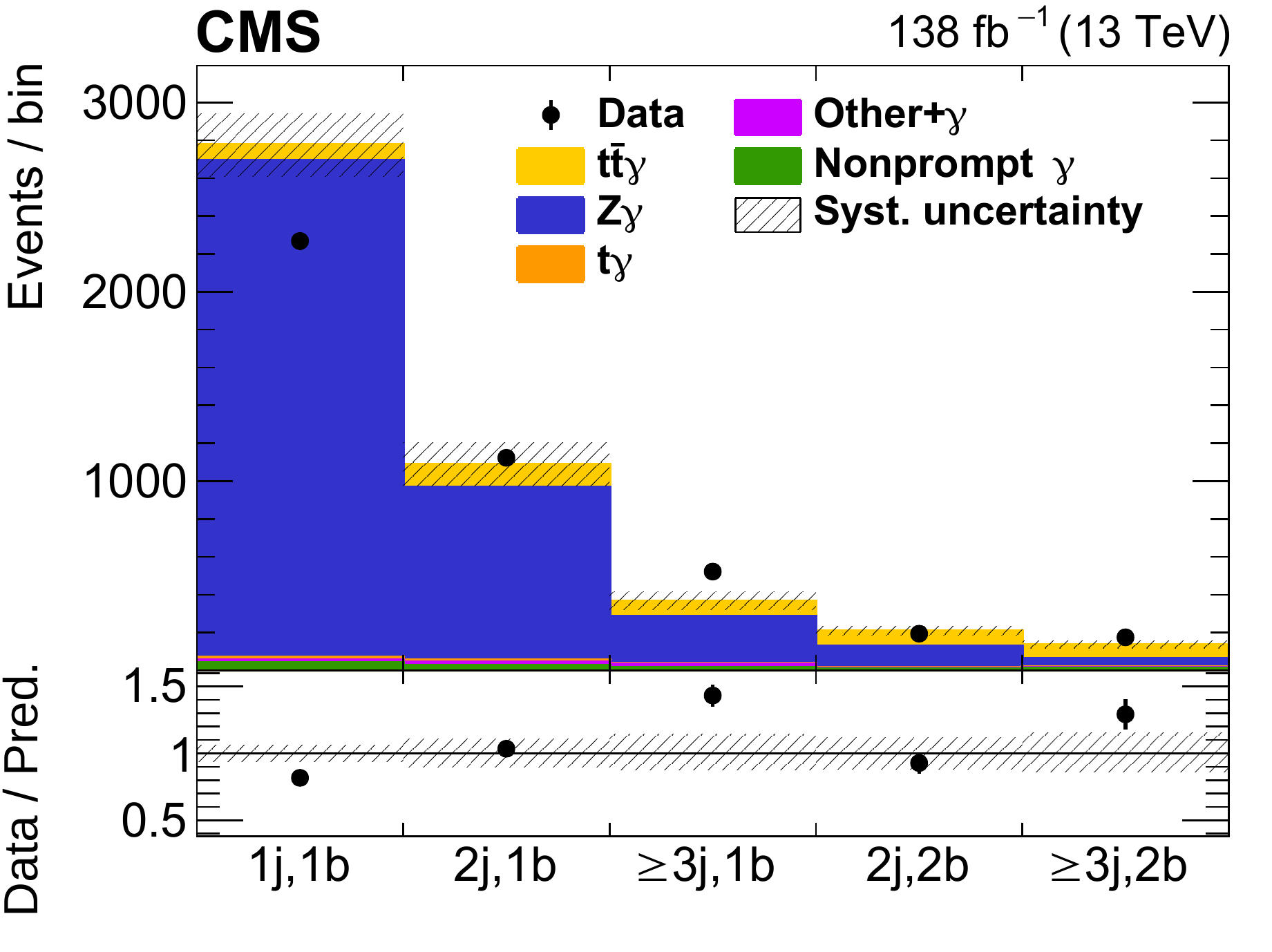}
\caption{The observed (points) and predicted (shaded histograms) event yields as a function of \mllg (upper left), \mll (upper right), photon \pt (lower left), and the number of jets j and \PQb-tagged jets \PQb (lower right), after applying the event selection for the \ZG control region.
The vertical lines on the points show the statistical uncertainties in the data, and the hatched bands the systematic uncertainty in the predictions.
The lower panels show the ratio of the event yields in data to the sum of the predictions.}
\label{fig:yieldZG}
\end{figure}

In Fig.~\ref{fig:yieldZG}, the measured and predicted yields are compared as functions of \mllg, \mll, the reconstructed photon \pt, and the numbers of jets and \PQb-tagged jets (\Njets and \Nbjets, respectively).
The photon \pt shape is well described, but a clear mismodelling is observed in the distributions of jet and \PQb jet multiplicity.
We derive correction factors as functions of \Njets and \Nbjets, which are applied in the signal selection to the \ZG background yields.
No correction factors are applied to \emu events since the \ZG yield is negligible for events with different-flavour leptons.

The data yields are found in bins of $(\Njets,\Nbjets)$, and the statistical uncertainty in each bin yield is propagated to the correction factor uncertainties.
Additionally, the normalization of the signal contamination in the \ZG control region is considered as a source of systematic uncertainty.
With a cross section normalization uncertainty of 18\% and a fraction of \ttG events in the control region of about 10\%, the resulting normalization uncertainty in the \ZG yield in the signal region is 1.8\%.

The correction factors significantly improve the precision of the prediction for FSR photons in \Zjets events.
After the signal selection, about 70\% of the \ZG background events are from FSR photons.
The contribution of ISR photons to the \ZG background is not constrained by using the control region, and thus retains a normalization uncertainty of 5\%~\cite{ATLAS:2016qjc}, resulting in an additional uncertainty of 1.5\% in the normalization of the total \ZG yield.

\subsection{Nonprompt-photon background}\label{sec:nonprompt}

The contribution of background processes with nonprompt photons is estimated from control samples in data with a ``tight-to-loose ratio'' method.
A transfer factor \TF is defined as the ratio of the number of nonprompt-photon events where the photon passes the identification criteria to the number where it fails.
It is used to predict the number \NumSR of nonprompt-photon events in the signal region, defined by the selection requirements discussed in Section~\ref{sec:selection}, from the observed number \NumSB of events in a sideband region enriched with nonprompt-photon events as $\NumSR=\TF\NumSB$.

The sideband region contains events that pass the signal selection except for having a photon that fails the \sieie identification criterion, which enriches the sideband region in nonprompt photon events.
To increase the fraction of nonprompt-photon events in the sideband region to more than 95\%, the photon is required to have $\sieie>0.012$.

For the evaluation of \TF, a measurement region with nonisolated-photon events is defined by inverting the signal region requirement on the charged particle isolation parameter of the reconstructed photon to $1.14<\chIso<15\GeV$.
To increase the amount of data in the measurement region, the $\Nbjets\geq1$ requirement is replaced by the looser requirement of $\Njets\geq1$ for \ee and \mumu events, and no jet requirement is imposed for \emu events.
The sideband region of the measurement region, defined by inverting both the \sieie and \chIso requirements of the signal region, has a fraction of nonprompt-photon events larger than 99.5\%.

The transfer factor is measured separately in bins of \pt and \abseta of the reconstructed photon.
Contributions from prompt-photon events in the measurement region and both sideband regions are estimated from simulated event samples and subtracted before the evaluation of \NumSR.

\begin{figure}[!ht]
\centering
\includegraphics[width=0.42\textwidth]{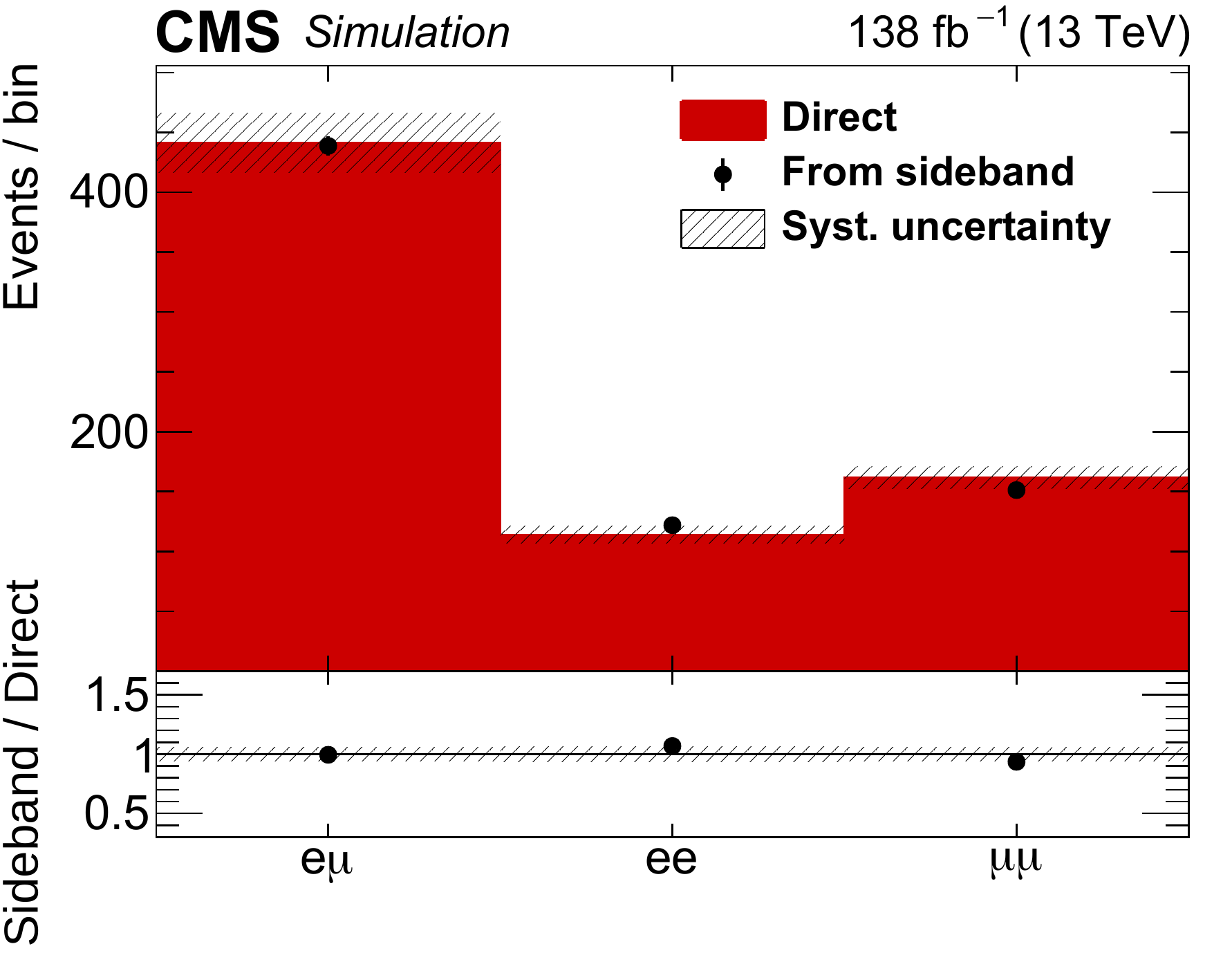}
\hspace{0.02\textwidth}
\includegraphics[width=0.42\textwidth]{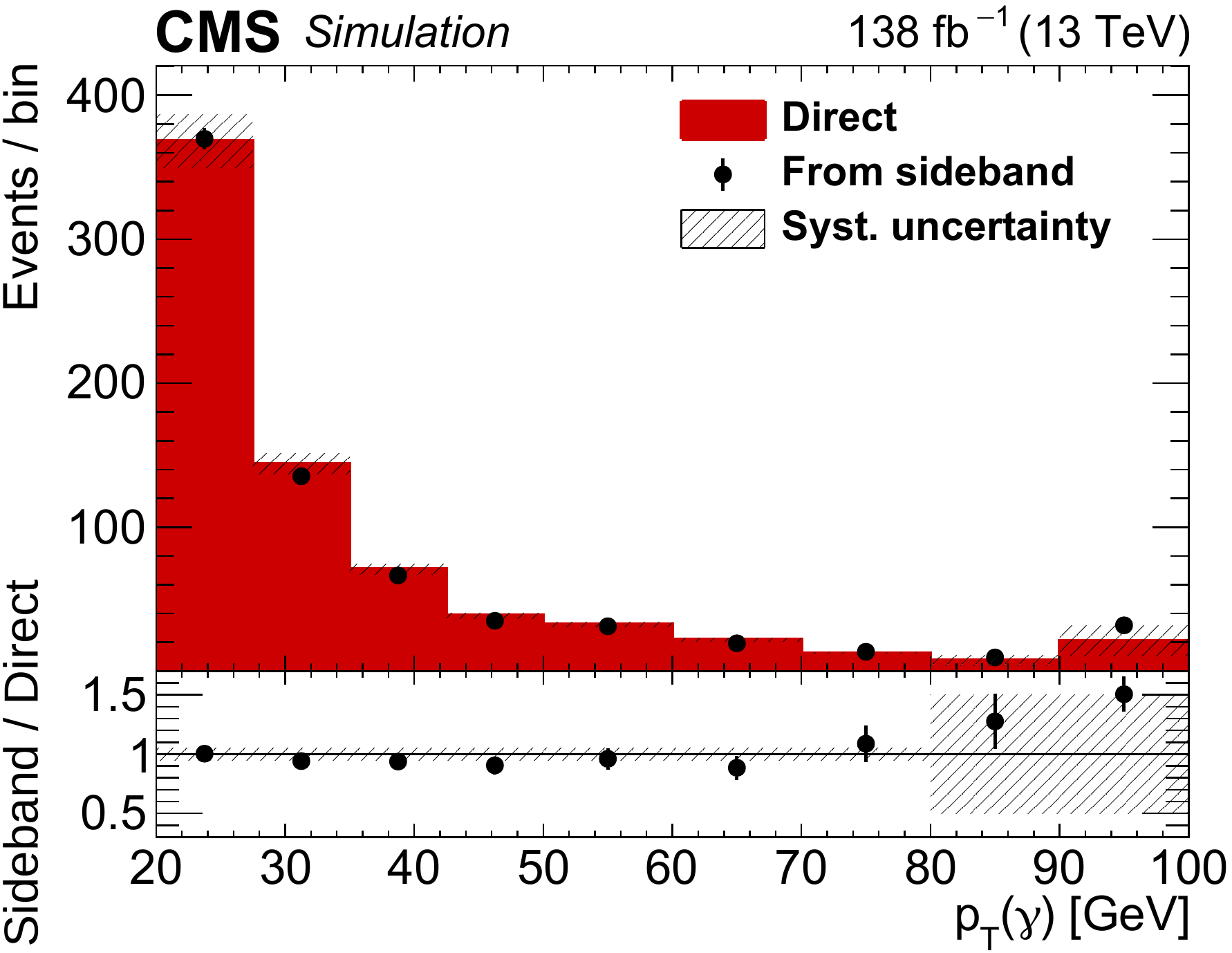}
\caption{Event yields in the signal region predicted from a simulated \ttbar event sample (shaded histogram) and estimated from applying the transfer factor to the event yields of the same sample in the sideband region (points), as a function of the lepton flavour (left) and the photon \pt (right).
The vertical lines on the points show the statistical uncertainties from the simulated event samples, and the hatched bands the total systematic uncertainty assigned to the nonprompt-photon background estimate.
The lower panels show the ratio of the two predictions.}
\label{fig:closure}
\end{figure}

We validate the performance of the nonprompt-photon background estimate with simulated event samples.
The transfer factors are measured from \ttbar and \Zjets samples since the measurement region is enriched in these two processes.
In the signal selection, however, the contribution from \Zjets production is minimal, and thus the nonprompt-photon background contribution in the signal selection is estimated from \ttbar samples alone in the sideband region, and compared to the direct prediction of the \ttbar samples for the signal region.
The comparison is displayed in Fig.~\ref{fig:closure}, and shows good agreement between the two predictions at the level of 5\%, which is assigned as a constant systematic uncertainty.
Only at large reconstructed photon \pt does the use of the transfer factor result in an overprediction of the nonprompt-photon background.
Hence, we assign an additional uncertainty of 50\% to the predicted event yields where the reconstructed photon has $\pt>80\GeV$.

\section{Systematic uncertainties}\label{sec:systematics}

Systematic uncertainties affect the signal selection efficiency, the predicted and measured background yields, and the measured distributions.
For each source of systematic uncertainty, variations of the predicted signal and background yields are evaluated in the relevant distributions, and either used to construct nuisance parameters in the fits employed for the inclusive cross section measurement and the EFT interpretation, or to repeat the differential cross section measurement and evaluate the uncertainty in the unfolded distribution.
A summary of all systematic uncertainties and the estimated impact on the measured inclusive cross section is given in Table~\ref{tab:systematics}.
Additionally, the table indicates the treatment of the uncertainties between the 2016, 2017, and 2018 data sets as uncorrelated, partially correlated, or fully correlated.

\begin{table}[!t]
\centering
\topcaption{Summary of the systematic uncertainty sources in the \ttG cross section measurements.
The first column lists the source of the uncertainty.
The second column indicates the treatment of correlations between the uncertainties in the three data-taking years, where \fullcorr means fully correlated, \partcorr means partially correlated, and \nocorr means uncorrelated.
For each systematic source, the ``prefit'' uncertainty is estimated from a cut-and-count analysis of the expected and observed event yields separately in bins of \ptgamma and for the three data-taking years using the input variations; the typical range across the three years is shown in the third column and can be compared between the different uncertainty sources.
The last column gives the impact of each uncertainty source on the measured inclusive \ttG cross section after the fit to the data (``postfit'').
The last two rows give the statistical and total uncertainty in the measured cross section.}
\renewcommand{\arraystretch}{1.15}
\begin{tabular}{clccc}
    & \multirow{2}{*}{Source} & \multirow{2}{*}{Correlation} & \multicolumn{2}{c}{Uncertainty [\%]} \\[-3pt]
    & & & Prefit range & Postfit \\ \hline
    \multirow{10}{*}{\rotatebox{90}{Experimental}}
    & Integrated luminosity             & \partcorr & 1.3--3.2 &    1.7 \\
    & Pileup                            & \fullcorr & 0.1--1.4 &    0.7 \\
    & Trigger efficiency                & \nocorr   & 0.6--1.7 &    0.6 \\
    & Electron selection efficiency     & \partcorr & 1.0--1.3 &    1.0 \\
    & Muon selection efficiency         & \partcorr & 0.3--0.5 &    0.5 \\
    & Photon selection efficiency       & \partcorr & 0.4--3.6 &    1.1 \\
    & Electron \& photon energy         & \fullcorr & 0.0--1.1 &    0.1 \\
    & Jet energy scale                  & \partcorr & 0.1--1.3 &    0.5 \\
    & Jet energy resolution             & \fullcorr & 0.0--0.6 & $<$0.1 \\
    & \PQb tagging efficiency           & \partcorr & 0.9--1.4 &    1.1 \\
    & L1 prefiring                      & \fullcorr & 0.0--0.8 &    0.3 \\[\cmsTabSkip]
    \multirow{6}{*}{\rotatebox{90}{Theoretical}}
    & Values of \muF and \muR           & \fullcorr & 0.3--3.5 &    1.3 \\
    & PDF choice                        & \fullcorr & 0.3--4.5 &    0.3 \\
    & PS modelling: ISR \& FSR scale    & \fullcorr & 0.3--3.5 &    1.3 \\
    & PS modelling: colour reconnection & \fullcorr & 0.0--8.4 &    0.2 \\
    & PS modelling: \PQb fragmentation  & \fullcorr & 0.0--2.2 &    0.7 \\
    & Underlying-event tune             & \fullcorr & 0.5      &    0.5 \\[\cmsTabSkip]
    \multirow{5}{*}{\rotatebox{90}{Background}}
    & \ZG correction \& normalization   & \fullcorr & 0.0--0.2 &    0.1 \\
    & \tG normalization                 & \fullcorr & 0.0--0.9 &    0.8 \\
    & Other+\PGg normalization          & \fullcorr & 0.3--1.0 &    0.8 \\
    & Nonprompt \PGg normalization      & \fullcorr & 0.0--1.8 &    0.7 \\
    & Size of background samples        & \nocorr   & 1.5--7.6 &    0.9 \\[\cmsTabSkip]
    & Total systematic uncertainty      &           &          &    3.6 \\
    & Statistical uncertainty           &           &          &    1.4 \\[\cmsTabSkip]
    & Total uncertainty                 &           &          &    3.9 \\
\end{tabular}
\label{tab:systematics}
\end{table}

The integrated luminosities of the 2016, 2017, and 2018 data-taking periods are individually known with uncertainties of 1.2, 2.3, and 2.5\%, respectively~\cite{CMS:2021xjt, CMS:2018elu, CMS:2019jhq}.
Some systematic effects in the calibration of the luminosity measurements are correlated, such that the uncertainty in the integrated luminosity of the combined data set is 1.6\%.
The uncertainty in the integrated luminosity affects both the normalization of the background contributions predicted from simulated event samples, as well as the extraction of the measured cross section from the final estimate of the number of signal events.

Simulated events are reweighted such that the simulated distribution of the number of interactions in each bunch crossing matches the expected distribution, assuming a total inelastic \pp cross section of 69.2\unit{mb}~\cite{CMS:2020ebo}.
Changes in this cross section by 4.6\% are used to produce varied signal and background predictions, fully correlated between the three data-taking years.

The efficiency of the trigger selection is corrected in simulated events to match the efficiency in data measured from two separate classes of independent trigger paths based on hadronic activity or missing transverse momentum signatures.
The scale factors depend on the momentum of the two selected leptons and deviate from unity by up to 20\% for low-\pt electrons.
In addition to the statistical uncertainty in the trigger efficiency measurement, the difference between the trigger efficiencies measured from the two classes of independent trigger paths is included as a source of systematic uncertainty.
Both the statistical and systematic uncertainty in the trigger efficiency measurement are treated as uncorrelated between the three data-taking years.

For the reconstruction, identification, and isolation of electrons, muons, and photons, the efficiencies are measured with the ``tag-and-probe'' method~\cite{CMS:2020uim, CMS:2018rym} separately in data and simulation.
Scale factors are applied to simulated events to correct for differences, and uncertainties are evaluated by varying the scale factors, separately for electrons, muons, and photons.
Additionally, systematic uncertainties in the energy scale and resolution of electrons and photons are measured with electrons from \PZ boson decays.
Statistical (systematic) sources of uncertainty in these measurements are treated as uncorrelated (correlated) between the three data-taking years.

Uncertainties in the jet energy scale and resolution are evaluated by varying the \pt of the reconstructed jets in simulated events separately for the several uncertainty sources described in Ref.~\cite{CMS:2016lmd}.
For each source, two separate variations that are either fully correlated or uncorrelated between the three data-taking years are considered.

Differences in the \PQb tagging efficiency between data and simulated events are corrected by applying scale factors to simulated events.
Uncertainties are evaluated by varying the scale factors separately for light- and heavy-flavour jets, where both correlated and uncorrelated variations between the three data-taking years are considered~\cite{CMS:2017wtu}.

During the 2016 and 2017 data-taking periods, the ECAL L1 trigger in the forward endcap region ($\abseta>2.4$) exhibited a gradual shift in the timing of its inputs, leading to a specific inefficiency known as ``prefiring''.
The effect was found to be most relevant in events with jets reconstructed with $2.4<\abseta<3.0$ and $\pt>100\GeV$, affecting also measurements that do not directly rely on such jets.
A correction determined from an unbiased data sample is applied, and 20\% of the correction is assigned as the associated uncertainty.

Several theoretical uncertainties are considered in the simulation of both signal and background contributions.
To evaluate the impact in the choice of the factorization scale \muF and the renormalization scale \muR, these two parameters are scaled up and down by a factor of 2, individually and simultaneously, and the envelope of the variations is taken to estimate the uncertainty.
Following the procedure described in Ref.~\cite{Butterworth:2015oua}, the choice of the PDF set is evaluated by using the MC replicas in the NNPDF PDFs~\cite{NNPDF:2014otw, NNPDF:2017mvq} and taking the root-mean-square of the variations as an estimate of the uncertainty.
The choice of \muF for ISR and FSR in the parton shower simulation is separately varied up and down by a factor of 2.
The default colour reconnection model in the parton shower simulation of the \ttG signal samples is replaced by three alternative models~\cite{Argyropoulos:2014zoa, Christiansen:2015yqa}, and only a small impact on the cross section results is found.
The uncertainty in the \PQb fragmentation function is evaluated by varying the parameters of the Bowler--Lund function~\cite{Bowler:1981sb}.
Finally, the uncertainty in the underlying-event tune~\cite{CMS:2019csb, CMS:2016kle} is evaluated for the \ttG signal sample, which results in a 0.5\% variation of the predicted signal yield.
All theoretical uncertainties are treated as correlated between the three data-taking years.

For all background predictions, normalization uncertainties as detailed in Section~\ref{sec:backgrounds} are included in the measurements, and treated as correlated between the three data-taking years.
The statistical uncertainty in the background samples, \ie both simulated event samples and data control samples, is included as a systematic uncertainty as well.

\section{Inclusive cross section measurement}\label{sec:inclusive}

The inclusive cross section is extracted following the statistical procedure described in Refs.~\cite{ATLAS:2011tau, Cowan:2010js} from a profile likelihood fit to the reconstructed photon \pt distribution.
The binned likelihood function $\likelihood(\sigstr,\nuisan)$ is constructed as the product of the Poisson probabilities to obtain the observed yields given the predicted signal and background estimates in each event category, and includes terms to account for the systematic uncertainties and their correlation pattern, as described in Section~\ref{sec:systematics}.
The signal strength modifier \sigstr scales the normalization of the predicted signal estimate, and \nuisan denotes a full set of nuisance parameters representing the systematic uncertainties, including the normalization of the background predictions.
The event categories are the bins of the photon \pt distribution separately for each year of data taking and for the three lepton-flavour channels.

The quantities \sigstrmax and \nuisanmax denote the signal strength and nuisance parameter set that simultaneously maximize the likelihood function.
Similarly, \nuisanmaxr maximizes the likelihood function for a fixed value of \sigstr.
The observed cross section is extracted using a limit calculation in which the test statistic $q(\sigstr)=-2\ln\likelihood(\sigstr,\nuisanmaxr)/\likelihood(\sigstrmax,\nuisanmax)$, which is based on the profile likelihood function, is simplified using an asymptotic approximation~\cite{ATLAS:2011tau, Cowan:2010js}.

\begin{figure}[!pt]
\centering
\includegraphics[width=0.42\textwidth]{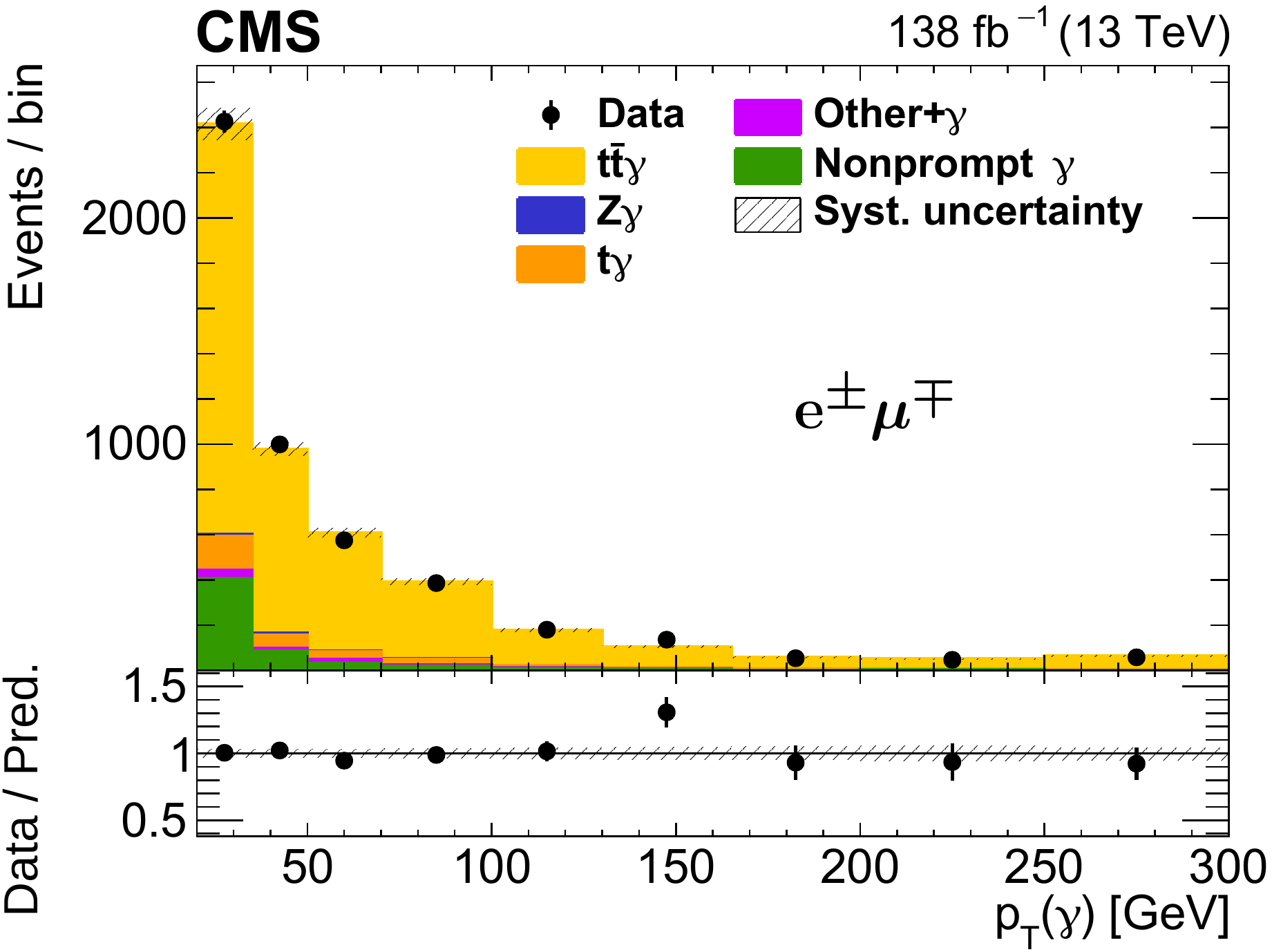}
\hspace{0.02\textwidth}
\includegraphics[width=0.42\textwidth]{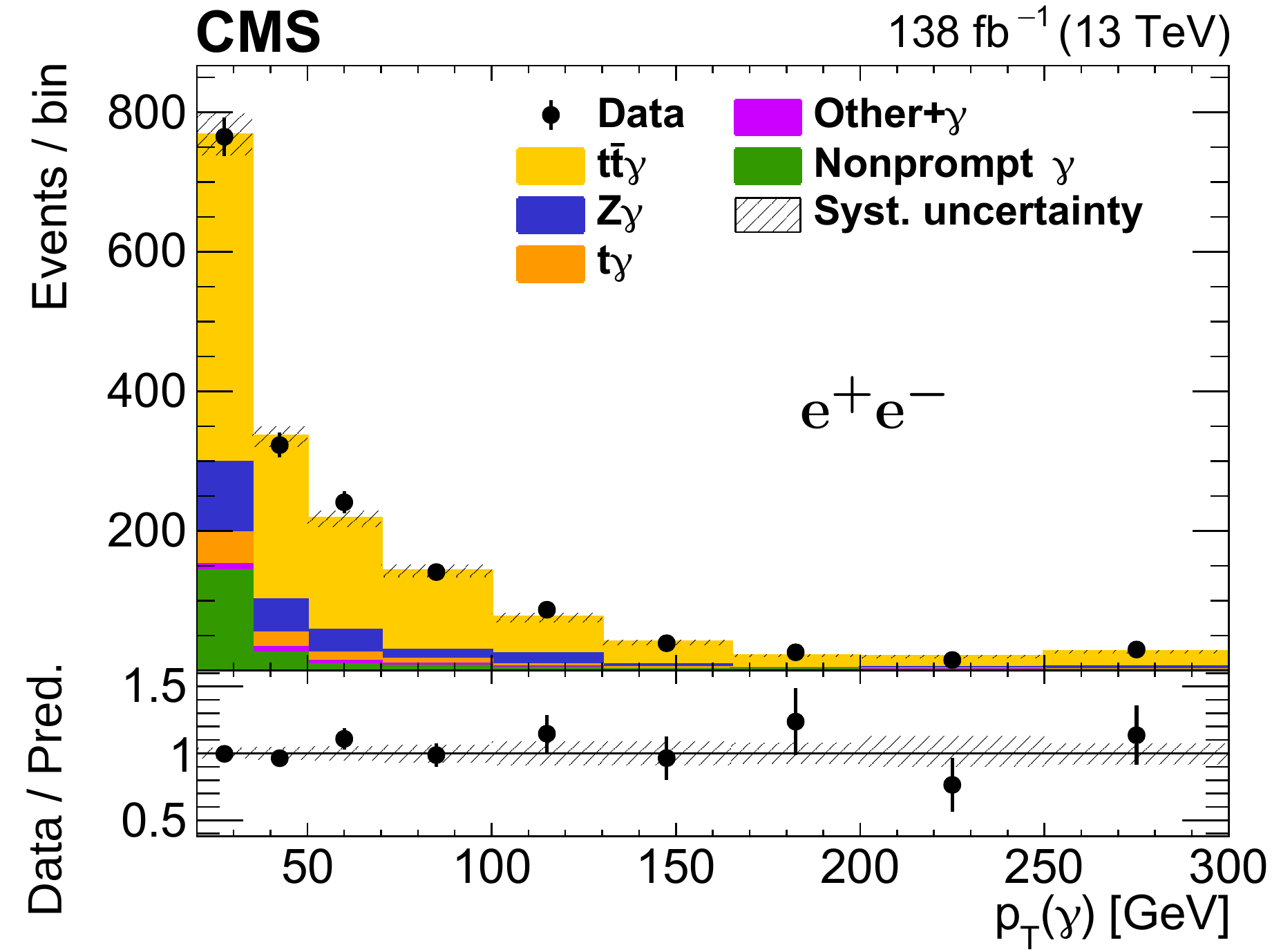}
\\
\includegraphics[width=0.42\textwidth]{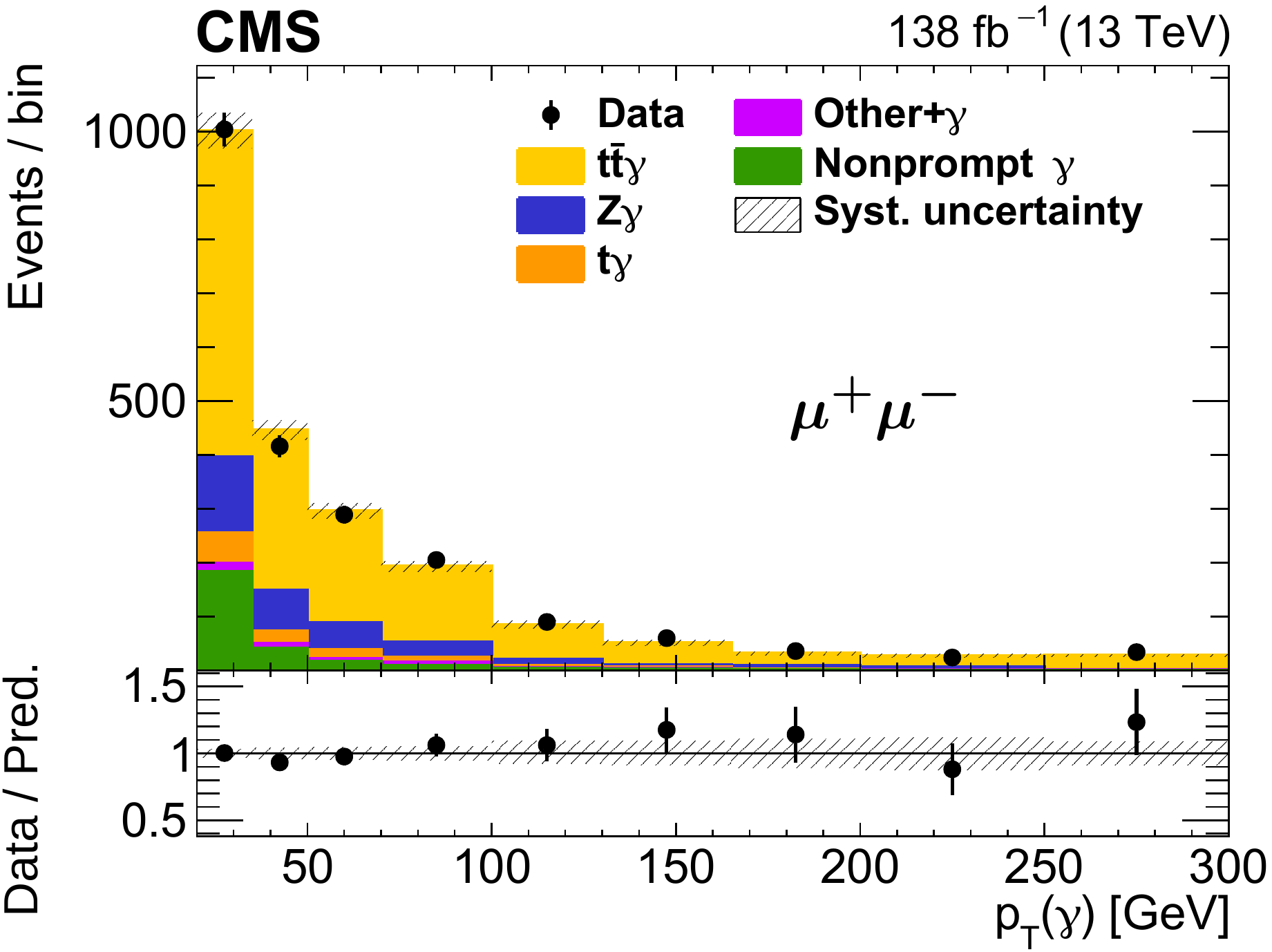}
\caption{The observed (points) and predicted (shaded histograms) event yields as a function of the reconstructed photon \pt after applying the signal selection, for the \emu (upper left), \ee (upper right), and \mumu (lower) channels, after the values of the normalizations and nuisance parameters obtained in the fit to the data are applied.
The vertical bars on the points show the statistical uncertainties in data, and the hatched bands the systematic uncertainty in the predictions.
The lower panels of each plot show the ratio of the event yields in data to the predictions.}
\label{fig:photonPt}
\end{figure}

\begin{figure}[!pt]
\centering
\includegraphics[width=0.6\textwidth]{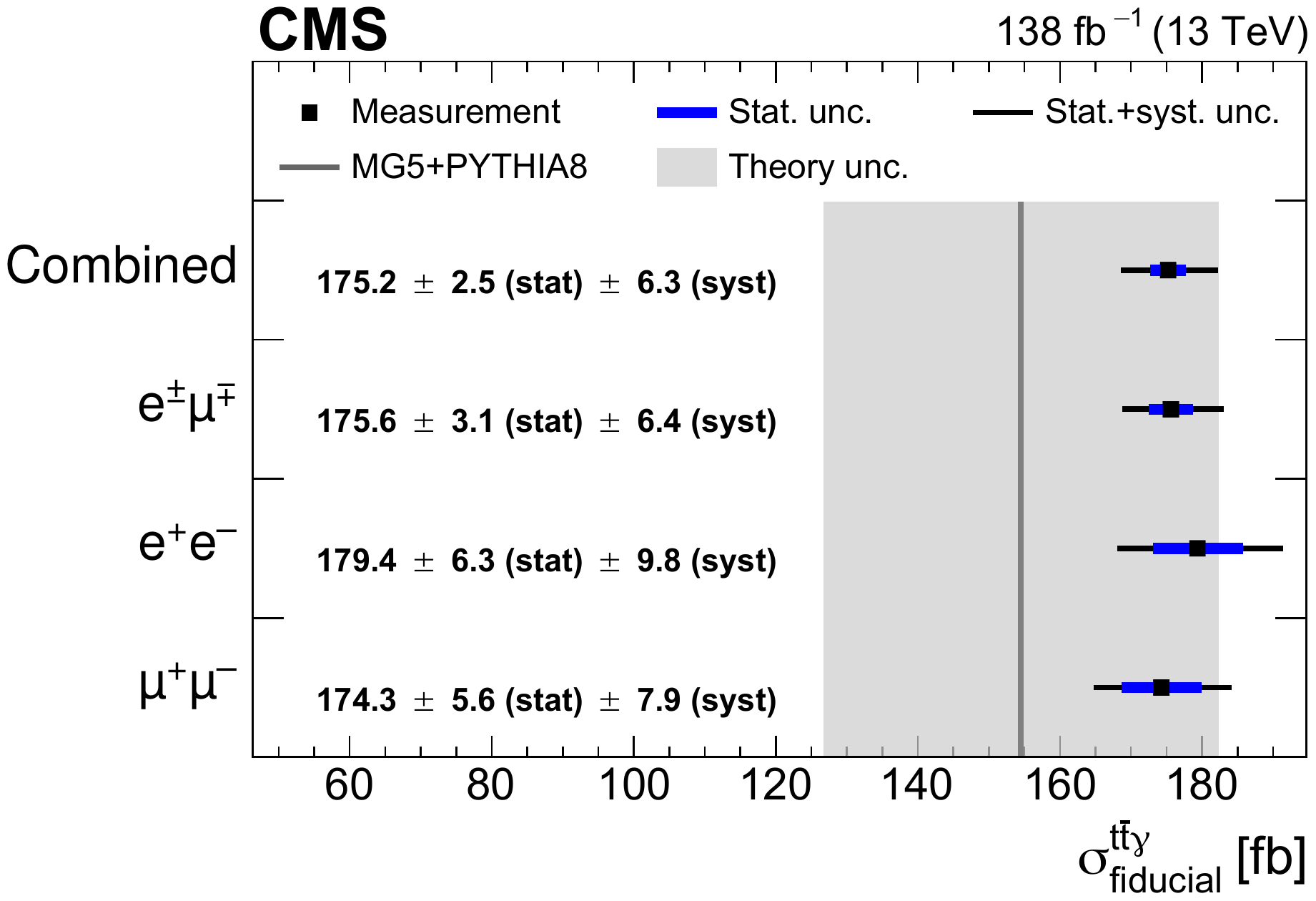}
\caption{The measured inclusive fiducial \ttG production cross section in the dilepton final state for the different dilepton-flavour channels and combined.
The thick and thin lines on the points show the statistical and total uncertainties, respectively.
The vertical line represents the SM prediction obtained with the \MGvATNLO (MG5) interfaced with \PYTHIAeight, as described in the text.
The shaded band shows the theoretical uncertainty in the prediction.}
\label{fig:channels}
\end{figure}

The reconstructed photon \pt distributions for each dilepton flavour, combined for the three data-taking years, are shown in Fig.~\ref{fig:photonPt} after the fit to the data.
The fiducial cross section for \ttG production in the dilepton final state is measured to be
\begin{linenomath}\begin{equation*}
    \sigfid=\XsecMeasuredFull.
\end{equation*}\end{linenomath}
Consistent results are found when fitting the flavour channels separately, as shown in Fig.~\ref{fig:channels}.

The impact of the different systematic uncertainty sources on the inclusive cross section measurements are summarized in Table~\ref{tab:systematics}.
The measurement from the fit of all the nuisance parameters describing the background yield normalizations and systematic uncertainty sources are consistent with their initial values within the obtained uncertainties.
The leading contributions to the overall systematic uncertainty arise from the uncertainty in the integrated luminosity, the values of \muF and \muR in the signal simulation, the PS modelling, and the \PQb tagging and electron selection efficiencies.

Compared to the CMS measurement of the inclusive \ttbar production cross section in the dilepton final state using 35.9\fbinv of data collected in 2016~\cite{CMS:2018fks}, this result profits significantly from improvements with respect to the two leading systematic uncertainties in the \ttbar measurement.
First, the uncertainty in the integrated luminosity was reduced from 2.5 to 1.7\%, mainly due to the improved luminosity measurement described in Ref.~\cite{CMS:2021xjt} and the combination of three data sets with only partially correlated luminosity uncertainties.
Second, the uncertainty in the lepton identification efficiency was improved from 2.0 to 1.1\%, due to the use of the identification discriminant presented in Ref.~\cite{CMS:2021rvz}.

The predicted cross section of $\sigSM=\XsecTheoryFull$ is in agreement with the measured cross section within the uncertainty, with a central value about 12\% smaller than the measurement, corresponding to a difference of 0.7 standard deviations.
As discussed in Section~\ref{sec:simulation}, the prediction is normalized with an NLO $K$-factor for the $2\to3$ process \pptottG, which does not take into account the radiation of photons from charged decay products of the top quarks.
This source of photons plays a significant role in the NLO calculation~\cite{Bevilacqua:2019quz} of the cross section, so neglecting it could be a cause for the theoretical prediction to be low.

The simulated \ttG samples used in the cross section measurement were generated assuming a top quark mass of 172.5\GeV.
With a higher top quark mass in the event generation, we would find a higher signal selection efficiency and thus a lower measured cross section.
For a quantitative evaluation of this effect, we reweight the simulated events to match the Breit--Wigner shapes corresponding to the varied top quark mass.
For an increase (decrease) of the top quark mass by the current experimental uncertainty of 0.3\GeV~\cite{ParticleDataGroup:2020ssz}, the measured cross section decreases (increases) by 0.4\%.

\section{Differential cross section measurement}\label{sec:differential}

The differential \ttG cross sections in the fiducial phase space are measured as functions of twelve observables defined in Table~\ref{tab:observables}.
For each observable, variable bin sizes are chosen to optimize the statistical uncertainty in the measured event yields and limit the number of bin-to-bin migrations.
In each bin of the reconstructed distributions, the yield is determined by subtracting the predicted background yields from the measured event yields, after which the expected fraction of \ttG events outside of the fiducial phase space is removed as well.

\begin{table}[!ht]
\centering
\renewcommand\arraystretch{1.15}
\topcaption{Definition of the observables used in the differential cross section measurement.}
\label{tab:observables}
\begin{tabular}{rl}
    Symbol & Definition \\
    \hline
    \ptgamma & Transverse momentum of the photon \\
    \etagamma & Absolute value of the pseudorapidity of the photon \\
    \drgammaclose & Angular separation between the photon and the closest lepton \\
    \drgammafirst & Angular separation between the photon and the leading lepton \\
    \drgammasecond & Angular separation between the photon and the subleading lepton \\
    \drgammabjet & Angular separation between the photon and the closest \PQb jet \\
    \detall & Pseudorapidity difference between the two leptons \\
    \dphill & Azimuthal angle difference between the two leptons \\
    \ptll & Transverse momentum of the dilepton system \\
    \sumptll & Scalar sum of the transverse momenta of the two leptons \\
    \drlepjet & Smallest angular separation between any of the selected leptons and jets \\
    \ptjet & Transverse momentum of the leading jet \\
\end{tabular}
\end{table}

Both detector response and acceptance effects are represented by response matrices derived from the \ttG \MGvATNLO simulated event samples,  using half the number of bins for the fiducial phase space observables as used in the reconstructed distributions to limit the bin-to-bin migrations, and including the same corrections and scale factors as used in the inclusive cross section measurement.
The differential cross section is then evaluated by applying an unfolding procedure.
The resolutions of the considered observables are found to be good, with a very small number of events migrating from one bin to another.
Under such conditions, matrix inversion without regularization is an unbiased and stable method to correct for detector response and acceptance~\cite{Cowan:1998ji}.
We apply the \TUnfold package~\cite{Schmitt:2012kp} to evaluate the differential cross sections from the background-subtracted measured event yields and the response matrices.

All systematic uncertainties described in Section~\ref{sec:systematics} are applied in the differential cross section measurement.
The theoretical uncertainties are evaluated by repeating the unfolding procedure with varied response matrices.
For the experimental uncertainties, the relative variation of the total signal and background predictions is evaluated, and this variation is then applied to the data before the subtraction of the varied background predictions.
The background normalization uncertainties are evaluated by varying the background subtraction.

\begin{figure}[!tp]
\centering
\includegraphics[width=0.42\textwidth]{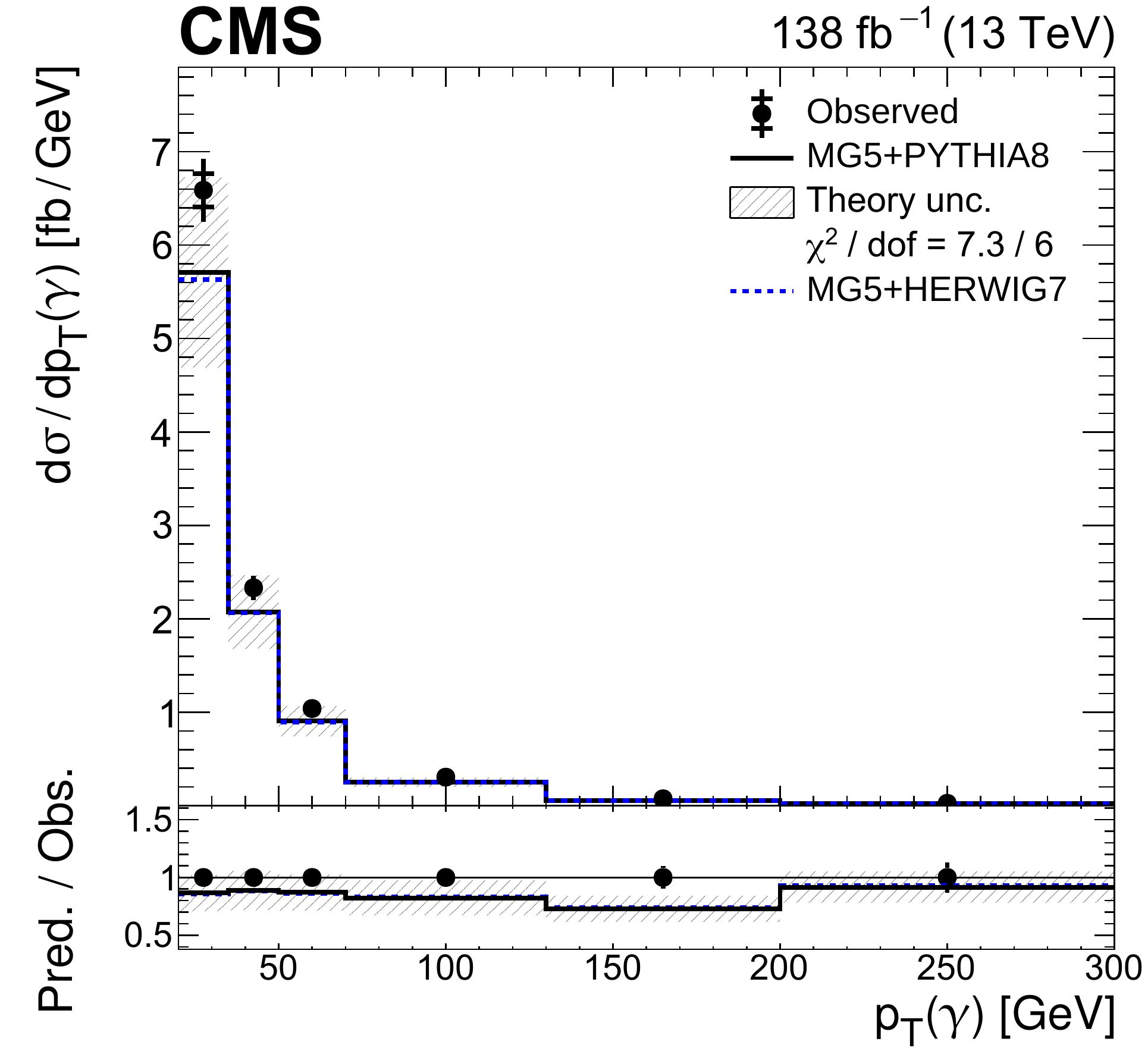}
\hspace{0.02\textwidth}
\includegraphics[width=0.42\textwidth]{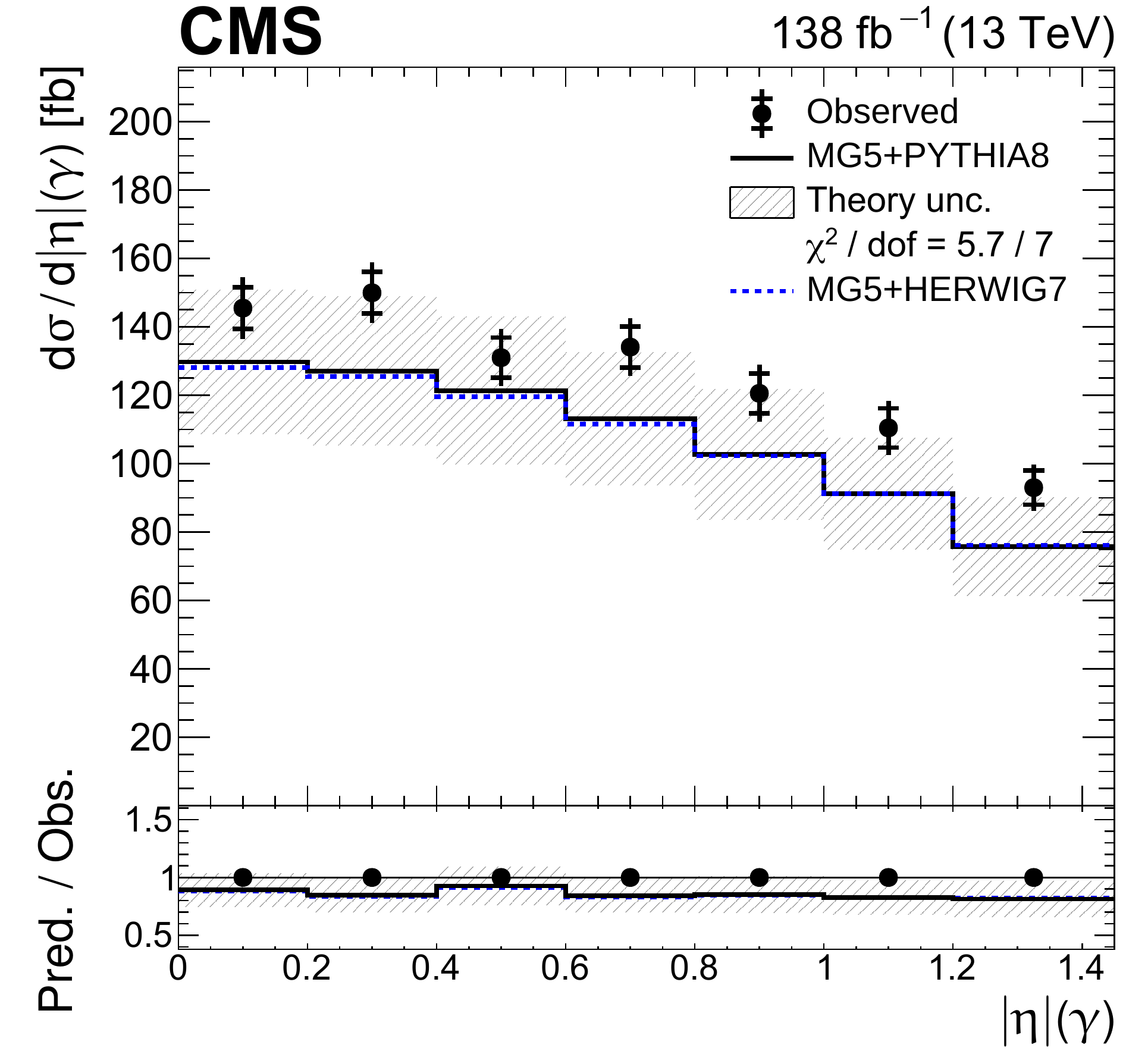}
\\
\includegraphics[width=0.42\textwidth]{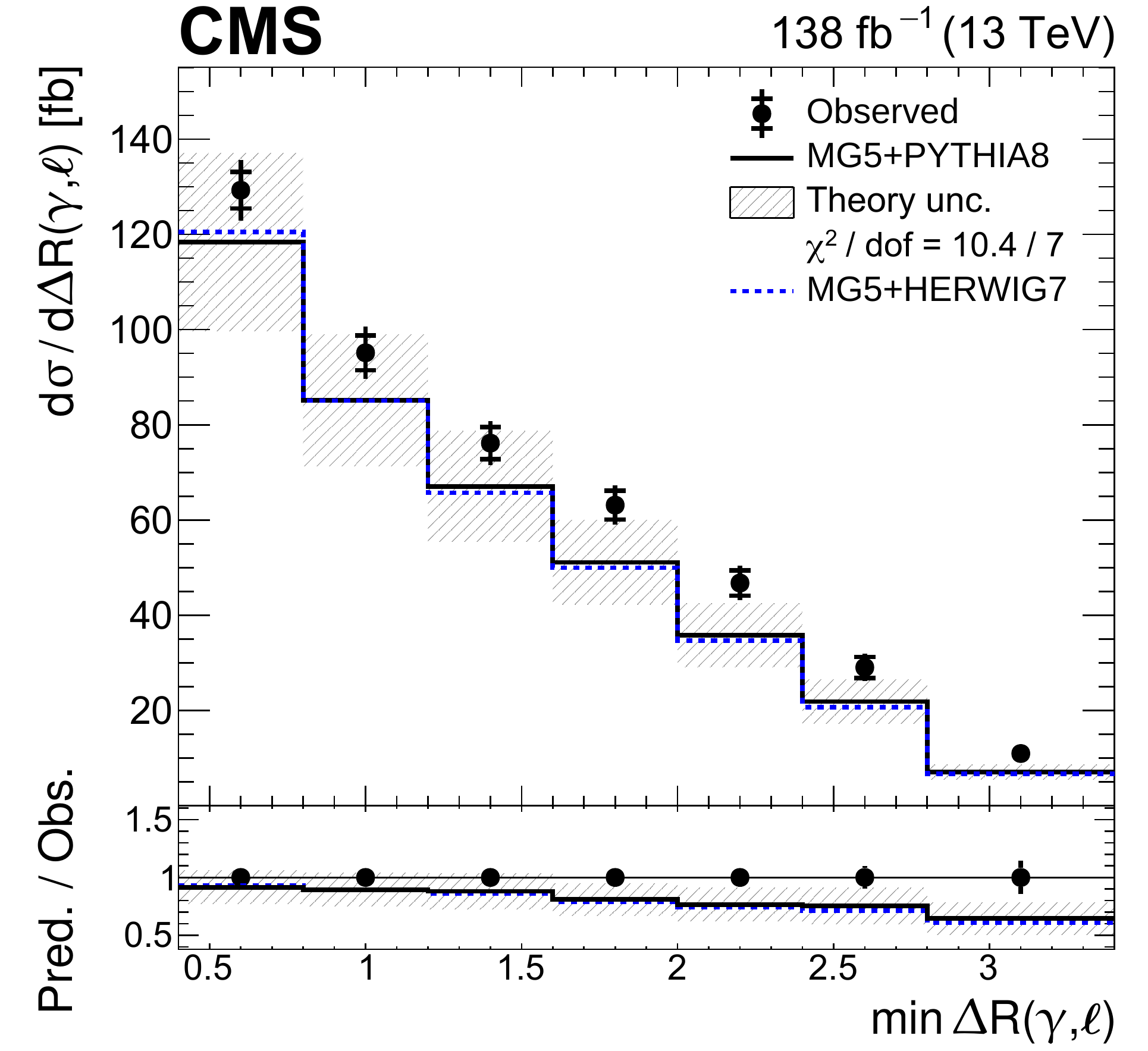}
\hspace{0.02\textwidth}
\includegraphics[width=0.42\textwidth]{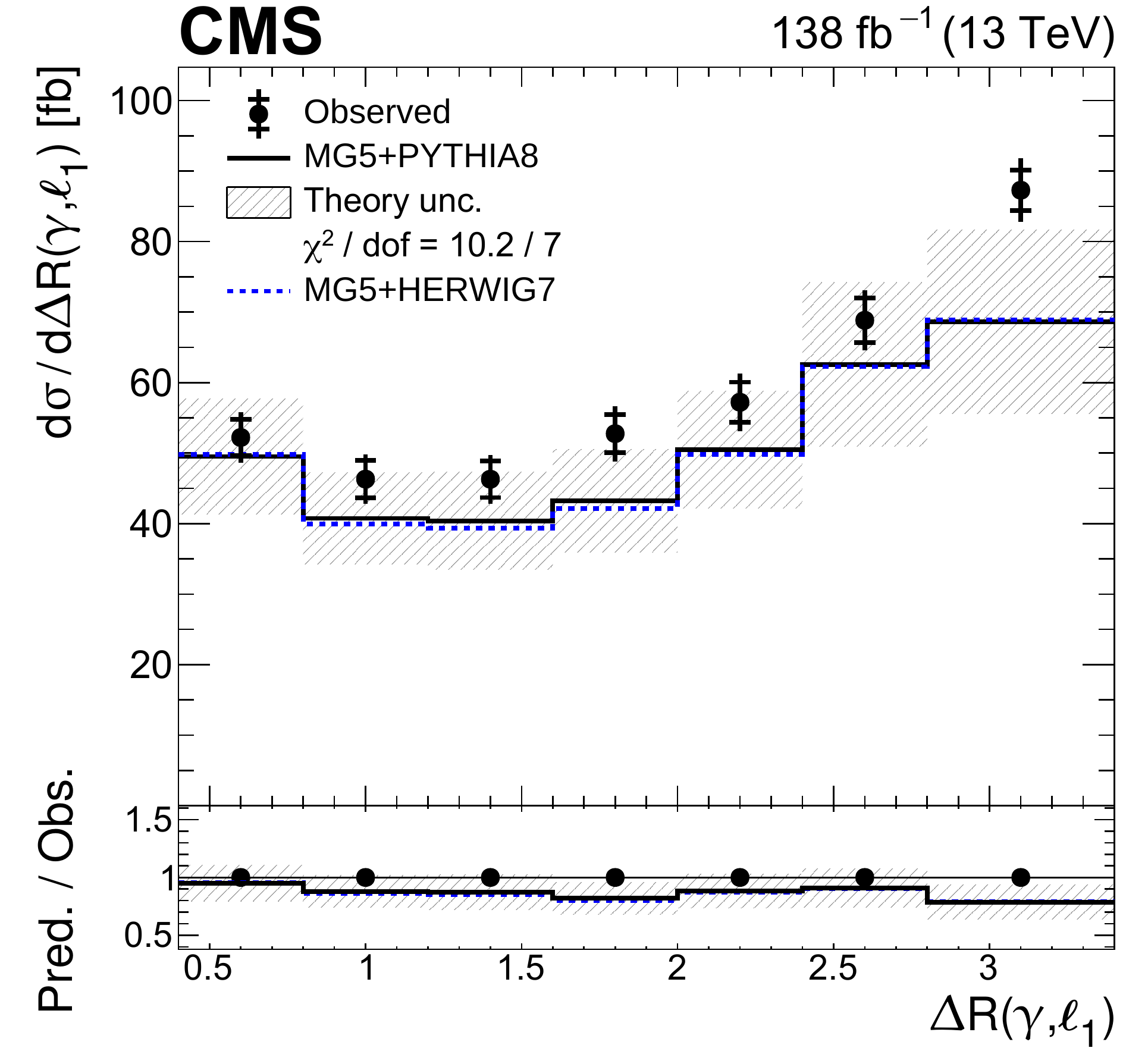}
\\
\includegraphics[width=0.42\textwidth]{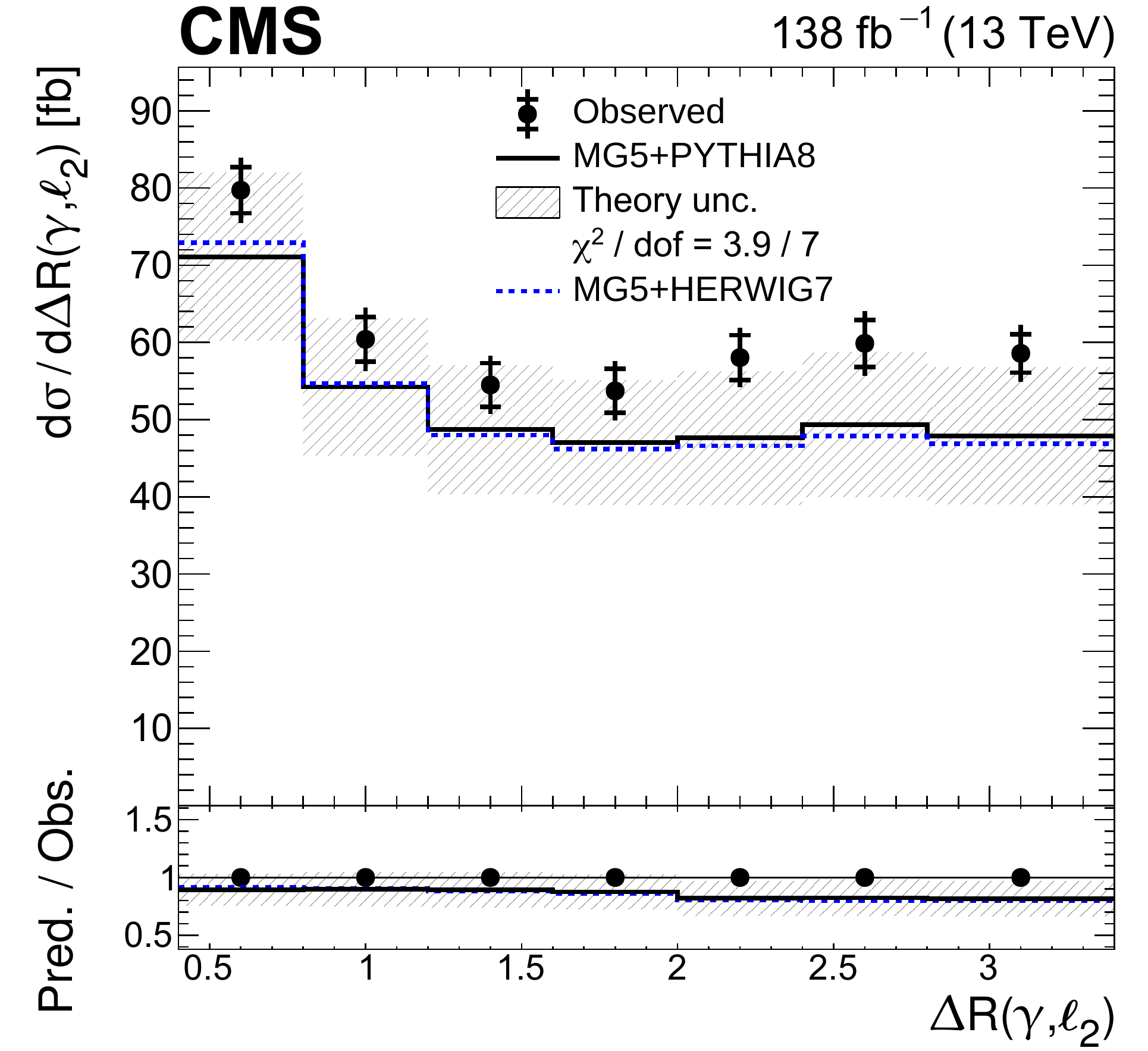}
\hspace{0.02\textwidth}
\includegraphics[width=0.42\textwidth]{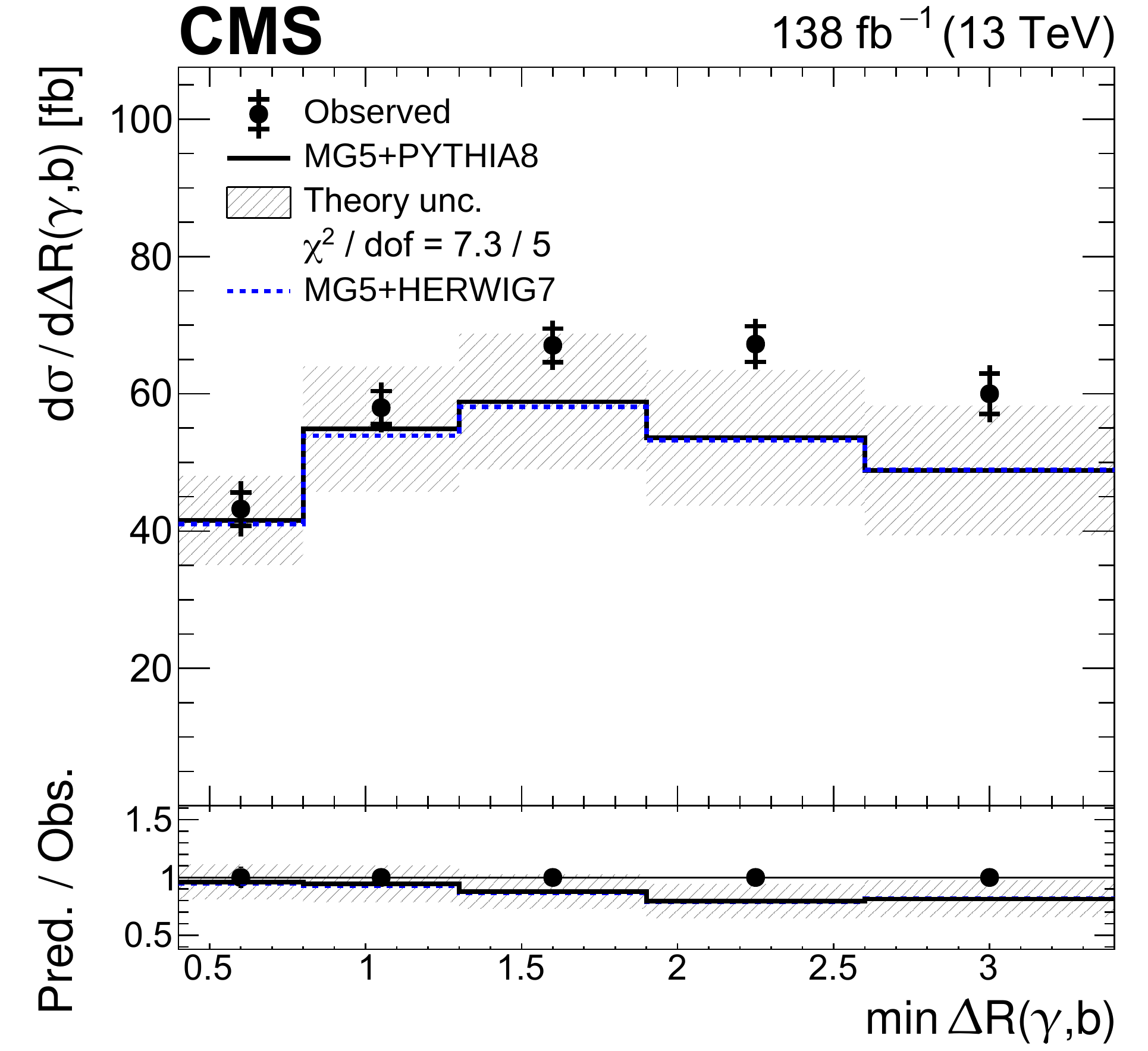}
\caption{Absolute differential \ttG production cross sections as functions of
\ptgamma (upper left),
\etagamma (upper right),
\drgammaclose (middle left),
\drgammafirst (middle right),
\drgammasecond (lower left),
and
\drgammabjet (lower right).
The data are represented by points, with inner (outer) vertical bars indicating the statistical (total) uncertainties.
The predictions obtained with the \MGvATNLO event generator interfaced with
\PYTHIAeight (solid lines) and $\HERWIG7$ (dotted lines) parton shower simulations are shown as horizontal lines.
The theoretical uncertainties in the prediction using \PYTHIAeight are indicated by shaded bands.
The lower panels display the ratios of the predictions to the measurement.
The values of the \chisq divided by the number of degrees of freedom (\dof) quantifying the agreement between the measurement and the prediction using \PYTHIAeight are indicated in the legends.}
\label{fig:unfoldedA}
\end{figure}

\begin{figure}[!tp]
\centering
\includegraphics[width=0.42\textwidth]{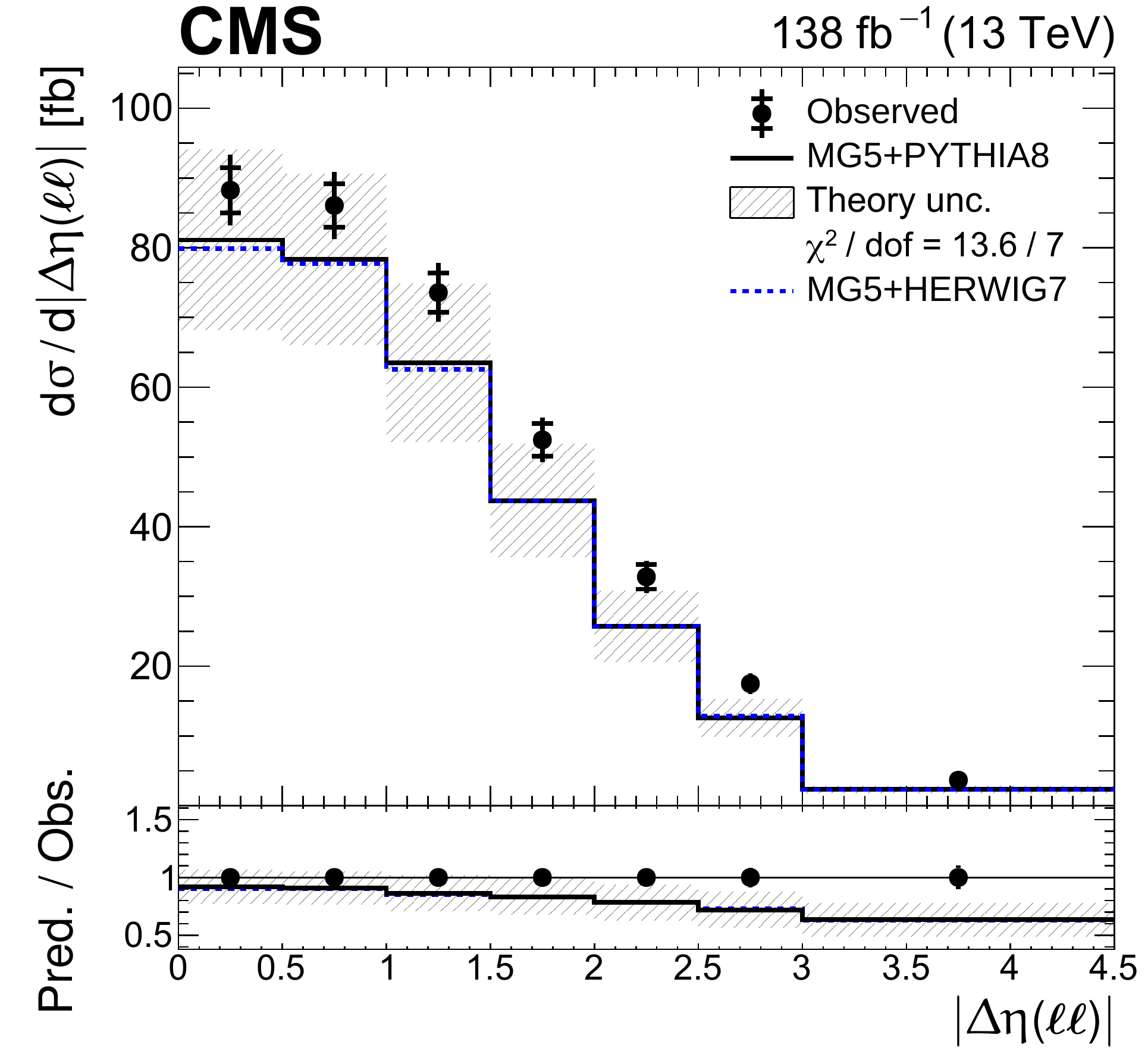}
\hspace{0.02\textwidth}
\includegraphics[width=0.42\textwidth]{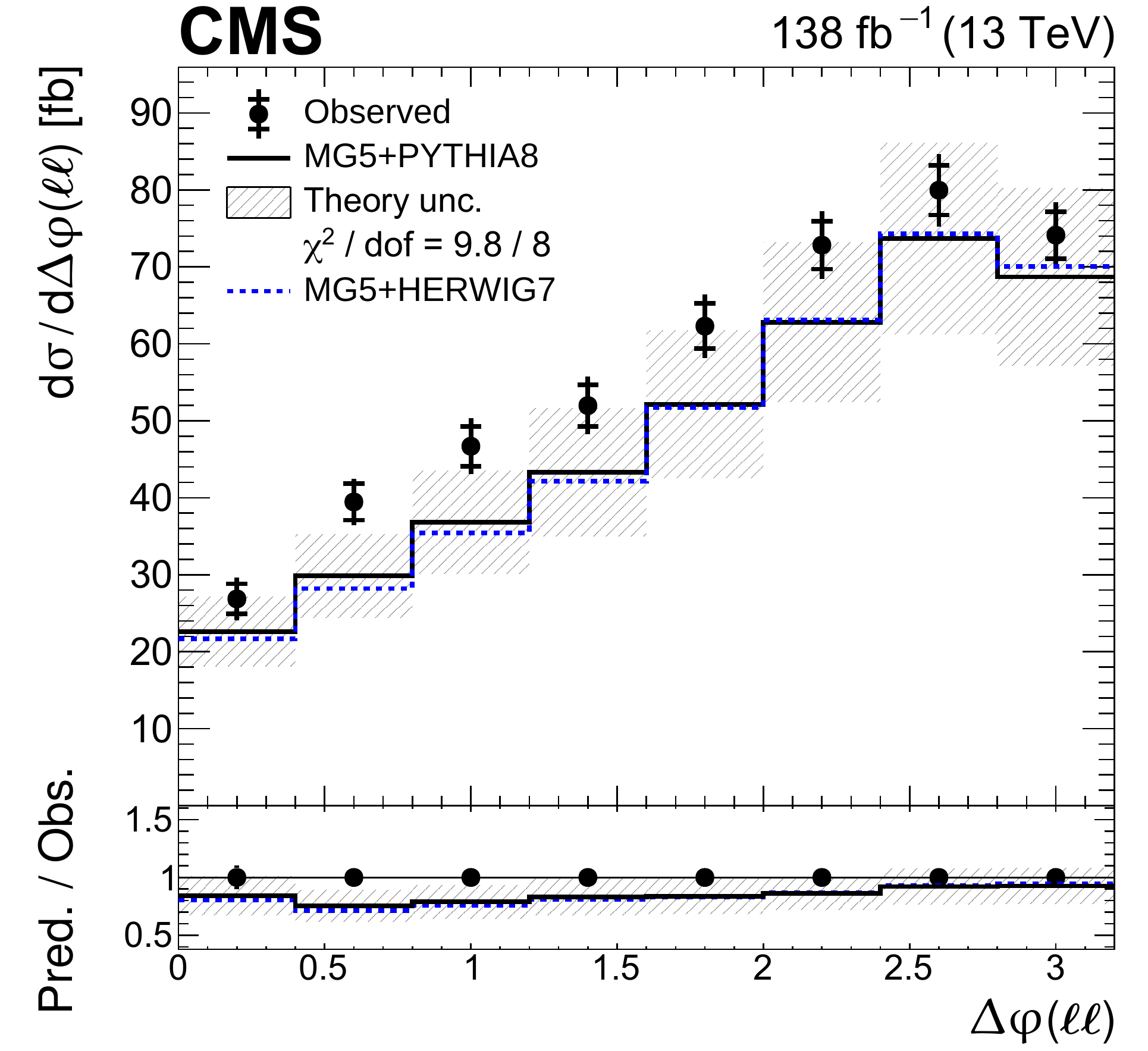}
\\
\includegraphics[width=0.42\textwidth]{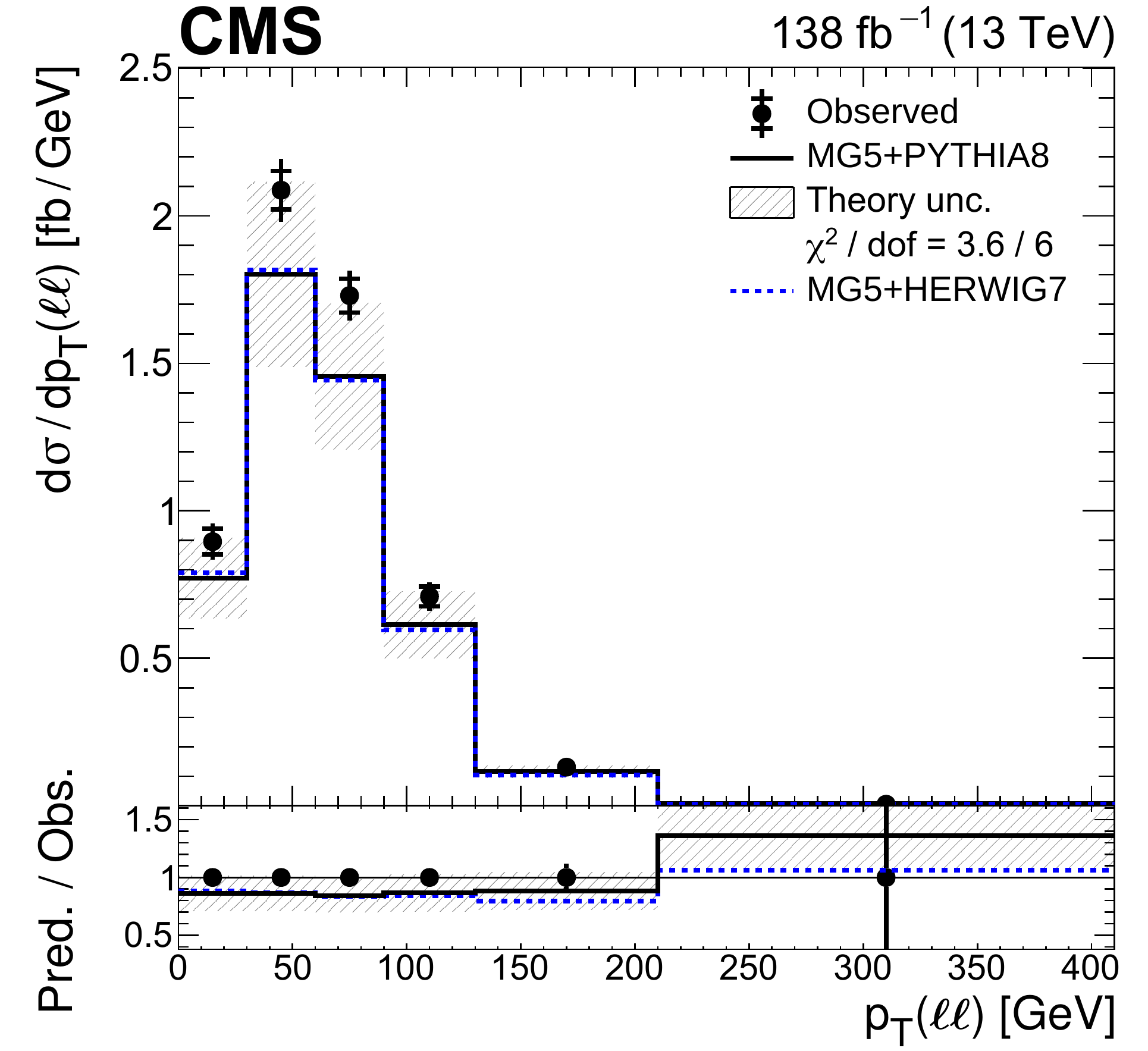}
\hspace{0.02\textwidth}
\includegraphics[width=0.42\textwidth]{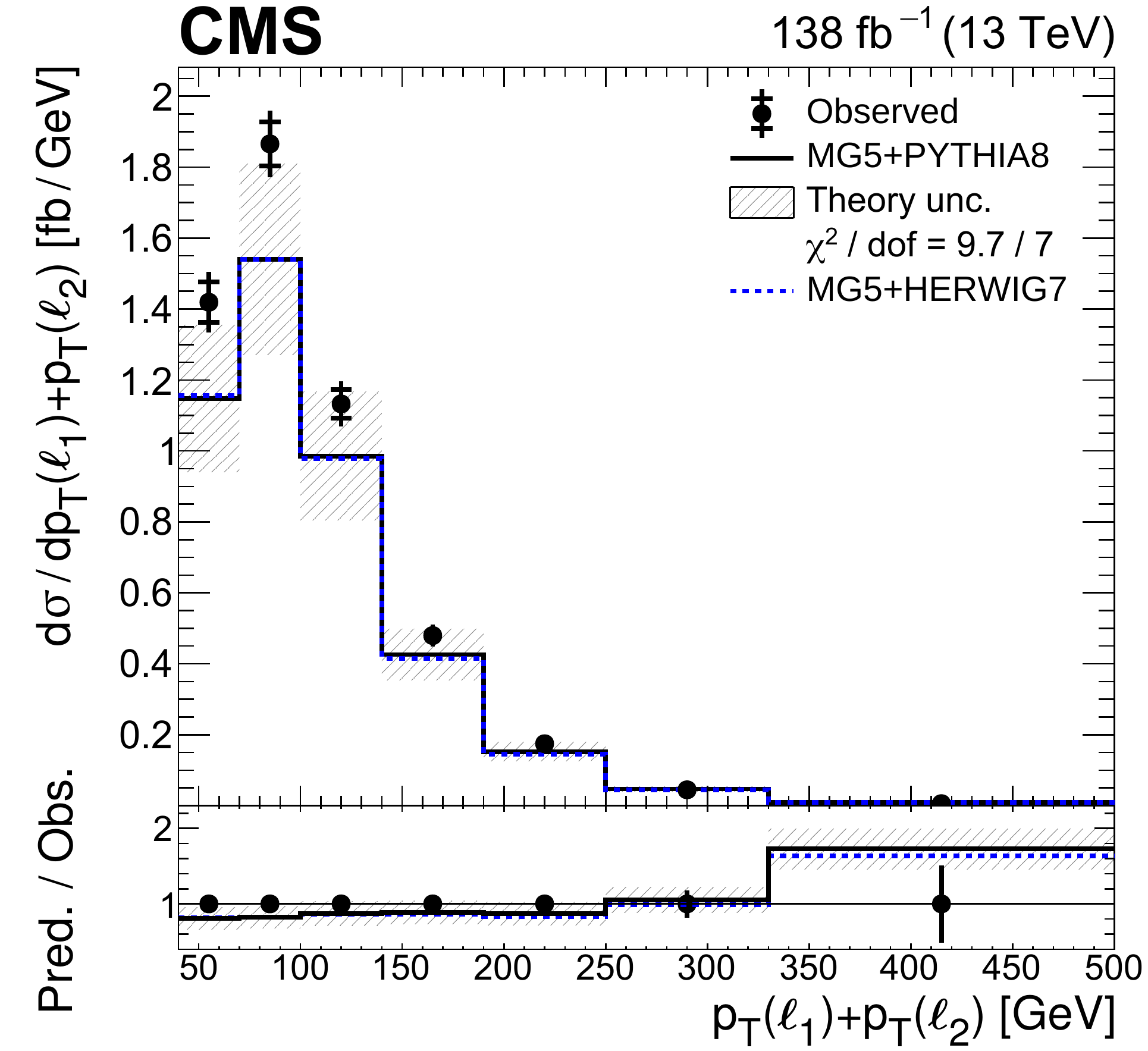}
\\
\includegraphics[width=0.42\textwidth]{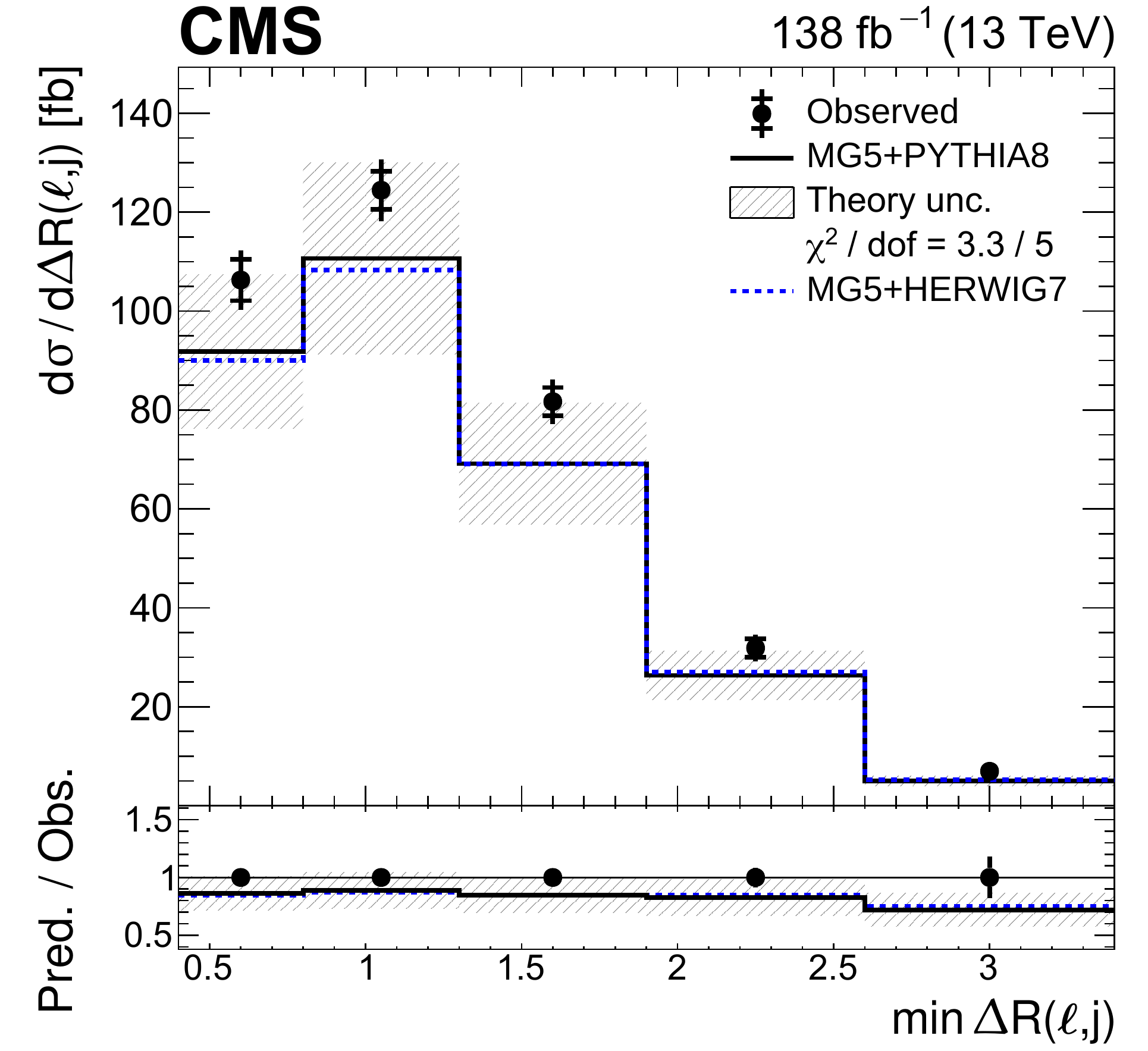}
\hspace{0.02\textwidth}
\includegraphics[width=0.42\textwidth]{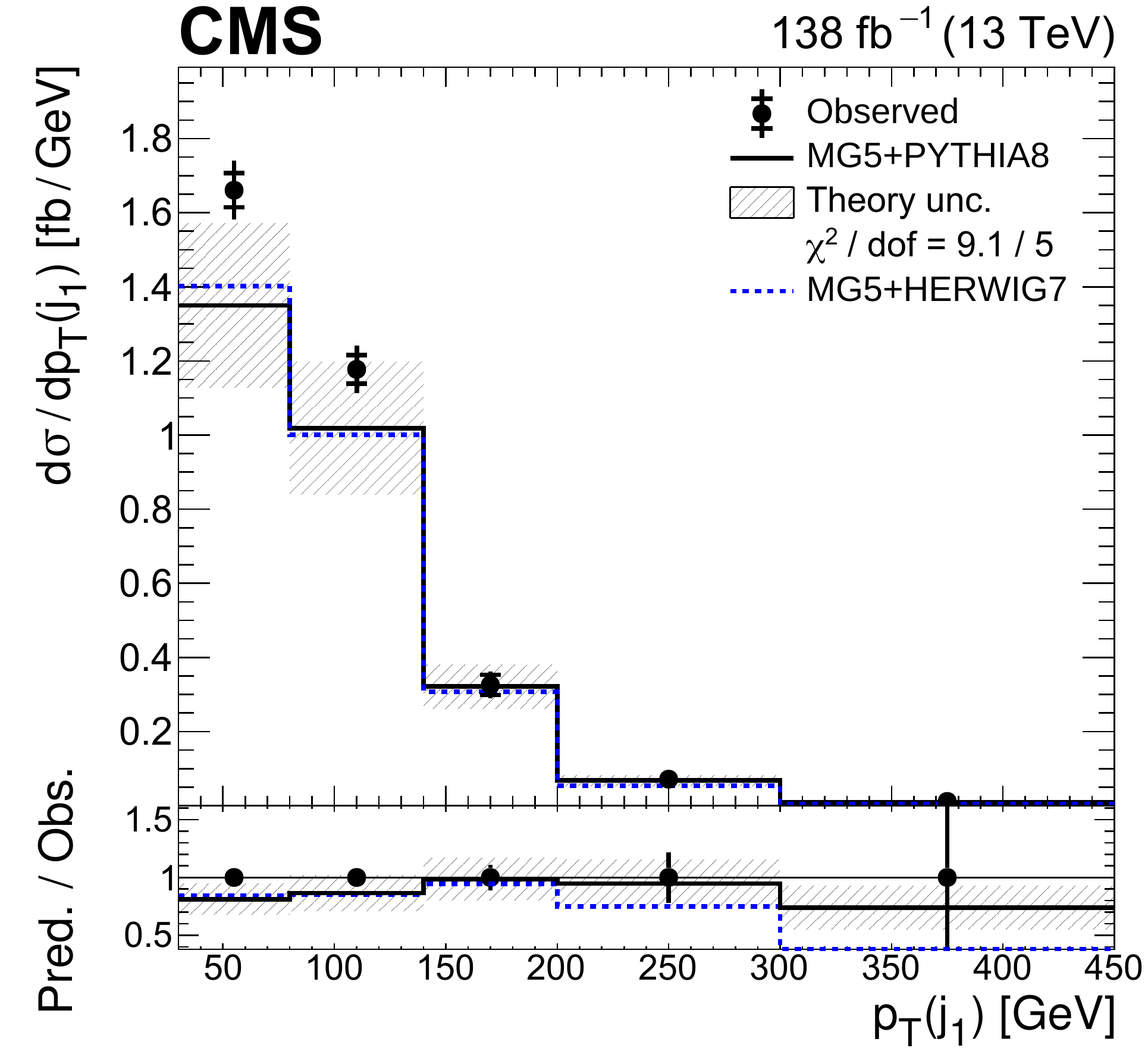}
\caption{Absolute differential \ttG production cross sections as functions of
\detall (upper left),
\dphill (upper right),
\ptll (middle left),
\sumptll (middle right),
\drlepjet (lower left),
and
\ptjet (lower right).
Details can be found in the caption of Fig.~\ref{fig:unfoldedA}.}
\label{fig:unfoldedB}
\end{figure}

\begin{figure}[!tp]
\centering
\includegraphics[width=0.42\textwidth]{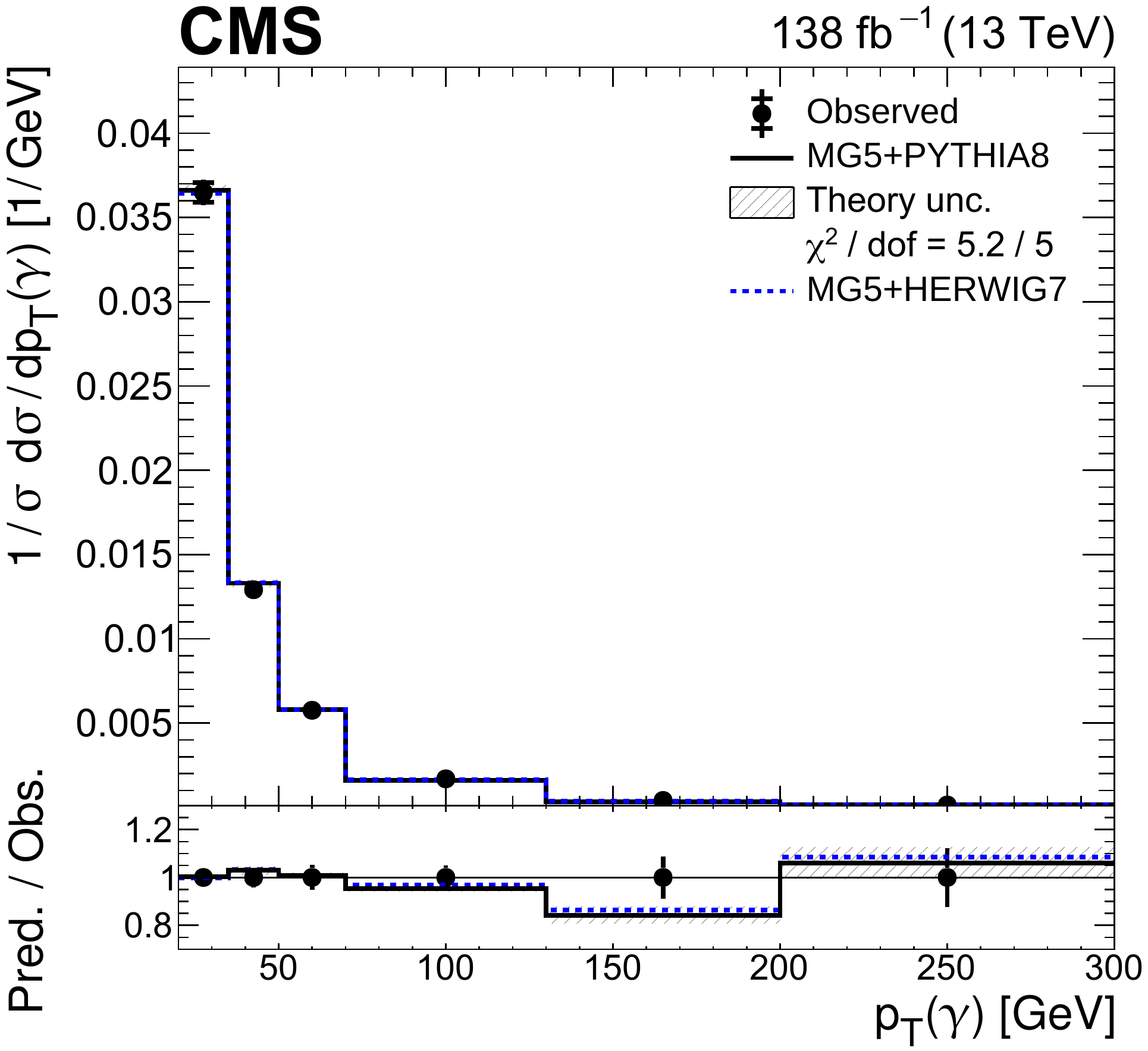}
\hspace{0.02\textwidth}
\includegraphics[width=0.42\textwidth]{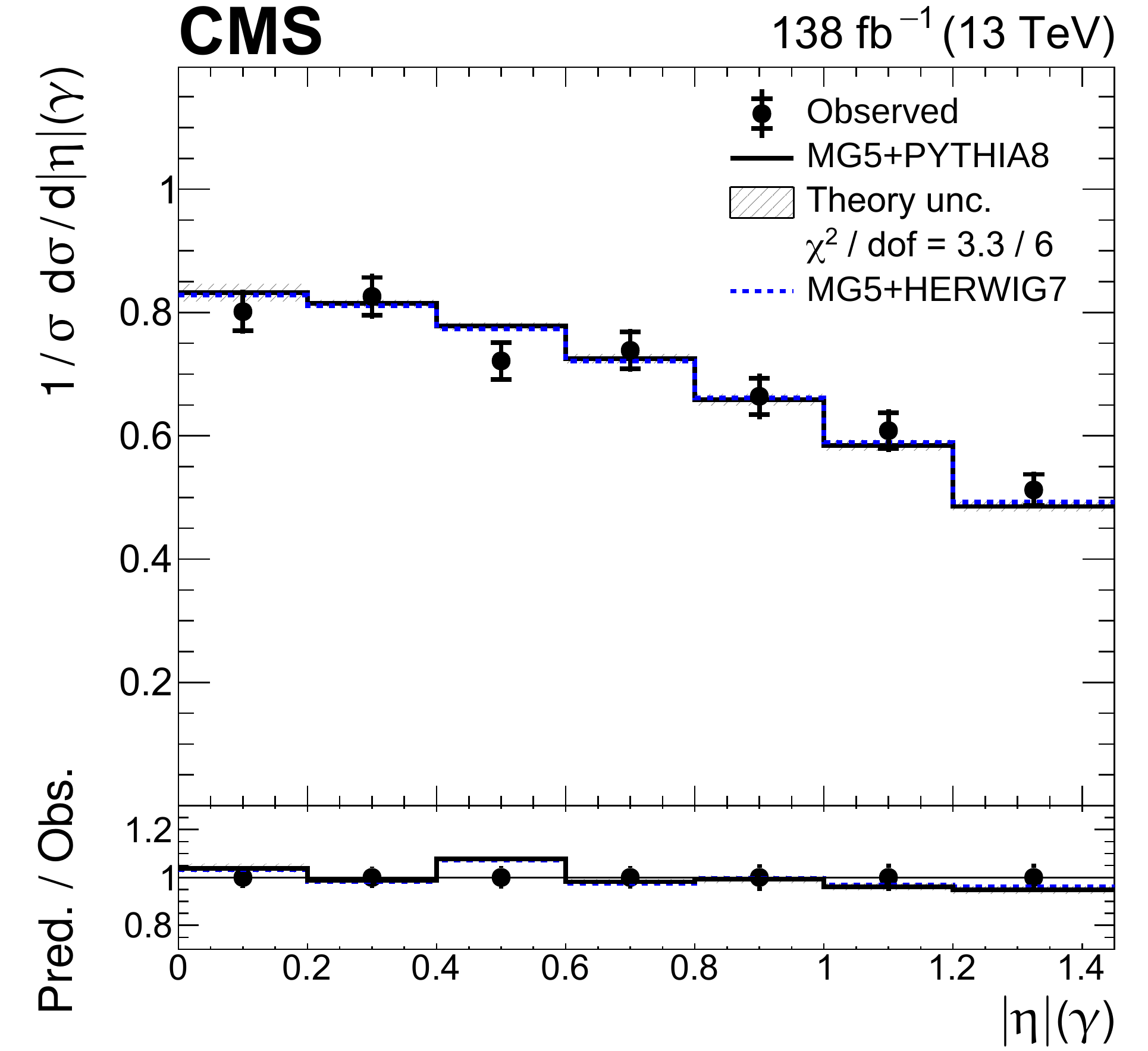}
\\
\includegraphics[width=0.42\textwidth]{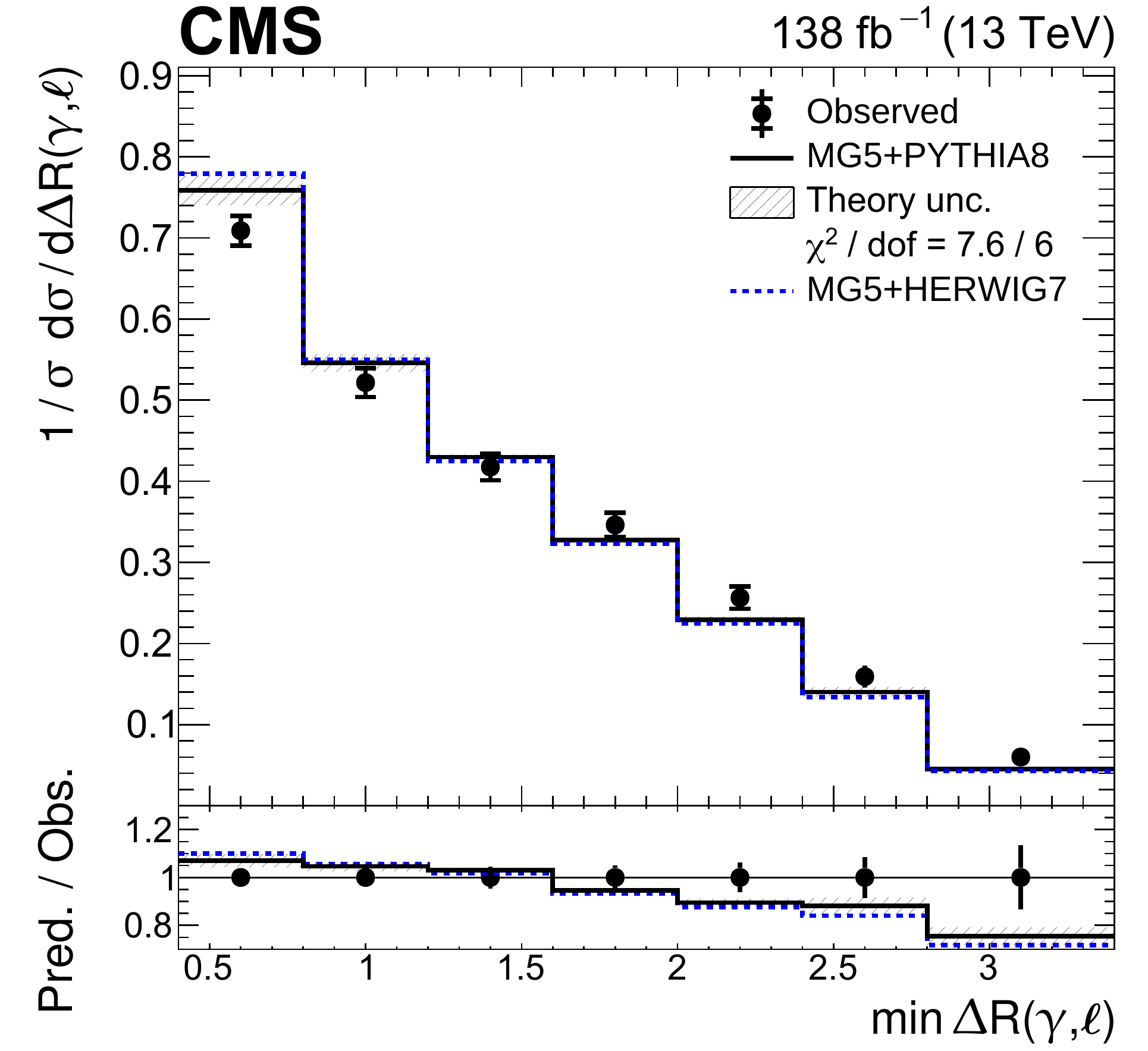}
\hspace{0.02\textwidth}
\includegraphics[width=0.42\textwidth]{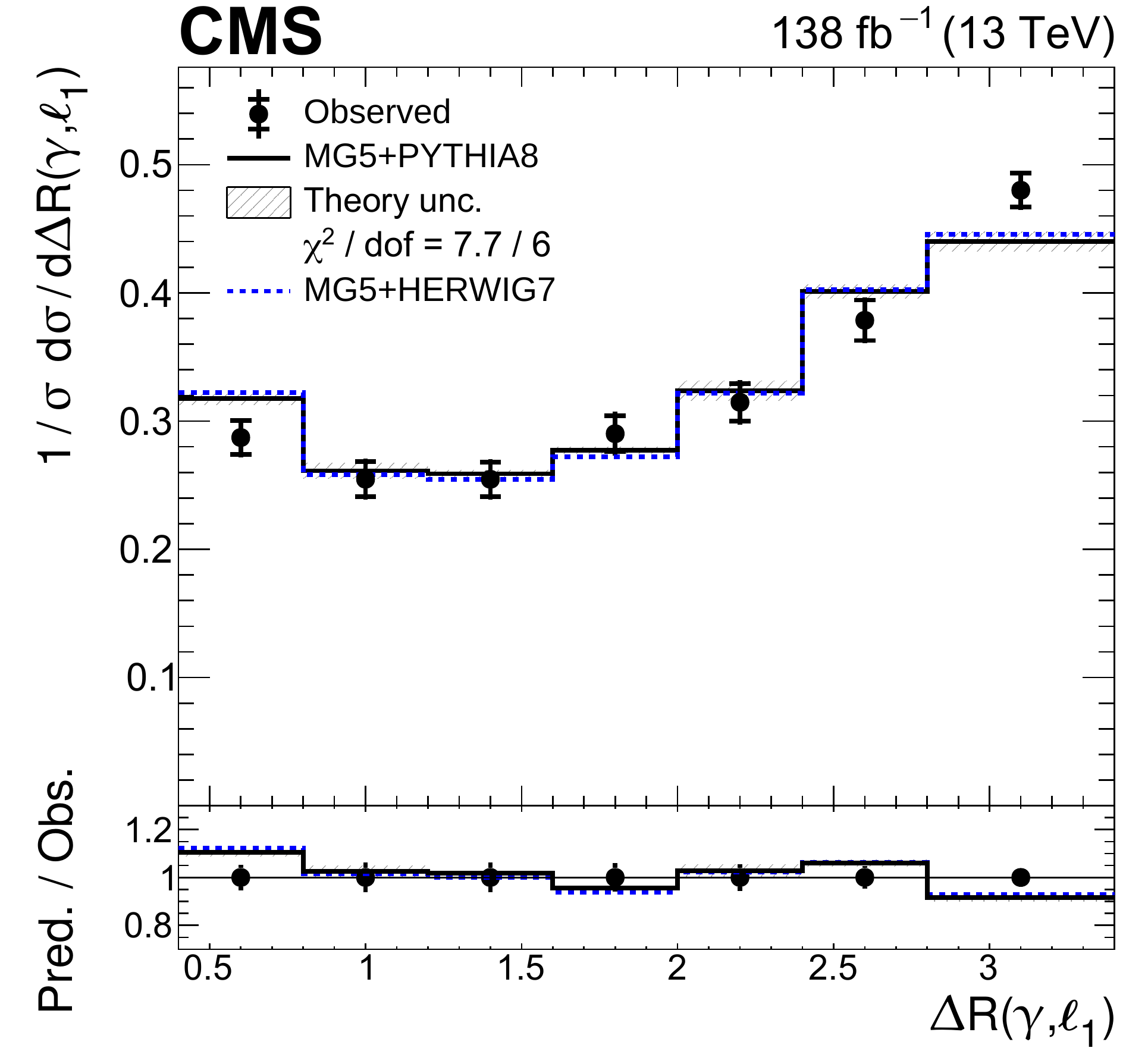}
\\
\includegraphics[width=0.42\textwidth]{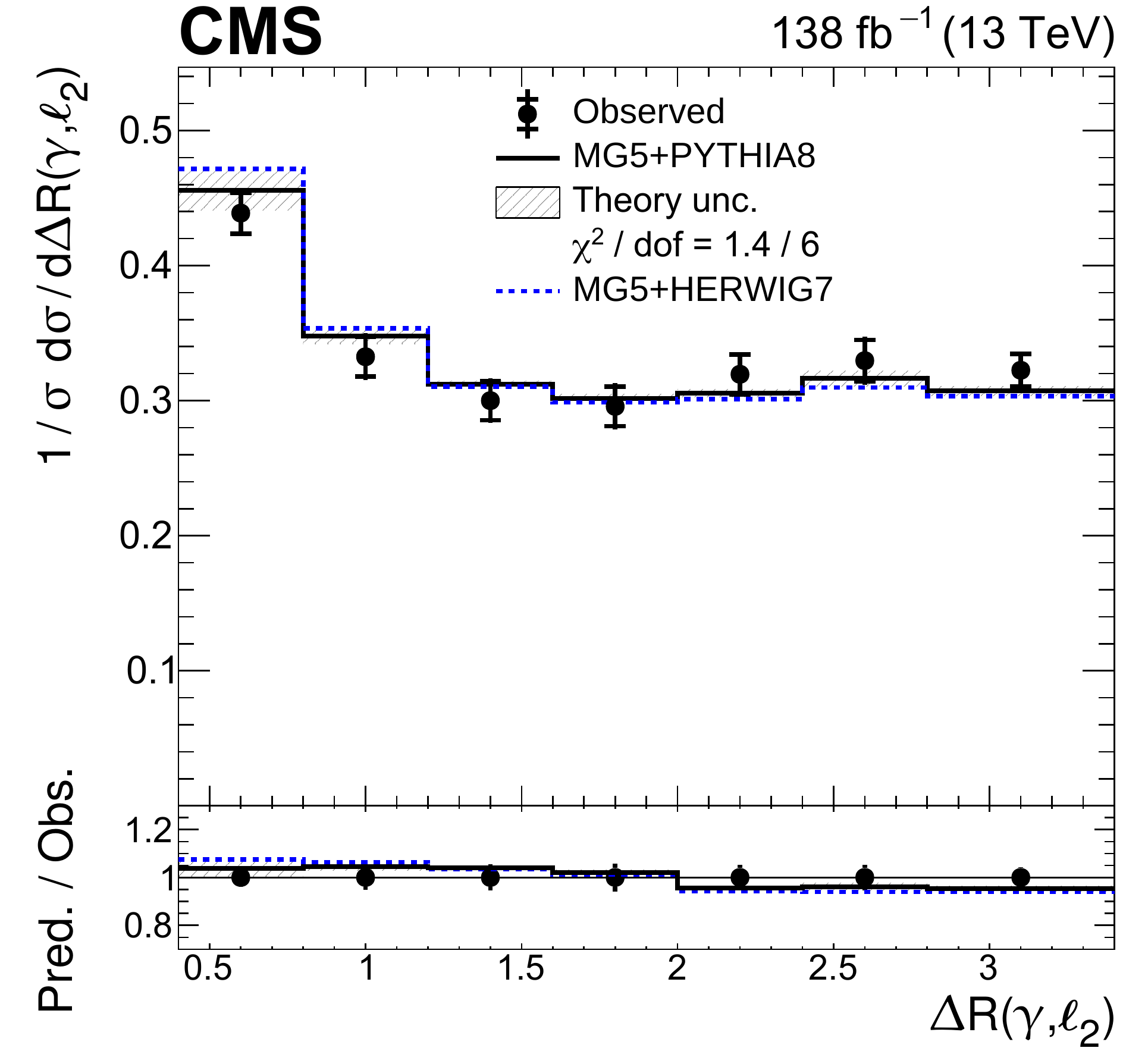}
\hspace{0.02\textwidth}
\includegraphics[width=0.42\textwidth]{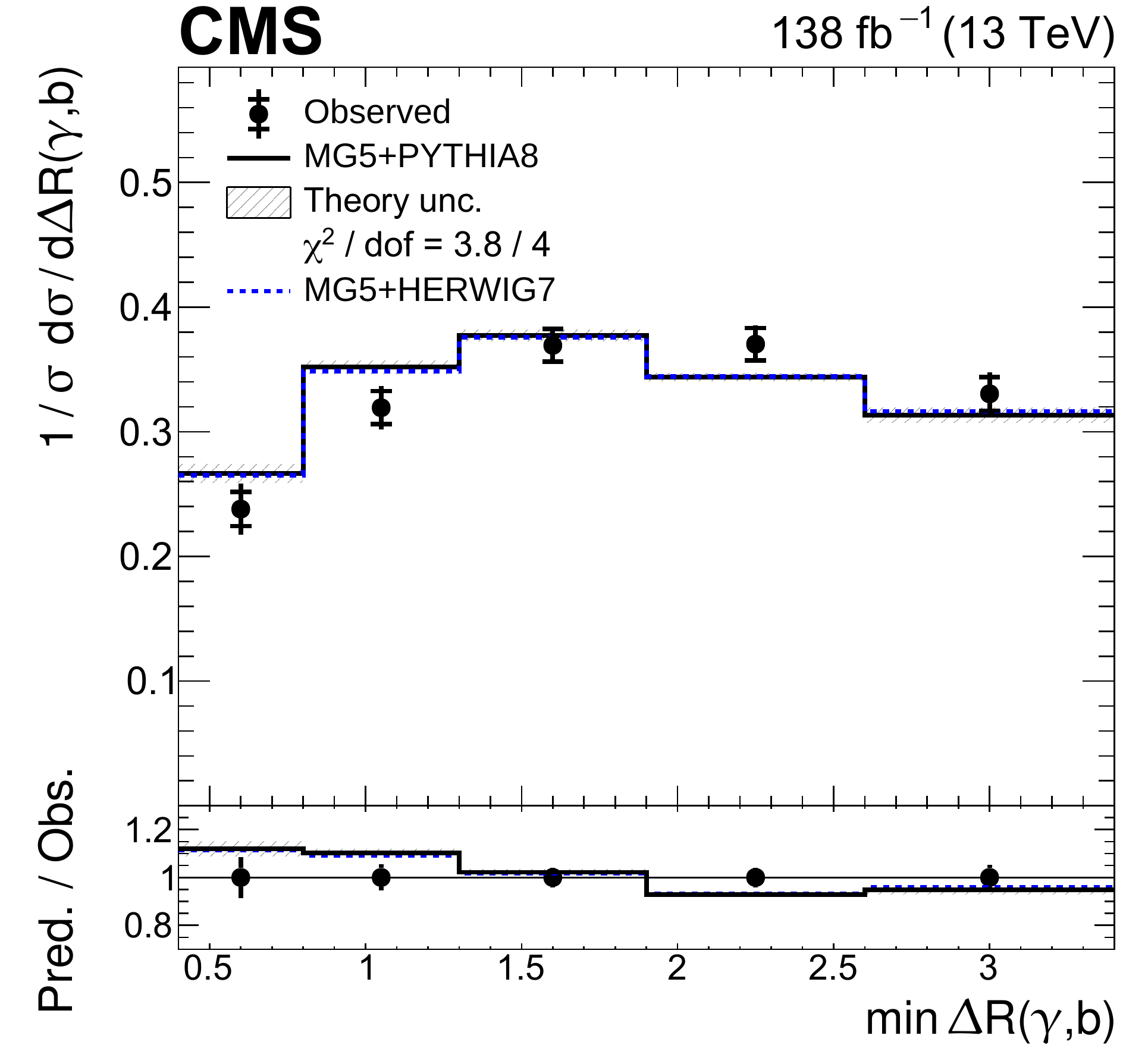}
\caption{Normalized differential \ttG production cross sections as functions of
\ptgamma (upper left),
\etagamma (upper right),
\drgammaclose (middle left),
\drgammafirst (middle right),
\drgammasecond (lower left),
and
\drgammabjet (lower right).
Details can be found in the caption of Fig.~\ref{fig:unfoldedA}.}
\label{fig:normalizedA}
\end{figure}

\begin{figure}[!tp]
\centering
\includegraphics[width=0.42\textwidth]{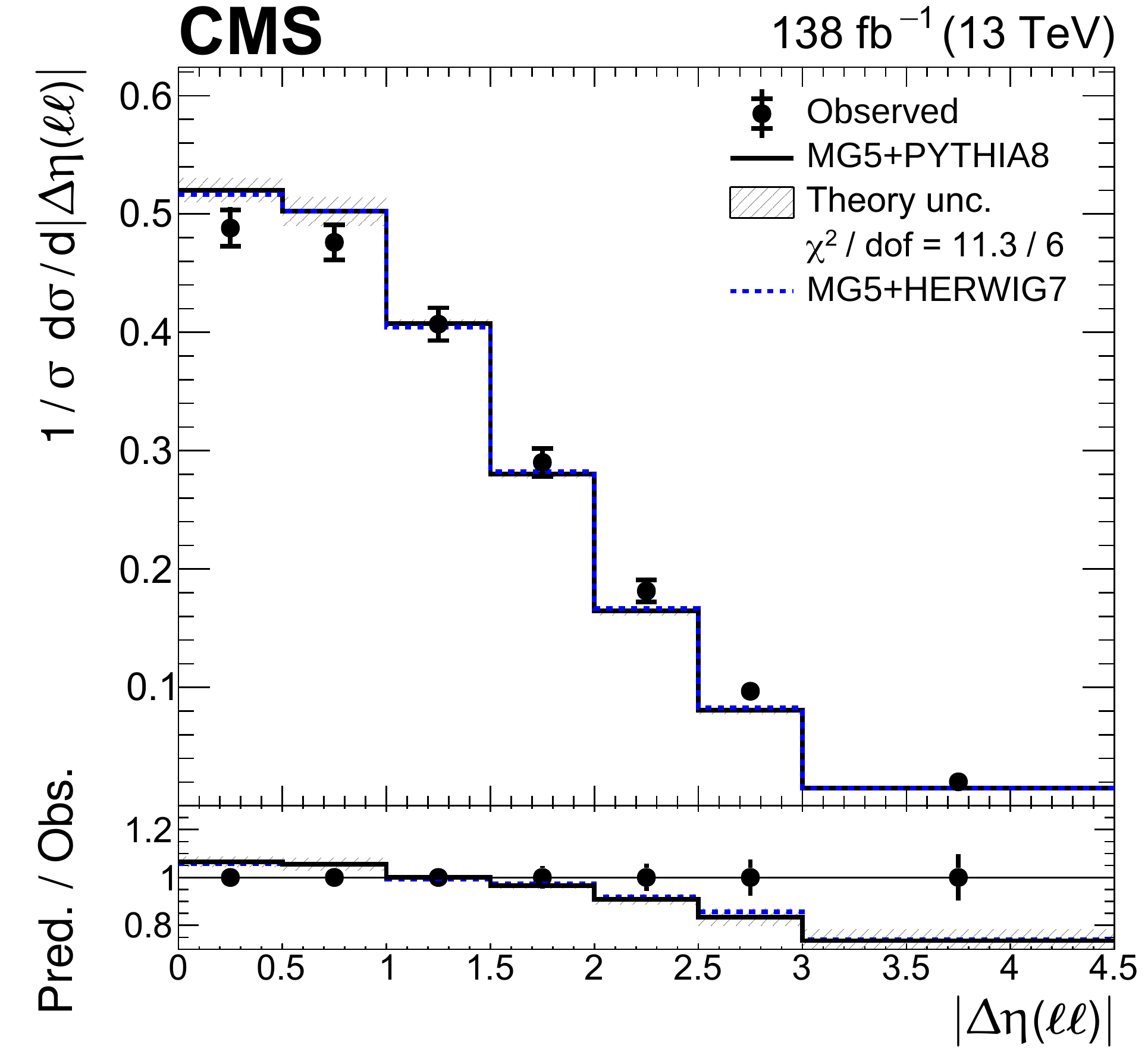}
\hspace{0.02\textwidth}
\includegraphics[width=0.42\textwidth]{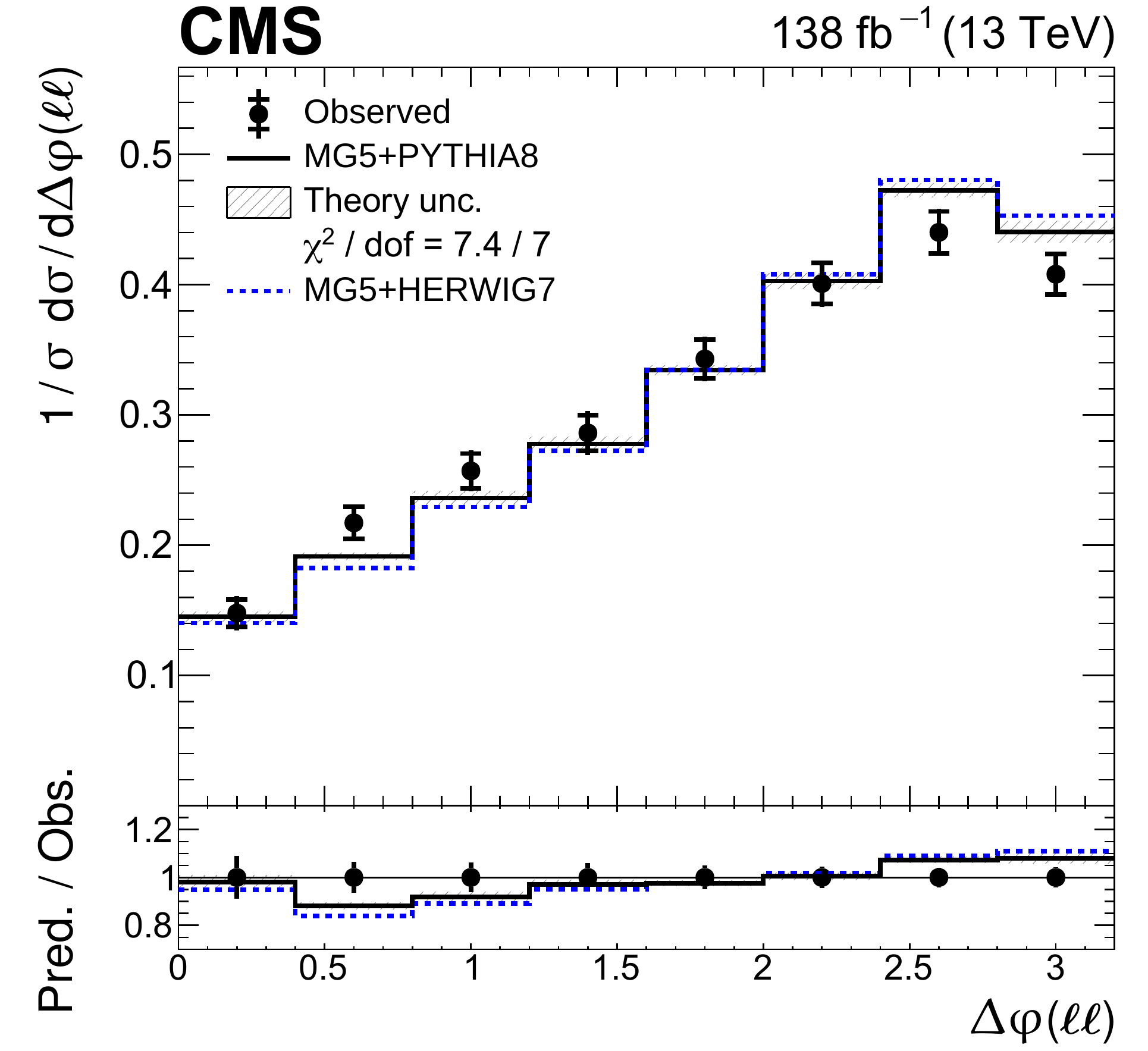}
\\
\includegraphics[width=0.42\textwidth]{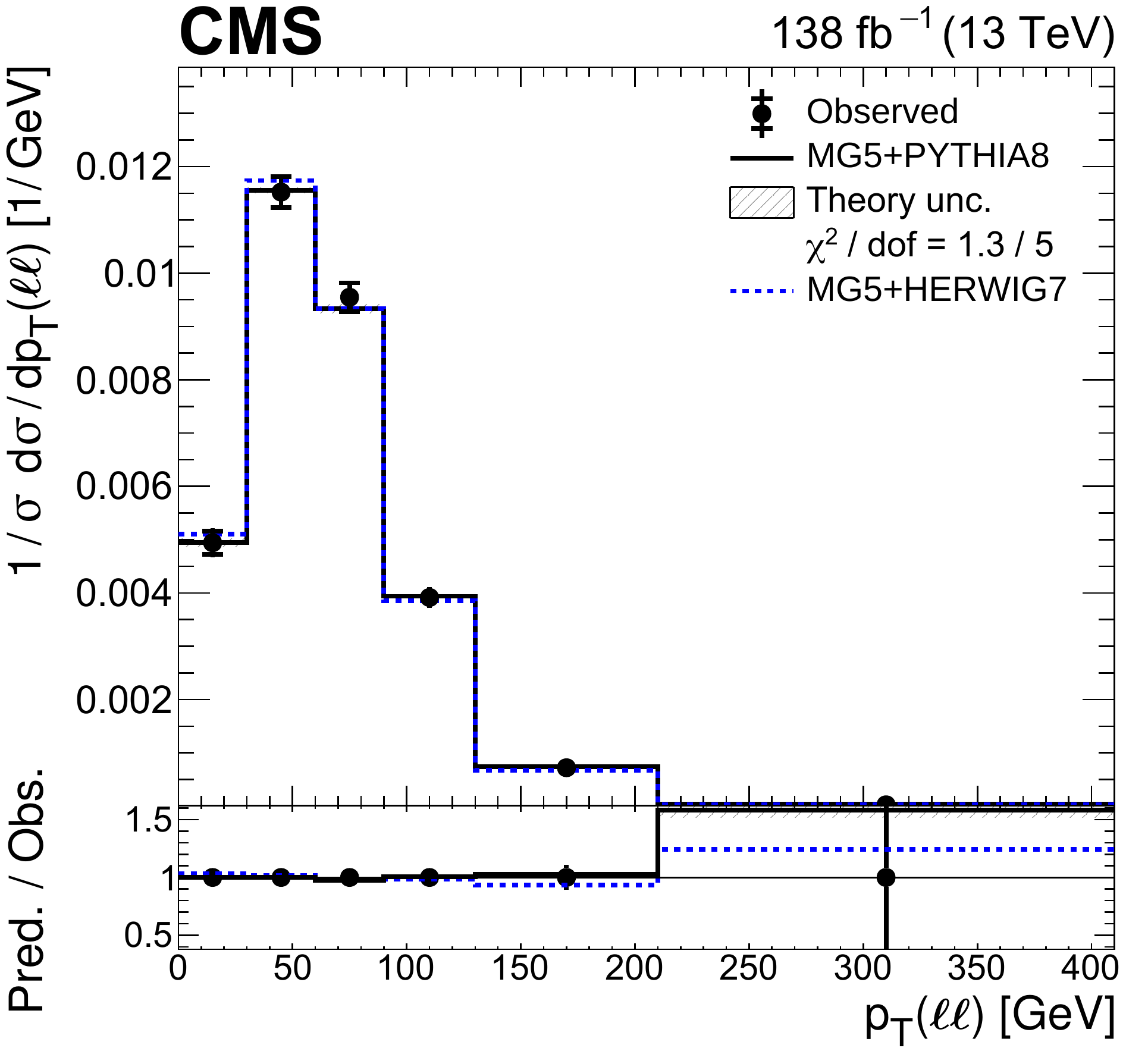}
\hspace{0.02\textwidth}
\includegraphics[width=0.42\textwidth]{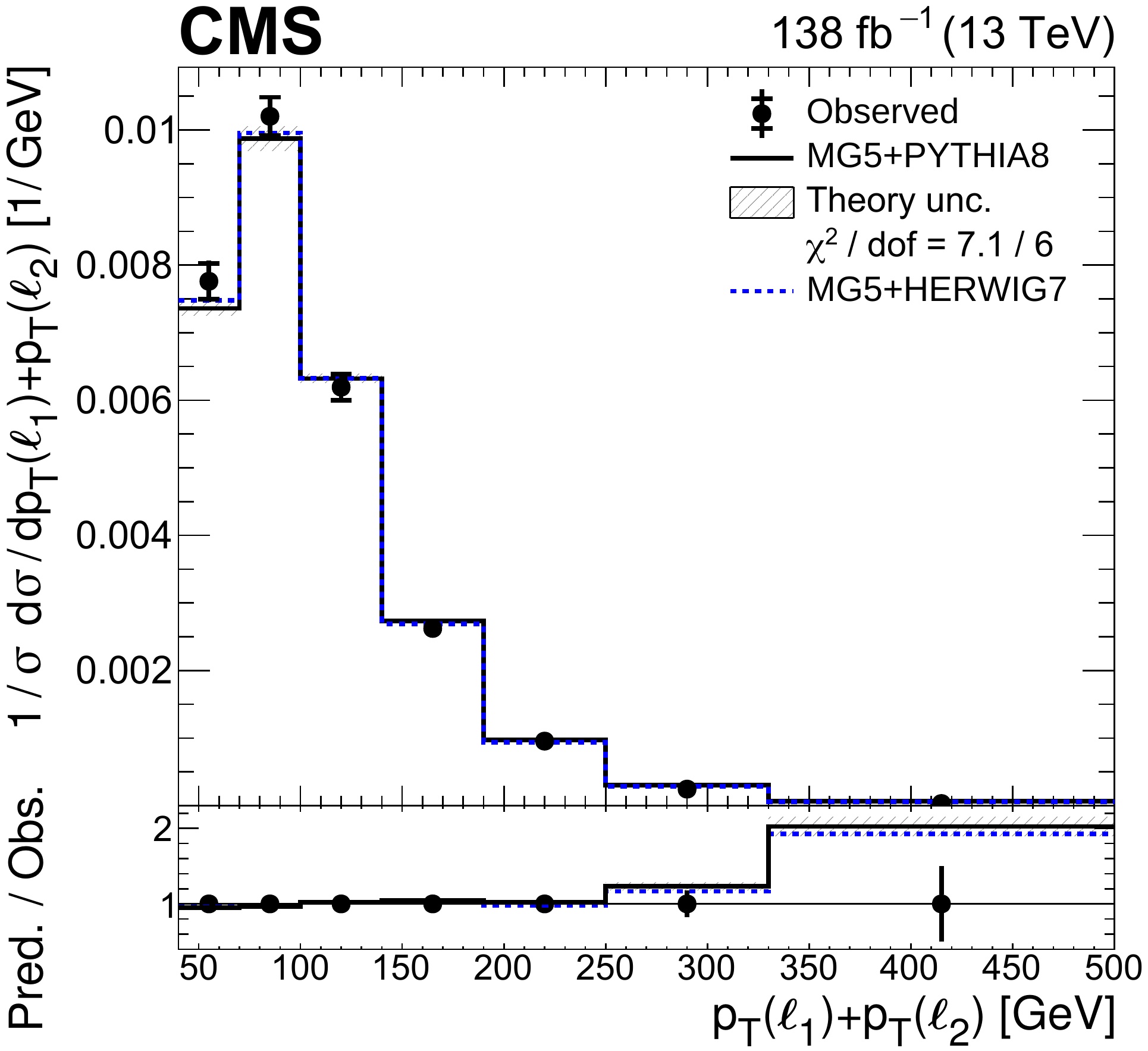}
\\
\includegraphics[width=0.42\textwidth]{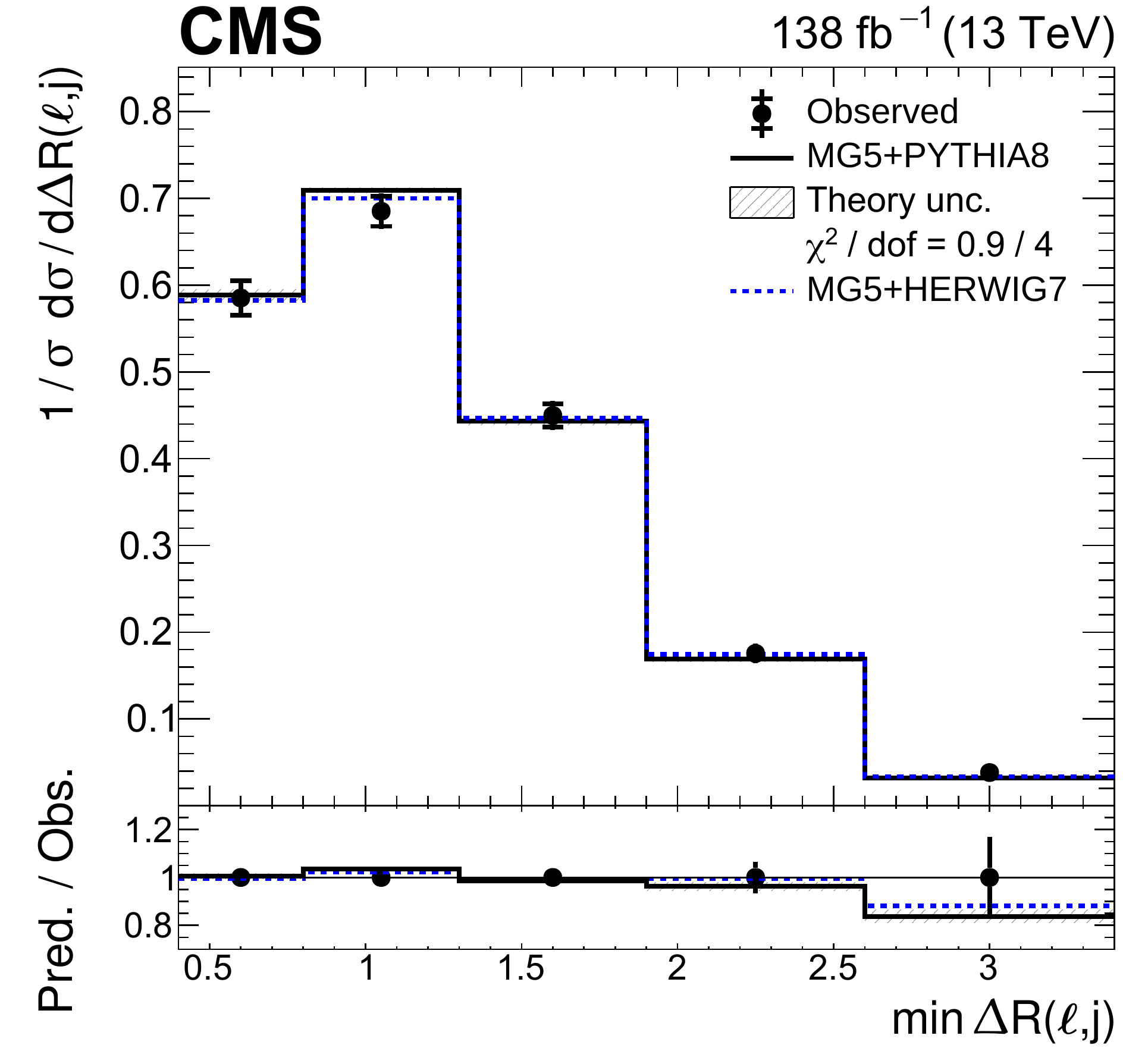}
\hspace{0.02\textwidth}
\includegraphics[width=0.42\textwidth]{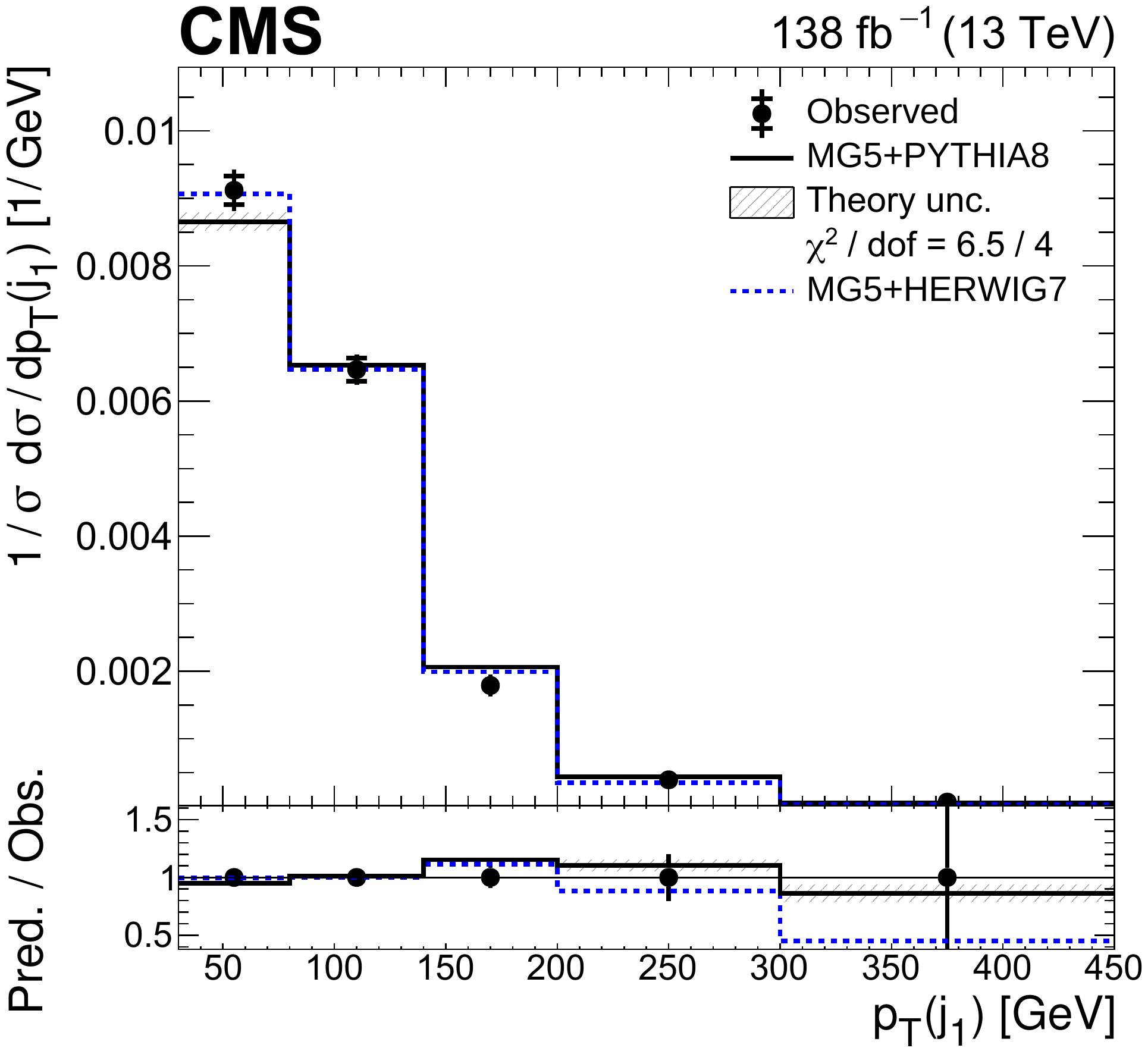}
\caption{Normalized differential \ttG production cross sections as functions of
\detall (upper left),
\dphill (upper right),
\ptll (middle left),
\sumptll (middle right),
\drlepjet (lower left),
and
\ptjet (lower right).
Details can be found in the caption of Fig.~\ref{fig:unfoldedA}.}
\label{fig:normalizedB}
\end{figure}

The resulting absolute differential cross sections are shown in Figs.~\ref{fig:unfoldedA} and~\ref{fig:unfoldedB}, and normalized differential cross sections are shown in Figs.~\ref{fig:normalizedA} and~\ref{fig:normalizedB}.
For a given observable, the normalized differential cross section is obtained by dividing the absolute differential cross section by the sum of the binned cross section measurements over all bins of the observable.
The measurements are compared to two cross section predictions obtained at particle level, generated with the \MGvATNLO event generator interfaced with parton shower simulations provided by \PYTHIAeight with the CP5 tune and by \HERWIG~\cite{Bellm:2015jjp} v7.1.4 with the CH3 tune~\cite{CMS:2020dqt}.
The prediction using \PYTHIAeight is shown with its uncertainty from scale variations and the PDF choice.
The scale variations affect the normalization of the predictions but have negligible impact on their shapes.
The agreement between the measured distribution and the prediction using \PYTHIAeight is evaluated by calculating a \chisq value, which is indicated on the figures.
The \chisq value computation takes into account the statistical and systematic covariance matrices of the measurement, as well as the uncertainty in the prediction.
For all distributions, no significant deviation between the measured and predicted distributions is observed.

The measurement of the kinematic properties of the photon, \ptgamma and \etagamma, provides sensitivity to the coupling between the top quark and the photon.
Both predictions show good agreement with the observed distribution, with only one bin in each observable having a disagreement larger than the uncertainty in the measurement.
The \chisqdof values of about $5/5$ ($3/6$) for the normalized differential cross sections of \ptgamma (\etagamma) confirm that there is no significant disagreement.

The angular separation variables between the photon and the top quark decay products provide sensitivity to the origin of the photon in \ttG events~\cite{Bevilacqua:2019quz}.
With \chisqdof values for the normalized distributions of about $8/6$ for \drgammaclose and \drgammafirst, $1/6$ for \drgammasecond, and $4/4$ for \drgammabjet, no statistically significant disagreement is observed.
The apparent trends in all four distributions for the separation to be smaller in the prediction than in the measurement hint at a mismodelling of the photon origins in the LO simulation of the \ttG signal process, but more precise measurements are required to draw a clear conclusion.

The observables that do not involve the photon provide sensitivity to the modelling of the top quark decay.
The pseudorapidity and azimuthal angle differences between the two leptons, \detall and \dphill, are of special interest due to their sensitivity to \ttbar spin correlations~\cite{CMS:2018adi, CMS:2019nrx}.
Their predictions show some difference with respect to the measurement that are, with a \chisqdof value of about $11/6$ for the normalized differential cross section of \detall, compatible within the uncertainties.
The predictions for \ptll, \sumptll, \drlepjet, and \ptjet show good agreement with the observed differential cross sections except for the last bins of the distributions where the statistical uncertainty in the measurement becomes very large.
The largest difference between the two MC predictions is seen in the \ptjet distribution.

\section{Effective field theory interpretation}\label{sec:eft}

Many theories of new physics beyond the SM predict the existence of new particles and mechanisms characterized by an energy scale $\Lambda$ that is well above the energy reach of the LHC.
Such theories may still lead to observable deviations in well-established processes through quantum-loop corrections.
In the SMEFT approach, such corrections are parameterized in a coherent and model-independent way by the extension of the SM Lagrangian with new, effective interactions between the SM fields, characterized by higher-order operators~\cite{Buchmuller:1985jz}.
The interaction strength of such an operator of dimension $d$ is proportional to $\Lambda^{4-d}$ and thus suppressed, which implies that the impact of new interactions can be approximated with a finite set of lower-order operators.
Under the assumption of lepton number conservation, the lowest order of SMEFT operators to contribute have dimension six~\cite{Kobach:2016ami}.
The SMEFT Lagrangian can thus be written as:
\begin{linenomath}\begin{equation*}
    \Lagrangian{SMEFT}=\Lagrangian{SM}+\sum_i\frac{\ci}{\Lambda^2}\Oi,
\end{equation*}\end{linenomath}
where \Lagrangian{SM} is the SM Lagrangian, \Oi are the dimension-six operators, and \ci the corresponding Wilson coefficients.

We follow the recommendations of Ref.~\cite{Aguilar-Saavedra:2018ksv} and adopt the Warsaw basis~\cite{Grzadkowski:2010es} with 59 baryon-number-conserving dimension-six operators.
Of those, 15 are found to be relevant for top quark interactions~\cite{Aguilar-Saavedra:2008nuh, Zhang:2010dr}.
While anomalous interactions between the top quark and the gluon are strongly constrained by measurements of \ttbar production~\cite{CMS:2018adi, CMS:2019nrx}, the measurement of \ttG production provides sensitivity to the electroweak dipole moments of the top quark, denoted by \CuB and \CuW.
Because of the SM gauge symmetries, the sensitivity to \ttG production is complementary to that provided by \ttZ production~\cite{Baur:2004uw, Bouzas:2012av, Schulze:2016qas}.
The coefficients describing the modifications of the \ttZ interaction vertex, \ctZ and \ctZI, and of the \ttG interaction vertex, \ctG and \ctGI, are expressed in the Warsaw basis as linear combinations of \CuB and \CuW:
\begin{linenomath}\begin{equation*}\begin{aligned}
     \ctZ&=\RE\left(-\sinw\CuB+\cosw\CuW\right), \\
    \ctZI&=\IM\left(-\sinw\CuB+\cosw\CuW\right), \\
     \ctG&=\RE\left( \cosw\CuB-\sinw\CuW\right), \\
    \ctGI&=\IM\left( \cosw\CuB-\sinw\CuW\right).
\end{aligned}\end{equation*}\end{linenomath}
The modification of the \Wtb vertex is better probed in measurements of \PW helicity fractions~\cite{CMS:2020ezf}.
Under the assumption of an SM \Wtb vertex with $\CuW=0$, the \ttZ and \ttG modifications are dependent.
We choose to parameterize the new-physics hypothesis in terms of \ctZ and \ctZI, and set the Wilson coefficients of all other operators to zero.

The effect of these modifications is probed in the measured distribution of the photon \pt at the reconstructed level, which is also used in the inclusive cross section measurement.
The other observables studied in the differential cross section measurement are found to be largely insensitive to these new-physics effects.
With a fixed SMEFT expansion parameter of $\Lambda=1\TeV$, the expected SMEFT modifications for nonzero values of \ctZ and \ctZI are estimated at the particle level and used to calculate per-event weights corresponding to the ratio of the predicted SM and SMEFT cross sections in bins of photon \pt.
In the parameterization used in the calculation, both linear and quadratic terms in the coefficients are included.
The nominal simulation is then reweighted after applying the full analysis selection criteria to the reconstructed events to estimate the expected SMEFT modifications at the reconstructed level.

The similar measurement by CMS using final states with one lepton and jets (\elljets) presented in Ref.~\cite{CMS:2021klw} probes an orthogonal fiducial phase space of the \ttG production process defined for the single-lepton decay channel of the \ttbar system.
While the measurement presented here profits from very small background contributions, the measurement in Ref.~\cite{CMS:2021klw} has a significantly larger number of signal events at large photon \pt values, which increases the sensitivity to modifications described by the studied Wilson coefficients.
To further improve the constraints on the Wilson coefficients, a combined EFT interpretation of both measurements is also performed.

The constraints on the Wilson coefficients are measured from a profile likelihood fit constructed in the same way as for the inclusive cross section measurement.
The set of nuisance parameters is extended by the uncertainty in the \ttG signal normalization.
The minimized likelihood value obtained in a fit using the SMEFT-predicted photon \pt distribution is compared to the corresponding likelihood value using the SM prediction.
A combined profile likelihood function is also constructed using the photon \pt distributions from both this analysis and the previous \elljets measurement, and the correlations between the nuisance parameters applied in both analyses are treated appropriately.

\begin{figure}[!htp]
\centering
\includegraphics[width=0.42\textwidth]{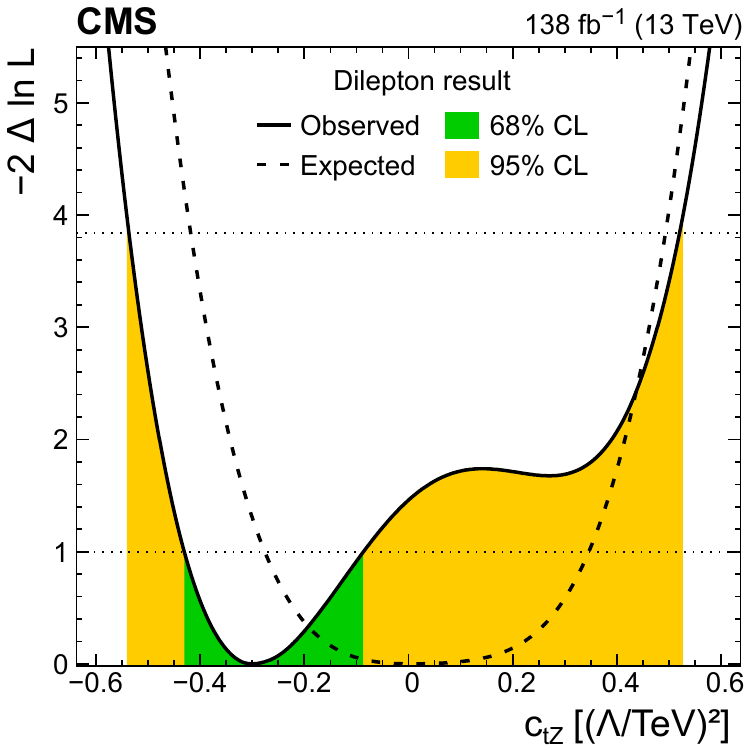}
\hspace{0.02\textwidth}
\includegraphics[width=0.42\textwidth]{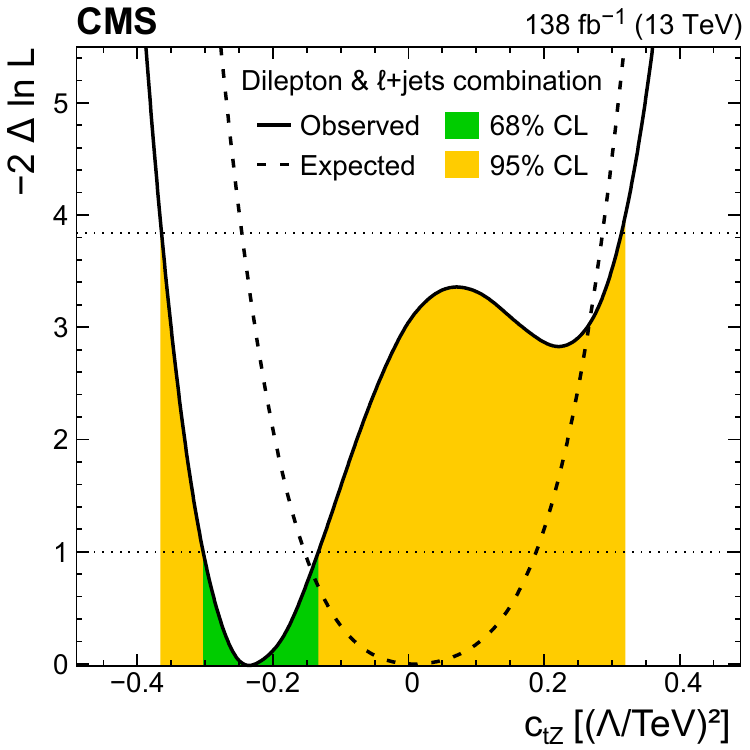}
\\
\includegraphics[width=0.42\textwidth]{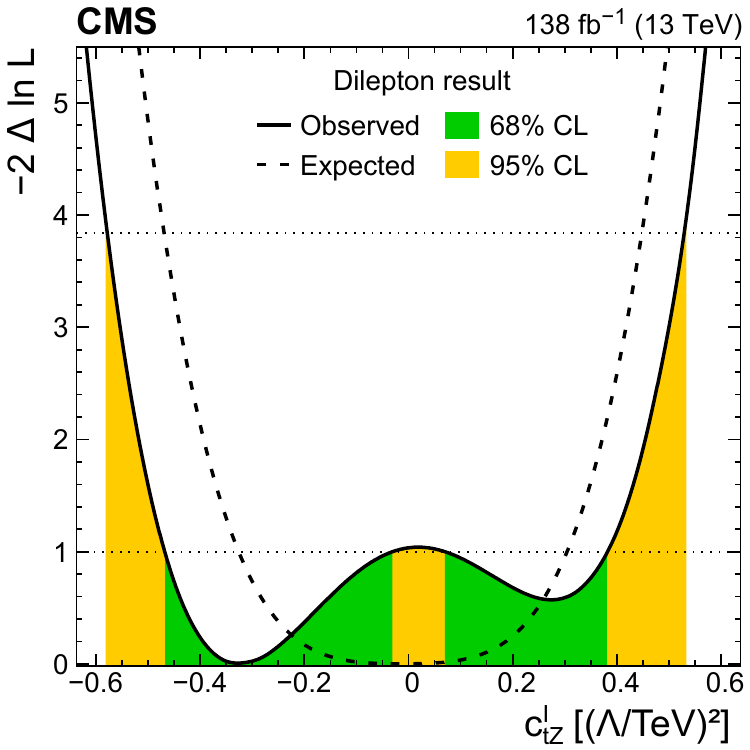}
\hspace{0.02\textwidth}
\includegraphics[width=0.42\textwidth]{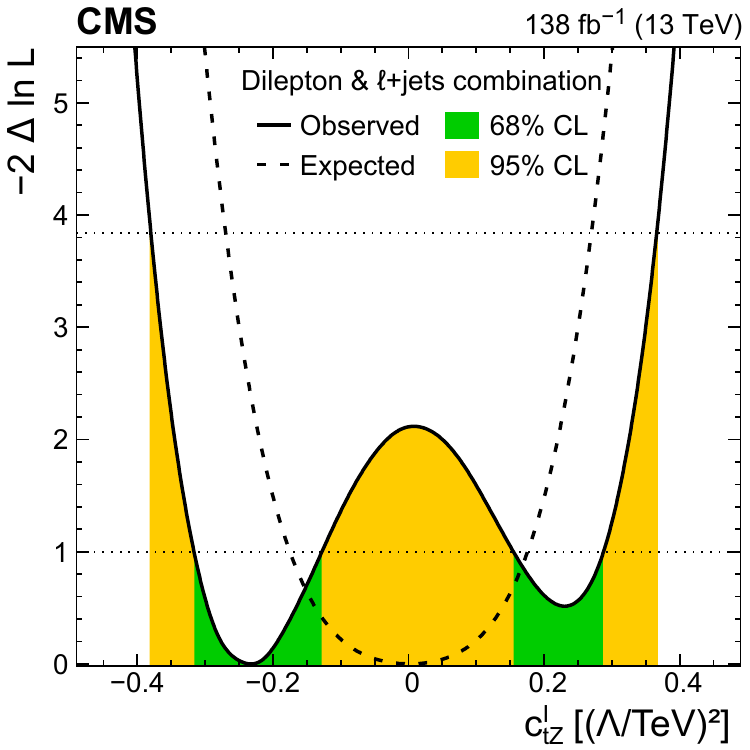}
\caption{Distributions of the observed (solid line) and expected (dashed line) negative log-likelihood difference from the best fit value for the one-dimensional scans of the Wilson coefficients \ctZ (upper) and \ctZI (lower), using the photon \pt distribution from this analysis (left) or the combination of this analysis with the \elljets analysis from Ref.~\cite{CMS:2021klw} (right).
In the scans, the other Wilson coefficient is set to zero.
The green (orange) bands indicate the 68 (95)\% \CL limits on the Wilson coefficients.}
\label{fig:eft1D}
\end{figure}

\begin{figure}[!htp]
\centering
\includegraphics[width=0.42\textwidth]{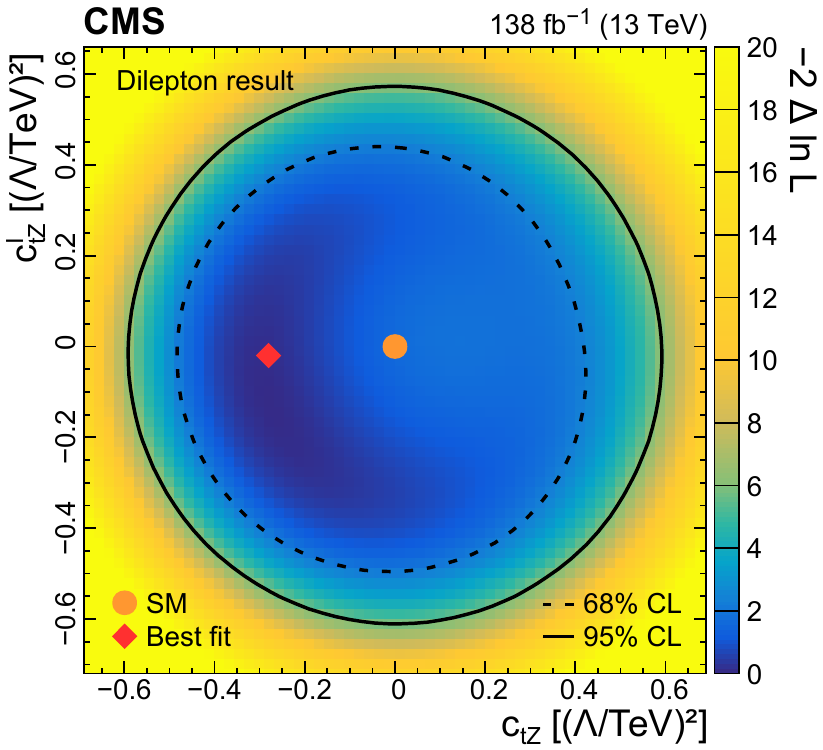}
\hspace{0.02\textwidth}
\includegraphics[width=0.42\textwidth]{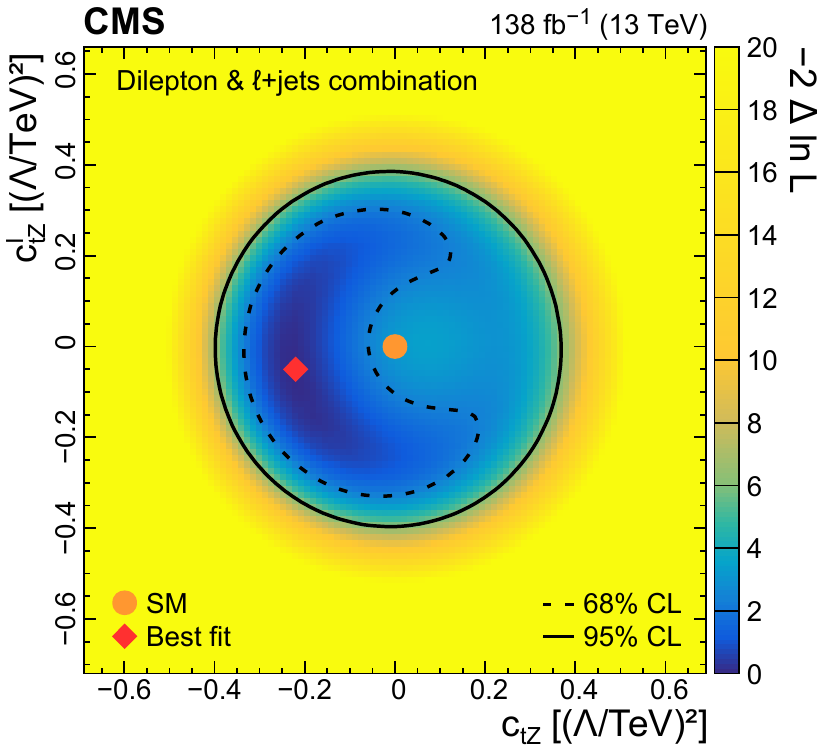}
\caption{Result from the two-dimensional scan of the Wilson coefficients \ctZ and \ctZI using the photon \pt distribution from this analysis (left) or the combination of this analysis with the \elljets analysis from Ref.~\cite{CMS:2021klw} (right).
The shading quantified by the colour scale on the right reflects the negative log-likelihood difference with respect to the best fit value that is indicated by the red diamond.
The 68\% (dashed curve) and 95\% (solid curve) \CL contours are shown for the observed result.
The orange circle indicates the SM prediction.}
\label{fig:eft2D}
\end{figure}

One-dimensional scans of the negative log-likelihood value difference to the best fit value, obtained either using the photon \pt distribution in dilepton events alone or from the combined interpretation, are shown in Fig.~\ref{fig:eft1D} for each Wilson coefficient, where the other Wilson coefficient is set to zero in the fit.
The corresponding 68 and 95\% confidence level (\CL) intervals are listed in Table~\ref{tab:eft1D}.
Fits are also performed where both Wilson coefficients are varied simultaneously in the fit.
The result of the two-dimensional scans and the corresponding 68 and 95\% \CL contours are shown in Fig.~\ref{fig:eft2D}.

\begin{table}[!htp]
\centering
\topcaption{Summary of the one-dimensional 68 and 95\% \CL intervals obtained for the Wilson coefficients \ctZ and \ctZI using the photon \pt distribution from this analysis or the combination of this analysis with the \elljets analysis from Ref.~\cite{CMS:2021klw}.
The profiled results correspond to the fits where the other Wilson coefficient is left free in the fit, otherwise it is set to zero.}
\renewcommand\arraystretch{1.2}
\newcommand\tworows[2]{\makebox[0pt]{\parbox{3cm}{\centering#1\\#2\strut}}}
\cmsTable{\begin{tabular}{lllcccc}
    & \multicolumn{2}{c}{\multirow{3}{*}{Wilson coefficient}} & \multicolumn{2}{c}{Dilepton result} & \multicolumn{2}{c}{Dilepton \& \elljets combination} \\
    & &          & 68\% \CL interval  & 95\% \CL interval & 68\% \CL interval  & 95\% \CL interval \\[-3pt]
    & &          & [\LambTevSq]       & [\LambTevSq]      & [\LambTevSq]       & [\LambTevSq]      \\ \hline
    \multirow{4}{*}{\rotatebox{90}{Expected}} & \multirow{2}{*}{\ctZ}
    & $\ctZI=0$  & [$-0.28$, $0.35$]  & [$-0.42$, $0.49$] & [$-0.15$, $0.19$]  & [$-0.25$, $0.29$] \\
    & & profiled & [$-0.28$, $0.35$]  & [$-0.42$, $0.49$] & [$-0.15$, $0.19$]  & [$-0.25$, $0.29$] \\[\cmsTabSkip]
    & \multirow{2}{*}{\ctZI}
    & $\ctZ=0$   & [$-0.33$, $0.30$]  & [$-0.47$, $0.45$] & [$-0.17$, $0.18$]  & [$-0.27$, $0.27$] \\
    & & profiled & [$-0.33$, $0.30$]  & [$-0.47$, $0.45$] & [$-0.18$, $0.18$]  & [$-0.27$, $0.27$] \\[\cmsTabSkip]
    \multirow{4}{*}{\rotatebox{90}{Observed}} & \multirow{2}{*}{\ctZ}
    & $\ctZI=0$  & [$-0.43$, $-0.09$] & [$-0.53$, $0.52$] & [$-0.30$, $-0.13$] & [$-0.36$, $0.31$] \\
    & & profiled & [$-0.43$, $0.17$]  & [$-0.53$, $0.51$] & [$-0.30$, $0.00$]  & [$-0.36$, $0.31$] \\[\cmsTabSkip]
    & \multirow{2}{*}{\ctZI}
    & $\ctZ=0$  &
    \tworows{[$-0.47$, $-0.03$]}{$\cup$\,[$0.07$, $0.38$]} & [$-0.58$, $0.52$] & \tworows{[$-0.32$, $-0.13$]}{$\cup$\,[$0.16$, $0.29$]} & [$-0.38$, $0.36$] \\
    & & profiled & [$-0.43$, $0.33$]  & [$-0.56$, $0.51$] & [$-0.28$, $0.23$]  & [$-0.36$, $0.35$] \\
\end{tabular}}
\label{tab:eft1D}
\end{table}

In the two-dimensional scans, the best fit point, shown by the red diamonds in Fig.~\ref{fig:eft2D}, is found at $(\ctZ,\ctZI)=(-0.28,-0.02)$ for the dilepton-only interpretation, and at $(-0.22,-0.05)$ for the combined interpretation.
The SM expectation of zero for both Wilson coefficients is 0.7 (1.2) standard deviations away from the best fit point for the dilepton-only (combined) interpretation.
The SM expectation is compatible with the 95\% \CL limits for both the one- and two-dimensional scans from this and the combined analysis.

\begin{figure}[!htp]
\centering
\includegraphics[width=0.6\textwidth]{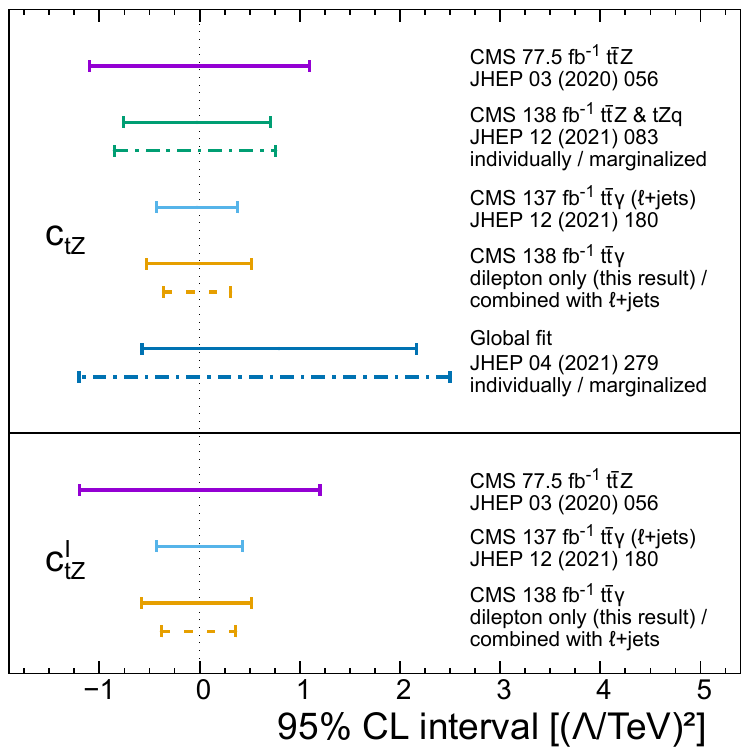}
\caption{Comparison of observed 95\% \CL intervals for the Wilson coefficients \ctZ (upper panel) and \ctZI (lower panel).
For the CMS \ttZ~\cite{CMS:2019too}, \ttG \elljets~\cite{CMS:2021klw} and \ttG dilepton results, the limits are shown for the case where all other considered Wilson coefficients are fixed to zero.
The dashed lines indicate the corresponding result of the \ttG \elljets and dilepton combination.
For the CMS result based on \ttZ and \tZq events~\cite{CMS:2021aly}, as well as from a global fit to the LHC, LEP, and Tevatron data~\cite{Ellis:2020unq}, the limits where all other considered Wilson coefficients are fixed to zero are shown with solid lines, and the marginalized limits from the full fits are shown with dashed-and-dotted lines.}
\label{fig:eftComparison}
\end{figure}

Other measurements~\cite{CMS:2017ugv, ATLAS:2019fwo, CMS:2019too, CMS:2020lrr, CMS:2021aly} and global SMEFT fits~\cite{Hartland:2019bjb, Brivio:2019ius, Ellis:2020unq, Bissmann:2020mfi, Ethier:2021bye} have also derived constraints on these Wilson coefficients.
A comparison of the 95\% \CL limits on \ctZ and \ctZI from a selection of these results and this analysis are shown in Fig.~\ref{fig:eftComparison}.

The CMS \ttZ measurement using an integrated luminosity of 77.5\fbinv~\cite{CMS:2019too} had 95\% \CL limits on \ctZ and \ctZI about three times less strict than the combined limits presented here, while the limit on \ctZ from a recent CMS result on \ttZ and \tZq production using 138\fbinv~\cite{CMS:2021aly} was a factor of two less strict.
Comparing with other experiments, the combined result presented here provides the best limits on the Wilson coefficients \ctZ and \ctZI yet published.

\section{Summary}\label{sec:summary}

A measurement of the inclusive and differential cross sections for top quark pair production in association with a photon (\ttG) has been presented, using 138\fbinv of proton-proton (\pp) collision data at \sqrts recorded with the CMS detector at the LHC.
The analysis is performed in a fiducial phase space defined at the particle level by the requirement of exactly one isolated photon, exactly two oppositely charged leptons, and at least one jet coming from the hadronization of a bottom quark, including the \emu, \ee, and \mumu channels of the \ttbar decay.
The inclusive cross section is extracted with a profile likelihood fit to the transverse momentum distribution of the reconstructed photon, and is measured to be $\sigfid=\XsecMeasuredFull$, in agreement with the standard model (SM) prediction of $\sigSM=\XsecTheoryFull$.

Differential cross sections are measured as functions of various kinematic properties of the photon, leptons, and jets, and unfolded to the particle level.
The comparison to SM predictions is performed using different parton shower algorithms.
No significant deviations from the SM predictions are found.

The measurements are also interpreted in terms of the SM effective field theory.
Constraints are derived on the Wilson coefficients \ctZ and \ctZI describing the modifications of the \ttZ and \ttG interaction vertices, from these results alone and in combination with another CMS measurement of \ttG production using the lepton+jets final state and the same data set.
From the combined interpretation, the best experimental limits on these Wilson coefficients to date are derived.

\begin{acknowledgments}
    We congratulate our colleagues in the CERN accelerator departments for the excellent performance of the LHC and thank the technical and administrative staffs at CERN and at other CMS institutes for their contributions to the success of the CMS effort. In addition, we gratefully acknowledge the computing  centers and personnel of the Worldwide LHC Computing Grid and other  centers for delivering so effectively the computing infrastructure essential to our analyses. Finally, we acknowledge the enduring support for the construction and operation of the LHC, the CMS detector, and the supporting computing infrastructure provided by the following funding agencies: BMBWF and FWF (Austria); FNRS and FWO (Belgium); CNPq, CAPES, FAPERJ, FAPERGS, and FAPESP (Brazil); MES and BNSF (Bulgaria); CERN; CAS, MoST, and NSFC (China); MINCIENCIAS (Colombia); MSES and CSF (Croatia); RIF (Cyprus); SENESCYT (Ecuador); MoER, ERC PUT and ERDF (Estonia); Academy of Finland, MEC, and HIP (Finland); CEA and CNRS/IN2P3 (France); BMBF, DFG, and HGF (Germany); GSRI (Greece); NKFIA (Hungary); DAE and DST (India); IPM (Iran); SFI (Ireland); INFN (Italy); MSIP and NRF (Republic of Korea); MES (Latvia); LAS (Lithuania); MOE and UM (Malaysia); BUAP, CINVESTAV, CONACYT, LNS, SEP, and UASLP-FAI (Mexico); MOS (Montenegro); MBIE (New Zealand); PAEC (Pakistan); MSHE and NSC (Poland); FCT (Portugal); JINR (Dubna); MON, RosAtom, RAS, RFBR, and NRC KI (Russia); MESTD (Serbia); MCIN/AEI and PCTI (Spain); MOSTR (Sri Lanka); Swiss Funding Agencies (Switzerland); MST (Taipei); ThEPCenter, IPST, STAR, and NSTDA (Thailand); TUBITAK and TAEK (Turkey); NASU (Ukraine); STFC (United Kingdom); DOE and NSF (USA).

    \hyphenation{Rachada-pisek} Individuals have received support from the Marie-Curie program and the European Research Council and Horizon 2020 Grant, contract Nos.\ 675440, 724704, 752730, 758316, 765710, 824093, 884104, and COST Action CA16108 (European Union); the Leventis Foundation; the Alfred P.\ Sloan Foundation; the Alexander von Humboldt Foundation; the Belgian Federal Science Policy Office; the Fonds pour la Formation \`a la Recherche dans l'Industrie et dans l'Agriculture (FRIA-Belgium); the Agentschap voor Innovatie door Wetenschap en Technologie (IWT-Belgium); the F.R.S.-FNRS and FWO (Belgium) under the ``Excellence of Science -- EOS" -- be.h project n.\ 30820817; the Beijing Municipal Science \& Technology Commission, No. Z191100007219010; the Ministry of Education, Youth and Sports (MEYS) of the Czech Republic; the Deutsche Forschungsgemeinschaft (DFG), under Germany's Excellence Strategy -- EXC 2121 ``Quantum Universe" -- 390833306, and under project number 400140256 - GRK2497; the Lend\"ulet (``Momentum") Program and the J\'anos Bolyai Research Scholarship of the Hungarian Academy of Sciences, the New National Excellence Program \'UNKP, the NKFIA research grants 123842, 123959, 124845, 124850, 125105, 128713, 128786, and 129058 (Hungary); the Council of Science and Industrial Research, India; the Latvian Council of Science; the Ministry of Science and Higher Education and the National Science Center, contracts Opus 2014/15/B/ST2/03998 and 2015/19/B/ST2/02861 (Poland); the Funda\c{c}\~ao para a Ci\^encia e a Tecnologia, grant CEECIND/01334/2018 (Portugal); the National Priorities Research Program by Qatar National Research Fund; the Ministry of Science and Higher Education, projects no. 0723-2020-0041 and no. FSWW-2020-0008, and the Russian Foundation for Basic Research, project No.19-42-703014 (Russia); MCIN/AEI/10.13039/501100011033, ERDF ``a way of making Europe", and the Programa Estatal de Fomento de la Investigaci{\'o}n Cient{\'i}fica y T{\'e}cnica de Excelencia Mar\'{\i}a de Maeztu, grant MDM-2017-0765 and Programa Severo Ochoa del Principado de Asturias (Spain); the Stavros Niarchos Foundation (Greece); the Rachadapisek Sompot Fund for Postdoctoral Fellowship, Chulalongkorn University and the Chulalongkorn Academic into Its 2nd Century Project Advancement Project (Thailand); the Kavli Foundation; the Nvidia Corporation; the SuperMicro Corporation; the Welch Foundation, contract C-1845; and the Weston Havens Foundation (USA).
\end{acknowledgments}

\bibliography{auto_generated}
\cleardoublepage \appendix\section{The CMS Collaboration \label{app:collab}}\begin{sloppypar}\hyphenpenalty=5000\widowpenalty=500\clubpenalty=5000\cmsinstitute{Yerevan~Physics~Institute, Yerevan, Armenia}
A.~Tumasyan
\cmsinstitute{Institut~f\"{u}r~Hochenergiephysik, Vienna, Austria}
W.~Adam\cmsorcid{0000-0001-9099-4341}, J.W.~Andrejkovic, T.~Bergauer\cmsorcid{0000-0002-5786-0293}, S.~Chatterjee\cmsorcid{0000-0003-2660-0349}, K.~Damanakis, M.~Dragicevic\cmsorcid{0000-0003-1967-6783}, A.~Escalante~Del~Valle\cmsorcid{0000-0002-9702-6359}, R.~Fr\"{u}hwirth\cmsAuthorMark{1}, M.~Jeitler\cmsAuthorMark{1}\cmsorcid{0000-0002-5141-9560}, N.~Krammer, L.~Lechner\cmsorcid{0000-0002-3065-1141}, D.~Liko, I.~Mikulec, P.~Paulitsch, F.M.~Pitters, J.~Schieck\cmsAuthorMark{1}\cmsorcid{0000-0002-1058-8093}, R.~Sch\"{o}fbeck\cmsorcid{0000-0002-2332-8784}, D.~Schwarz, S.~Templ\cmsorcid{0000-0003-3137-5692}, W.~Waltenberger\cmsorcid{0000-0002-6215-7228}, C.-E.~Wulz\cmsAuthorMark{1}\cmsorcid{0000-0001-9226-5812}
% ...
\vskip\cmsinstskip
% ...
1:~Also at TU Wien, Wien, Austria\\
% ...
\end{sloppypar}
\end{document}